\documentclass[letterpaper]{JHEP3}

\newcommand{\beq}{\begin{equation}}
\newcommand{\eeq}{\end{equation}}

\usepackage{graphicx}
\usepackage{amsmath}
\usepackage{amsfonts}
\usepackage{cite}

\title{A new approach to static numerical relativity, and its application to Kaluza-Klein black holes}

\author{Matthew Headrick \\
Martin Fisher School of Physics, Brandeis University, Waltham MA 02454, USA \\
\email{mph@brandeis.edu}
}

\author{Sam Kitchen and Toby Wiseman \\
Theoretical Physics Group, Blackett Laboratory, Imperial College, London SW7 2AZ, UK \\
\email{sam.kitchen03@imperial.ac.uk}, \email{t.wiseman@imperial.ac.uk}
}

\preprint{BRX-TH 608}

\date{May 2009}

\abstract{
We propose a framework for solving the Einstein equation for static and Euclidean metrics. 
First, we address the issue of gauge-fixing by borrowing from the Ricci-flow literature the so-called DeTurck trick, which renders the Einstein equation strictly elliptic and generalizes the usual harmonic-coordinate gauge. We then study two algorithms, Ricci-flow and Newton's method, for solving the resulting Einstein-DeTurck equation. We illustrate the use of these methods by studying localized black holes and non-uniform black strings in five-dimensional Kaluza-Klein theory, improving on previous calculations of their thermodynamic and geometric properties. We study spectra of various operators for these solutions, in particular finding the negative modes of the Lichnerowicz operator. We classify the localized solutions into two branches that meet at a minimum temperature. We find good evidence for a merger between the localized and non-uniform solutions. We also find a narrow window of localized solutions that possess negative modes yet have positive specific heat.
}

\begin{document}

%
\section{Introduction}
%

In four-dimensional general relativity, the space of physically relevant geometries, and the methods for finding them, are well understood. The stationary solutions are privileged, being possible end states of dynamical processes, and in vacuum the uniqueness theorems imply that the most general solution is the Kerr black hole, whose solution is known analytically. The problem of constructing dynamical spacetimes is in general a numerical one, as exact solutions are often unknown for situations of interest, in particular those with few symmetries such as generic gravitational collapse or black hole mergers. The tools for solving the dynamical Einstein equations are well understood and have enjoyed intense study both theoretically and computationally.
 
The consideration of general relativity in more than four dimensions can be motivated in a variety of ways. Such studies date back almost a century to Kaluza and Klein, and in a modern context are given weight by string theory and other phenomenological and formal considerations, such as Randall-Sundrum and large-extra-dimensions scenarios \cite{ADD,ADD2,Randall:1999ee,Randall:1999vf} and the AdS-CFT correspondence \cite{AdSCFTreview}. However the space of static and stationary solutions is considerably more complicated in these contexts, and many solutions known or believed to exist currently have no analytic solution.

For asymptotically flat $D$-dimensional gravity in the absence of matter, with $D>4$, static black holes are given uniquely by the Schwarzschild-Tangherlini black hole \cite{Tangherlini:1963bw,Gibbons:2002av,Gibbons:2002bh}. However, for stationary solutions in $D = 5$, in addition to the Myers-Perry solution which generalizes Kerr \cite{MP}, the Emparan-Reall black ring is found \cite{Ring}; these may even be combined into a `black Saturn'  \cite{Elvang:2007rd,Elvang:2007hg}. Whilst these are understood analytically, in $D > 5$ the black ring solution is not known and it is thought that the structure of stationary solutions is yet more complicated than in $D = 5$ \cite{Emparan:2003sy,Emparan:2007wm}, with little hope of admitting closed-form solutions.

Now consider Kaluza-Klein theory, by which we mean gravity in five dimensions such that the asymptotic geometry is a product of four-dimensional Minkowski space and a circle. This is the canonical toy model for theories with compact dimensions. In this case the space of static vacuum solutions, which need not have an isometry in the compact direction, is not analytically understood, and the solutions that have been found numerically so far have a rich structure \cite{Kol:2004ww,Harmark:2005pp,Harmark:2007md}.\footnote{In the simplest examples of the AdS-CFT correspondence the situation is the same, with the $5$-sphere playing the role of the Kaluza-Klein circle, and an analogous structure of solutions is thought to exist \cite{Hubeny:2002xn,Delsate:2008kw}. The structure of static solutions here is particularly interesting, as the phase diagram of the black holes is dual to the phase diagram of the finite temperature gauge theory, and gravitational phase transitions can be translated into field theory ones  \cite{Witten:1998zw,Susskind:1997dr,Li:1998jy,Martinec:1998ja,Aharony:2004ig,Harmark:2004ws}. Even black holes that are homogeneous on the $5$-sphere may have a complicated form, such as the `plasma-ball' black holes in simple confining models of AdS-CFT \cite{Aharony:2005bm}, which have interesting links to formalizing the membrane paradigm \cite{Cardoso:2006ks,Caldarelli:2008mv}.} A numerical approach exists in the literature to find cohomogenity-two static vacuum metrics, including black holes, employed  first in the context of brane worlds \cite{star} and then Kaluza-Klein theory to find non-uniform and localized black holes \cite{Wiseman:2002zc,Kudoh:2003ki,Sorkin:2003ka}. 
A similar approach was used independently in four-dimensional Einstein-Yang-Mills systems in the earlier work of \cite{Kleihaus:1996vi,Kleihaus:1997ic} (including extension to four-dimensional stationary systems \cite{Kleihaus:2000kg}) although the `constraint equations' and their consistency was not discussed. The method has been used effectively in a variety of contexts \cite{Kudoh:2003xz,Kudoh:2003vg,Kudoh:2004kf,Sorkin:2006wp,Kleihaus:2006ee,Kleihaus:2007cf,Yoshino:2008rx,Delsate:2009bd}, but is fundamentally limited in that it requires cohomogenity two, a particular conformal coordinate system, and a certain `wizardry' in actually making it work---for example, the `trick' employed when treating a symmetry axis \cite{star,Kol:2004ww}. 

In Kaluza-Klein theory, the static vacuum solution that preserves four-dimensional Poincar\'e invariance is trivial to write down analytically. However with additional compact dimensions even this is not true. For example, in string theory phenomenology the vacua of the theory that preserve supersymmetry typically involve products of four-dimensional Minkowski space with complicated manifolds such as Calabi-Yaus. In the supergravity approximation, and in the simplest cases with trivial fluxes and dilaton, the metric on the compact Calabi-Yau should be Ricci-flat, and such metrics are not known analytically. In general, finding four-dimensional Poincar\'e invariant vacua can be thought of as the problem of solving the Euclidean Einstein equation on the compactification manifold with specific matter. The Euclidean problem of finding Calabi-Yau metrics that are Ricci-flat has received attention recently, with an approach based on Ricci-flow \cite{Headrick:2005ch} and a spectral approach based on an algebraic representation of the metric \cite{Donaldson,Douglas:2006rr,Douglas:2006hz,Braun:2008jp}.\footnote{The related case of Einstein metrics on del Pezzo surfaces has also been studied \cite{Doran:2007zn,Keller}. } However, these are fundamentally tied to working with K\"ahler geometries, and using the associated complex coordinates.

A natural question is then whether there exist numerical methods to find general stationary and static black holes in more than four dimensions, and to find solutions to the Euclidean Einstein equation such as those governing phenomenological vacuum compactifications in string theory. In these higher dimensional settings the dynamical numerical methods that are so well explored in four dimensions simply generalize. Thus in principle a numerical algorithm exists: one chooses some initial data whose late-time evolution asymptotes to our stationary or static vacuum solution of interest. The first point to make is that this will never allow one to find dynamically unstable solutions, which may be of interest for theoretical reasons. However, the second more serious obstruction to this being a practical method is that dynamical evolutions, particularly involving horizons, are very complicated, partly due to the constraints that must be initially solved---already a hard elliptic problem in general---and then ensuring that these constraints are properly preserved under evolution. For these reasons we believe that such a dynamical approach is not a viable one in practice.

The challenge then is to find a more direct but unified method to solve the Einstein equation in stationary, static, and Euclidean contexts. In this paper we provide a partial answer to this challenge. We present a general covariant framework in which to address the problem of finding static and Euclidean vacuum solutions in arbitrary dimensions, and with no particular symmetry properties (i.e.\ any cohomogenity). The geometry may contain no horizon, or one horizon component, in which case the algorithm only yields the spacetime exterior to the horizon. We employ the fact that such static Lorentzian spacetimes, exterior to a horizon, may be analytically continued in time to smooth Euclidean geometries. The Lorentzian Killing vector field continues to a vector field which generates a $U(1)$ isometry, and vanishes only at the point that continues to the Lorentzian horizon. Thus the problem of finding static vacuum solutions without horizon, or exterior to a single horizon, is included in the general  Ricci-flat smooth Euclidean problem. The heart of the paper is to provide an algorithm to find smooth Ricci-flat Euclidean geometries in a fully covariant manner by directly solving the Ricci-flatness condition.

We begin, in Section 2, by finding a general and covariant method for gauge-fixing that generalizes the harmonic-coordinate gauge condition. Specifically, we borrow from the study of Ricci-flow the so-called DeTurck trick \cite{MR697987}, in which, using a fixed reference connection, the Ricci-flow equation is rendered strictly parabolic without changing its geometric content. In our context the DeTurck trick renders the Einstein equation into a strictly elliptic system of partial differential equations, which we refer to as the Einstein-DeTurck equation.
In Section 3, we then consider two types of algorithms for solving this elliptic system.

The simplest algorithm is to exploit the ellipticity of the Einstein-DeTurck equation and solve it via a diffusive flow. The canonical flow is Ricci-DeTurck flow (diffeomorphic to Ricci-flow), and the idea is to provide an initial guess and then simulate the flow, which hopefully asymptotes to the desired Ricci-flat fixed point. However for vacuum black hole spacetimes the Ricci-flat solution may not be an attractive fixed point of Ricci-flow, as such black holes usually possess  Euclidean negative modes \cite{Headrick:2006ti}---in vacuum we expect that any black hole with a negative specific heat capacity is likely to have one or more negative modes \cite{Whiting:1988qr,Whiting:1988ge,Prestidge:1999uq,Reall:2001ag,Gregory:2001bd}. Other interesting Euclidean geometries also possess negative modes, and show analogous behaviour to black holes under Ricci-flow \cite{Holzegel:2007zz}. We therefore present an algorithm in which a finite-dimensional family of initial guess geometries must be chosen, and the parameters of this family varied in order to reach the fixed point. Since the Ricci-DeTurck flow is diffeomorphic to Ricci-flow, which provides a flow not only in the space of metrics but also in the space of diffeomorphism classes (or `geometries'), this method has the attractive feature that the algorithm is independent of the choice of reference connection. We denote this approach the `Ricci-flow method'.

A more straightforward approach to solve an elliptic PDE is to iteratively linearize it, i.e.\ apply Newton's method. The point here is that there exist direct methods to solve the resulting linear PDE. For example, using real-space finite differencing this PDE appears as a sparse system of linear equations, for which fast solvers exist such as the biconjugate gradient method. Such solvers are not affected by negative modes (although they are by zero modes).
Whilst more complicated to implement in practice, Newton's method avoids having to tune families of initial data---just one guess metric in the basin of attraction of the desired solution is sufficient. Unlike the Ricci-flow method, the trajectory of Newton's method in the space of geometries does depend on the reference connection, and hence so does the basin of attraction of a solution. In principle this makes the method somewhat less elegant than Ricci-flow, although in practice it appears to make little difference.

In summary, finding Euclidean Ricci-flat geometries and static vacuum Lorentzian solutions, including black holes, can be phrased as the same problem and solved using the same unified, covariant elliptic approach. The problem of dealing with Euclidean negative modes in either context can be overcome either by using Ricci-flow and tuning initial data, or by using Newton's method. In the Lorentzian context, solutions may be found independently of their dynamical stability properties. Whilst these methods do not assume any isometries, if such isometries exist, it is natural to exploit them to effectively reduce the dimensionality of the problem.

It is instructive to see the two methods described in Section \ref{sec:algorithms} demonstrated in a non-trivial example. Thus in Section \ref{sec:KK} we test both these methods by finding the various static black holes in five-dimensional Kaluza-Klein theory---the `localized black holes' and `non-uniform black strings'.  We find that Newton's method gives excellent results in practice and is considerably easier to use than the Ricci-flow method with its tuning of initial data. We compute various properties of these black holes, such are  their thermodynamic properties and horizon geometry. We are able to provide good evidence for a merger between the localized and non-uniform solutions \cite{Kol:2004ww,Harmark:2005pp,Harmark:2007md}, computing the localized branch closer to the potential merger point than previously done in \cite{Kudoh:2004hs} even though we have used more modest resources. 

New results include finding that  the localized solutions have a minimum temperature which divides the localized solutions into two branches which we term the \emph{small} and \emph{large} localized black holes.\footnote{This nomenclature follows that of black holes in AdS \cite{Hawking:1982dh} or in a cavity \cite{York:1986it} which also possess a minimum temperature. However the thermodynamic behaviour of the localized black holes is rather different from these cases, with the large localized solutions having negative specific heat capacity unlike for AdS or a cavity.} The Euclidean negative modes of the localized and non-uniform solutions are computed in the class of eigenfunctions which preserve the isometry of the background, and we find small localized solutions have a single negative mode, and the large localized and non-uniform strings computed have two. These negative modes are consistent with merger of the large localized and the non-uniform string branches, with one eigenvalue remaining finite and the other apparently diverging to minus infinity at the merger point. We also provide additional evidence for merger by considering low lying positive eigenvalues of various operators. Perhaps the most intriguing result is the observation of a rather narrow window of small localized solutions that appear to have positive specific heat capacity, but possess a negative mode. This will require more rigorous numerical work to confirm. Assuming it holds true, this provides the first static counter-example to any claim that one might be able to reverse the link between local thermodynamic stability and the existence of negative modes discussed in \cite{Whiting:1988qr, Whiting:1988ge,Prestidge:1999uq,Reall:2001ag,Gregory:2001bd}, complementing the interesting recent stationary counter-example of \cite{Monteiro:2009tc}.

%
\section{The Einstein-DeTurck equation}
\label{sec:setup}
%

We wish to solve the vacuum Einstein equation,
\begin{equation}\label{Einstein}
R_{\mu\nu} = 0\,,
\end{equation}
for a Euclidean-signature metric on a given $D$-dimensional manifold $M$, with appropriate boundary conditions if $M$ has a boundary. For simplicity and concreteness, we are restricting our attention to the vacuum Einstein equation with zero cosmological constant. However, most of our considerations will be local, and including a non-zero cosmological constant and matter fields should not present any significant additional difficulties.

As we will explain below, equation \eqref{Einstein} is \emph{weakly elliptic}; roughly speaking, it is elliptic for the physical degrees of freedom in the metric. Most methods for solving elliptic PDEs are relaxation methods, meaning that one starts with an initial guess for the unknown function, which presumably does not solve the PDE, and then iteratively deforms it to obtain functions that solve it to a better and better approximation. This is to be contrasted with methods for solving initial-value problems, associated with hyperbolic or parabolic PDEs, which usually proceed by building up the solution time-slice by time-slice.

Before being able to apply any of the standard numerical methods for elliptic PDEs, one must deal with the issue of gauge invariance. The problem is that \eqref{Einstein} is underdetermined. Naively it appears to consist of $D(D+1)/2$ equations, per point, for $D(D+1)/2$ unknowns (the metric components). However, as a consequence of diffeomorphism invariance, the Ricci tensor for any metric satisfies the Bianchi identity, $\nabla^\mu R_{\mu\nu} - \frac12\partial_\nu R = 0$, which has $D$ components. Hence \eqref{Einstein} actually represents only $D(D-1)/2$ non-trivial equations. Stated more precisely, the principal symbol of the Ricci tensor (considered as a nonlinear operator on the metric) annihilates pure gauge modes of the metric (corresponding to small diffeomorphisms), while it is positive definite on physical modes. This is the technical meaning of ``weakly elliptic".

The same issue, of course, arises in the Lorentzian context, where the Einstein equation is only weakly hyperbolic, and our discussion will to a large extent parallel the standard treatment of the latter problem using harmonic coordinates (see for example section 10.2 of \cite{Wald}). As we will explain in subsection 2.1, the weak ellipticity makes the Einstein equation unsuitable for numerical solution without some form of gauge fixing. Of course, there are many different possible gauge-fixing schemes, many of them adapted to particular problems. Here we are looking for one that can be applied as generally as possible. A local analysis of the problem of gauge-fixing leads us to a generalized form of harmonic gauge that is somewhat more covariant than the textbook version. Then in subsection 2.2 we advocate an \emph{a posteriori} method of implementing the gauge-fixing, in which the gauge-fixing condition is not imposed from the outset but rather solved for at the same time as the Einstein equation. Specifically, we propose solving not \eqref{Einstein} itself but another PDE (see \eqref{EDT}) that we call the Einstein-DeTurck equation. The Einstein-DeTurck equation simultaneously encodes the Einstein equation and the gauge-fixing condition, and is strictly elliptic. In the next section, we will discuss algorithms for solving the Einstein-DeTurck equation.

\subsection{Gauge condition}

If we make a small perturbation $\delta g_{\mu\nu} = h_{\mu\nu}$ to the metric, the change in the Ricci tensor is, to first order,
\begin{equation}\label{deltaR}
\delta R_{\mu\nu} \equiv \Delta_Rh_{\mu\nu} = 
\Delta_Lh_{\mu\nu} + \nabla_{(\mu}v_{\nu)}\,,
\end{equation}
where
\begin{equation}\label{Lichvdef}
\Delta_Lh_{\mu\nu} \equiv 
-\frac12\nabla^2h_{\mu\nu} - R_\mu{}^\kappa{}_\nu{}^\lambda h_{\kappa\lambda} + {R_{(\mu}}^\kappa h_{\nu)\kappa} \,,\qquad v_\mu \equiv \nabla_\nu {h^\nu}_\mu - \frac12\partial_\mu h\,.
\end{equation}
For the moment we are concerned with the local properties of $\Delta_R$, so we consider perturbations with wavelength much smaller than any curvature scale of the metric, and therefore take $g_{\mu\nu}$ to be constant; this yields the principal symbol $\Delta_R^{\rm principal}$ (the part of $R_{\mu\nu}$ with only second derivatives of the metric):
\begin{equation}
\Delta_R^{\rm principal}h_{\mu\nu} =
- \frac12\nabla^2h_{\mu\nu} - \frac12\partial_\mu\partial_\nu h + \partial_\lambda\partial_{(\nu} {h_{\mu)}}^\lambda\,.
\end{equation}
Any such perturbation can be decomposed as $h_{\mu\nu} = \hat h_{\mu\nu} + \partial_{(\mu}w_{\nu)}$, where $\partial_\nu\hat h^\nu{}_\mu -\frac12\partial_\mu\hat h = 0$; $\hat h_{\mu\nu}$ is the transverse (physical) part, and $\partial_{(\mu}w_{\nu)}$ is the longitudinal (pure gauge) part. Specifically, the perturbation $\partial_{(\mu}w_{\nu)}$ is the change in the metric under the small diffeomorphism generated by the vector field $w^\mu$, and is annihilated by $\Delta_R^{\rm principal}$. On the other hand, $\Delta_R^{\rm principal}$ acts as $-\frac12\nabla^2$ on $\hat h_{\mu\nu}$.

We will now briefly explain why the weak ellipticity of \eqref{Einstein} is problematic for numerical purposes. For concreteness, suppose that we represent the metric by the values of its components on a set of lattice points in each coordinate patch (the argument does not change if we use some other representation). In general, to be representable the metric components must be smooth on the (coordinate) scale of the lattice spacing. However, this smoothness condition is not preserved by diffeomeorphisms. Thus, even if our lattice is sufficiently fine to represent some particular metric that solves \eqref{Einstein}, other metrics in its diffeomorphism class will not be representable. (In some sense, ``most" such metrics will not be.) Therefore we need to control which representative of the diffeomorphism class we aim for. What will go wrong in practice if we do not? During the relaxation process, since short-wavelength pure-gauge modes are not damped out, small errors will accumulate, leading to a loss of control.

There are of course many different ways to fix a gauge in general relativity. One can fix certain metric components; for example, in the presence of spherical symmetry one can set the coordinate $r$ to equal the proper radius of the sphere. Alternatively, one can take advantage of certain special properties of the metric. If one is solving for a K\"ahler metric, then one can choose to employ complex coordinates; the only allowed gauge transformations are then holomorphic diffeomorphisms, which have no local degrees of freedom. (This is actually a special case of harmonic coordinates, which we will discuss below.) Our aim in this paper, however, is to find an all-purpose method for solving \eqref{Einstein}, one that requires as little special knowledge of the metric one is trying to find as possible.

Let us return to our local analysis. We wish to forbid the pure-gauge fluctuations; in other words we wish to require $\partial_\nu {h^\nu}_\mu - \frac12\partial_\mu h = 0$. We need to de-linearize this equation, i.e.\ to find a one-form $\xi_\mu$ depending on the metric (and of course \emph{not} diffeomorphism-invariant) whose variation $\delta\xi_\mu$ about a constant $g_{\mu\nu}$ equals $\partial_\nu {h^\nu}_\mu - \frac12\partial_\mu h$. Fixing $\xi_\mu$ will then project out the pure-gauge modes. The simplest such one-form is
\begin{equation}\label{xidef1}
\xi_\mu = g^{\lambda\nu}\left(
\partial_\lambda g_{\nu\mu} - \frac12\partial_\mu g_{\lambda\nu}
\right)\,.
\end{equation}
Here are three expressions for the dual vector field $\xi^\mu$:
\begin{equation}
\xi^\mu = g^{\lambda\nu}\Gamma^\mu_{\lambda\nu} = -\nabla^2x^\mu = -g^{-1/2}\partial_\nu\left(g^{1/2}g^{\mu\nu}\right) \,,
\end{equation}
where $x^\mu$ are the coordinates considered as scalar functions and $\nabla^2$ is the scalar Laplacian. The gauge choice $\xi^\mu = 0$ is the \emph{harmonic gauge}, commonly employed in Lorentzian numerical relativity.

Although it does the job, the harmonic gauge is a bit inelegant insofar as it refers to a particular coordinate system. In particular, $\xi_\mu$ transforms in a complicated way on chart overlaps, and a metric that is in harmonic gauge in one patch will generally not be in that gauge in a neighboring one. This problem is easily fixed by replacing the coordinate derivatives in \eqref{xidef1} with any fixed covariant derivative. We thus choose an arbitrary background metric $\tilde g_{\mu\nu}$ on $M$, and define
\begin{equation}\label{xidef2}
\xi_\mu = g^{\lambda\nu}\left(
\tilde\nabla_\lambda g_{\nu\mu} - \frac12\tilde\nabla_\mu g_{\nu\lambda}\right), \qquad
\xi^\mu = g^{\lambda\nu}(\Gamma^\mu_{\lambda\nu} - \tilde\Gamma^\mu_{\lambda\nu}) = 
-\left(\frac g{\tilde g}\right)^{-1/2}\tilde\nabla_\nu\left(\left(\frac g{\tilde g}\right)^{1/2}g^{\mu\nu}\right),
\end{equation}
where $\tilde\Gamma^\mu_{\lambda\nu}$ is the Levi-Civita connection for $\tilde g_{\mu\nu}$ and $\tilde\nabla_\mu$ is the associated covariant derivative.\footnote{Note that this is different from the gauge choice referred to as ``generalized harmonic coordinates" in the numerical relativity literature \cite{Garfinkle:2001ni,Pretorius:2005gq}, which involves setting $\nabla^2 x^\mu = H^\mu$, where $H^\mu$ is a fixed vector field. Of course one could combine the two generalizations by setting $\xi^\mu = H^\mu$.} In the theory of harmonic maps between Riemannian manifolds, $\xi^\mu$ is referred to as the \emph{tension field} of the identity map from $(M,g_{\mu\nu})$ to $(M,\tilde g_{\mu\nu})$, and the gauge-fixing condition
\begin{equation}\label{gfc}
\xi^\mu = 0
\end{equation}
is equivalent to the requirement that the identity map be harmonic.

Ideally a gauge-fixing condition should intersect each gauge orbit exactly once: given a metric $\hat g_{\mu\nu}$, there should be a diffeomorphism (unique up to isometries) that maps it to a metric $g_{\mu\nu}$ satisfying \eqref{gfc}. Such a diffeomorphism can be composed with the above identity map from $(M,g_{\mu\nu})$ to $(M,\tilde g_{\mu\nu})$ to obtain a harmonic map from $(M,\hat g_{\mu\nu})$ to $(M,\tilde g_{\mu\nu})$. In physicists' language, this harmonic map is a solution to the classical equation of motion for the sigma model from $(M,\hat g_{\mu\nu})$ into $(M,\tilde g_{\mu\nu})$:
\begin{equation}\label{sigmamodel}
S[y] = \int_M \hat g^{1/2}\hat g^{\mu\nu}\tilde g_{\alpha\beta}(y)\partial_\mu y^\alpha\partial_\nu y^\beta\,,
\end{equation}
where $y^\alpha$ are the coordinates on the target space $(M,\tilde g_{\mu\nu})$. At the local level, it is easy to see the existence and uniqueness of solutions to \eqref{gfc}. Under a small diffeomorphism, generated by a vector field $w^\mu$, the change in $\xi^\mu$ is $\delta\xi^\mu = -\Delta_Vw^\mu$, where
\begin{equation}\label{deltaxi}
\Delta_Vw^\mu \equiv -\frac12\nabla^2w^\mu - \frac12{R^\mu}_\nu w^\nu + (\Gamma^\mu_{\lambda\nu} - \tilde\Gamma^\mu_{\lambda\nu})\nabla^\lambda w^\nu\,.
\end{equation}
$\Delta_V$ is a strictly elliptic operator, so locally $\xi^\mu + \delta\xi^\mu = 0$ will have a unique solution for $w^\mu$. The global existence and uniqueness problems are more difficult, of course. On a compact manifold the operator defined by \eqref{deltaxi} will have at most a finite-dimensional kernel, so the space of harmonic maps is at most finite-dimensional (see also \cite{EellsLemaire}). As far as existence is concerned, there is no general proof, but there are many partial results (including for the case of manifolds with boundary); for example, Eells and Sampson \cite{EellsSampson} proved existence under the assumption of non-positive sectional curvatures for $\tilde g_{\mu\nu}$. As far as we are aware, there are no known counterexamples.\footnote{See \cite{EellsBook} for a summary of the literature on harmonic maps through 1991.}

\subsection{A posteriori gauge fixing}

We now turn to the problem of implementing our gauge-fixing condition $\xi^\mu = 0$. This is a differential equation for the metric, in contrast to algebraic gauge-fixing conditions such as fixing the values of certain metric components. In the latter case, when implementing a relaxation method, one may simply impose the gauge-fixing condition from the outset. Assuming the gauge fixing is complete, the Einstein equation will be a strictly elliptic operator on the space of metrics obeying the gauge-fixing condition. In our case, however, the space of gauge-fixed metrics is not so easy to describe explicitly.

Instead of imposing the gauge-fixing condition a priori, we therefore propose to solve $\xi^\mu = 0$ simultaneously with $R_{\mu\nu} = 0$. This can be done quite elegantly by combining the two equations as follows:
\begin{equation}\label{EDT}
R^H_{\mu\nu} \equiv R_{\mu\nu} - \nabla_{(\mu}\xi_{\nu)} = 0\,,
\end{equation}
which we will refer to as the \emph{Einstein-DeTurck equation}. The tensor $R^H_{\mu\nu}$ appears in work of DeTurck \cite{MR697987}; he gave a simple proof of short-time existence for the Ricci-flow, $\partial g_{\mu\nu}/\partial\lambda = - 2R_{\mu\nu}$, by pointing out that it is diffeomorphically equivalent to the strictly parabolic PDE $\partial g_{\mu\nu}/\partial\lambda = -2R^H_{\mu\nu}$. Roughly speaking, the term $\nabla_{(\mu}\xi_{\nu)}$ supplies the longitudinal components that are missing in the Ricci tensor. The linearization of $R^H_{\mu\nu}$ is
\begin{equation}\label{deltaRH}
\delta R^H_{\mu\nu} \equiv \Delta_Hh_{\mu\nu} = 
\Delta_Lh_{\mu\nu} -\frac12\pounds_\xi h_{\mu\nu} + \nabla_{(\mu}u_{\nu)} \,, \qquad
u^\mu \equiv (\Gamma^\mu_{\nu\lambda}-\tilde\Gamma^\mu_{\nu\lambda})h^{\nu\lambda} \,.
\end{equation}
The second and third terms on the right-hand side do not contain second derivatives of $h_{\mu\nu}$, so the principal symbol of $R^H_{\mu\nu}$ is simply $-\frac12\nabla^2h_{\mu\nu}$, and \eqref{EDT} is strictly elliptic. In effect, rather than projecting out the pure-gauge modes (which would be the result of working in the space of gauge-fixed metrics), we have given them a kinetic term, putting them on the same footing with the physical transverse modes.

We now turn to the question, does a solution to \eqref{EDT} necessarily solve \eqref{Einstein}? We begin with a local analysis. Equation \eqref{EDT} implies
\begin{equation}\label{xieqn}
\nabla^\mu R^H_{\mu\nu} - \frac12\partial_\nu R^H = 
-\frac12\left(\nabla^2\xi_\nu + {R_\nu}^\mu\xi_\mu\right) = 
0\,,
\end{equation}
which locally has only the trivial solution $\xi^\mu = 0$. Thus $R^H_{\mu\nu} = 0$, which is $D(D+1)/2$ equations for the same number of metric components, is locally equivalent to the $D(D-1)/2$ equations of $R_{\mu\nu} = 0$ and the $D$ equations of $\xi^\mu = 0$.

As usual, the global question is more difficult. However, Perelman has shown that, on a compact manifold without boundary, \eqref{EDT} implies \eqref{Einstein} \cite{Perelman1}. (In Perelman's theorem, $\xi_\mu$ is a general one-form; it does not refer to the particular form \eqref{xidef2}.) It would be interesting to generalize his theorem to manifolds with boundary. (One would have to impose some boundary condition on $\xi^\mu$, such as the vanishing of its normal component. Otherwise one can easily construct a counterexample by taking any compact region of any of the known non-compact steady Ricci solitons.) The theorem also covers the case with negative cosmological constant: $R_{\mu\nu} - \nabla_{(\mu}\xi_{\nu)} = \kappa g_{\mu\nu}$, with $\kappa$ a negative constant, implies $R_{\mu\nu} = \kappa g_{\mu\nu}$. On the other hand, for positive $\kappa$ there are known counterexamples (called \emph{shrinking Ricci solitons}), for example on the first and second del Pezzo surfaces \cite{MR1145263, MR2084775,MR2243675,HeadrickdP2}. However, the existence of such metrics should not present a major problem as long as they are isolated, since it is easy enough to check whether one's algorithm is converging to a true solution to the Einstein equation or merely to a soliton, and in the latter case to restart it with a different initial guess. (Similar comments presumably apply in the presence of matter fields, but we have not studied this case in detail.)

Let us now discuss what boundary conditions are appropriate for equation \eqref{EDT}. It is well known that the Einstein equation, being effectively $D(D-1)/2$ equations for the same number of variables, requires the same number of boundary conditions. For example, one can fix the induced metric on the boundary or its extrinsic curvature, which are respectively Dirichlet and Neuman boundary conditions for the tangential components of the metric. For the Einstein-DeTurck equation, we have an additional $D$ equations and therefore need an additional $D$ boundary conditions. Since we want to obtain a solution with $\xi^\mu=0$, and since \eqref{xieqn} is strictly elliptic, it seems clear that we should require $\xi^\mu|_{\partial M} = 0$. If we denote the normal and tangential components of $\xi^\mu$ by $\xi^n$ and $\xi^i$ respectively, then it is easy to show that $\xi^n|_{\partial M} = 0$ fixes the normal derivative of $g_{nn}$, while $\xi^i|_{\partial M} = 0$ fixes the normal derivative of $g_{ni}$, in terms of their values, and the values and derivatives of the other metric components. (This is the same issue as imposing boundary conditions for the parabolic PDE mentioned above, $\partial g_{\mu\nu}/\partial\lambda = -2R^H_{\mu\nu}$. In \cite{Headrick:2006ti} it was argued that $\xi^\mu|_{\partial M} = 0$ are appropriate boundary conditions for that PDE, and uniqueness has been proved in that context \cite{Holzegel:2007zz}.)

%
\section{Algorithms for solving the Einstein-DeTurck equation}
\label{sec:algorithms}
%

The Einstein-DeTurck equation \eqref{EDT} is a strictly elliptic nonlinear PDE, and there exist many standard methods for solving such equations. As mentioned above, most of these are relaxation methods, meaning that one starts with a trial function that does not solve the equation, and then iteratively deforms it to obtain functions that solve the equation to a better and better approximation. There are two basic strategies for how to relax the metric. If we iteratively replace $g_{\mu\nu}$ with $g_{\mu\nu} + h_{\mu\nu}$, then one can either deform in the direction of $-R^H_{\mu\nu}$, 
\begin{equation}\label{RF}
h_{\mu\nu} = -\epsilon R^H_{\mu\nu}\,,
\end{equation}
or solve the linearized equation,
\begin{equation}\label{Newton}
h_{\mu\nu} = -\Delta_H^{-1}R^H_{\mu\nu}\,.
\end{equation}
The first is closely related to Ricci-flow, and the second is a Newton method. In this section we will discuss these two approaches in turn, trying to understand their general properties (rather than the specifics of how to implement them). Our discussion should not be considered as exhaustive, and these strategies can be generalized, refined, and combined in various ways. We will for the most part work in the continuum limit, although it should be understood that in practice the metric and operators will be discretized in some form. The choice of discretization scheme---finite differencing, finite element, spectral, etc.---is a separate issue that we will not address in this section. (For the application discussed in Section 4, we have chosen finite differencing, because it is the easiest to implement. One intriguing other possibility would be to implement a finite-element method on the basis of the Regge calculus.)

Before proceeding, it will be useful to understand the spectrum of the operator $\Delta_H$---the linearization of $R^H_{\mu\nu}$---about a metric satisfying $R_{\mu\nu}=0$ and $\xi^\mu=0$.

\subsection{Spectrum of $\Delta_H$}
\label{sec:spectrum}

With $R_{\mu\nu}=0$ and $\xi^\mu=0$, the operator $\Delta_H$ simplifies as follows; we will also need $\Delta_V$ (see \eqref{deltaR}, \eqref{Lichvdef}, \eqref{deltaxi}, and \eqref{deltaRH}):
\begin{eqnarray}
\Delta_Hh_{\mu\nu} &=& \Delta_Rh_{\mu\nu} + \nabla_{(\mu}t_{\nu)}
\,, \qquad
t^\mu \equiv -\delta\xi^\mu = -\nabla_\nu h^{\nu\mu} + \frac12\partial^\mu h + (\Gamma^\mu_{\nu\lambda}-\tilde\Gamma^\mu_{\nu\lambda})h^{\nu\lambda}\,.\label{deltaH} \\
\Delta_Vw^\mu &=& -\frac12\nabla^2w^\mu + (\Gamma^\mu_{\lambda\nu} - \tilde\Gamma^\mu_{\lambda\nu})\nabla^\lambda w^\nu\,.
\end{eqnarray}

$\Delta_R$ is a self-adjoint operator on the space $\mathcal{H}$ of metric fluctuations, with respect to the inner product $\int_M g^{1/2}(h_{\mu\nu}h^{\mu\nu} - \frac12h^2)$, and therefore has a real spectrum. Let $\mathcal{H}_l\subset\mathcal{H}$ be the space of pure-gauge fluctuations, i.e.\ those of the form $h_{\mu\nu} = \nabla_{(\mu}w_{\nu)}$ for some vector field $w^\mu$. 
$\mathcal{H}_l$ is annihilated by $\Delta_R$, since the Ricci tensor remains zero under a gauge transformation. The quotient space $\mathcal{H}/\mathcal{H}_l$ is the space of physical metric fluctuations. On this space $\Delta_R$ may have a finite number of negative or zero eigenvalues. For example, if there is a moduli space of solutions, then each modulus will give rise to a zero mode (although the converse does not hold: not every zero mode can be ``exponentiated" to produce a continuous family of solutions). The Euclidean Schwarzschild metric has a single negative mode, concentrated near the horizon \cite{GPY}, as does the ``small black hole" solution for gravity in a cavity \cite{Gregory:2001bd}. 

We now turn to the spectrum of $\Delta_H$. Note that this operator is not self-adjoint, and may have complex eigenvalues. It maps the pure-gauge fluctuation $\nabla_{(\mu}w_{\nu)}$ to $\nabla_{(\mu}\Delta_Vw_{\nu)}$. It therefore has the same spectrum as $\Delta_V$ acting on the space of vector fields (modulo Killing fields), which may include a finite number of eigenvalues with negative or zero real part. From \eqref{deltaH} we see that the action of $\Delta_H$ equals $\Delta_R$ when acting on the quotient $\mathcal{H}/\mathcal{H}_l$, so they have the same spectrum.

In the presence of a boundary, there is one technical subtlety we should address. For concreteness we assume that we are imposing Dirichlet boundary conditions on the metric. This puts a boundary condition $h_{ij}|_{\partial M} = 0$ on the metric fluctuations in $\mathcal{H}$. Furthermore, the vector fields $w^\mu$ generating elements of $\mathcal{H}_l$ should vanish on the boundary. Now, when we study the operator $\Delta_H$, we impose an extra boundary condition, $t^\mu|_{\partial M} = 0$ (the linearization of $\xi^\mu = 0$). Let $\mathcal{H}'\subset\mathcal{H}$ be the space of fluctuations obeying this boundary condition, and $\mathcal{H}'_l = \mathcal{H}_l\cap\mathcal{H}'$ be the space of such pure-gauge fluctuations. If, as we have been assuming, for each metric there is a gauge transformation taking it into the gauge $\xi^\mu = 0$, then $\mathcal{H}'/\mathcal{H}'_l = \mathcal{H}/\mathcal{H}_l$.

All in all, the spectrum of $\Delta_H$ is the union of the spectrum of $\Delta_R$ (on $\mathcal{H}/\mathcal{H}_l$) and the spectrum of $\Delta_V$ (on $\mathcal{H}'_l$).

\subsection{Ricci-flow method}

We consider equation \eqref{RF}. Here we are following the strategy of solving an elliptic PDE by simulating the associated parabolic one, for example, solving the Poisson equation by simulating diffusion. In the limit that we take the parameter $\epsilon$ very small, we are simulating Ricci-DeTurck flow \cite{MR697987},
\begin{equation}\label{RDTflow}
\frac{\partial g_{\mu\nu}}{\partial\lambda} = -2R^H_{\mu\nu}.
\end{equation}
The sign is chosen to make this a parabolic flow. Since locally the operator $\Delta_H$ is $-\frac12\nabla^2$, locally we are indeed simulating diffusion, and the same issues arise as in that case. For example, in order to avoid short-wavelength instabilities $\epsilon$ must be chosen positive but not too large; the higher the spatial resolution the smaller $\epsilon$ must be.\footnote{This limitation can be circumvented by  implementing an implicit updating scheme. However, this is probably not worth the trouble unless one is interested in accurately simulating the Ricci-flow per se.} The method can be sped up by various strategies, such as Gauss-Seidel updating, successive over-relaxation, and the multi-grid method.

An important property of the flow \eqref{RDTflow} is that it is equivalent, modulo a time-dependent diffeomorphism, to the pure Ricci-flow $\partial g_{\mu\nu}/\partial\lambda = -2R_{\mu\nu}$. Therefore the diffeomorphism class of the metric evolves independently of the choice of background metric $\tilde g_{\mu\nu}$.

Numerical simulation of Ricci-flow has been performed previously in various contexts \cite{Hori:2001ax,Garfinkle:2003an,MR2115754,Headrick:2006ti,Doran:2007zn,HeadrickdP2,Holzegel:2007zz,Holzegel:2007ud}. In \cite{Doran:2007zn} it was used as a method for solving the Einstein equation (with positive cosmological constant), on the third del Pezzo surface. There the flow was performed in the space of K\"ahler metrics, and a theorem of Tian-Zhu guaranteed that any initial K\"ahler metric (in the correct class) would flow to the K\"ahler-Einstein metric \cite{TianZhuConvergence}.

What can we say in the absence of such a theorem, for example in the case of black hole metrics? Supposing that a solution to $R_{\mu\nu}=0$ (and $\xi^\mu=0$) exists on the given manifold with the given boundary conditions, can we expect Ricci-flow starting from a generic initial metric to converge to it? The necessary and sufficient condition for generic convergence is that the fixed point be attractive. The stability of a fixed point under the flow $\partial g_{\mu\nu}/\partial\lambda = -2R^H_{\mu\nu}$ is controlled by the spectrum of the operator $\Delta_H$, which we studied in the previous subsection. We found that the spectrum of $\Delta_H$ is the union of the spectrum of $\Delta_V$ (acting on the space of vector fields) and the spectrum of $\Delta_R$ (acting on the space of physical metric fluctuations). The former depends on the choice of background metric $\tilde g_{\mu\nu}$ as well as $g_{\mu\nu}$, while the latter depends only on $g_{\mu\nu}$. Either operator may have a finite number of negative eigenvalues (or, in the case of $\Delta_V$, eigenvalues with negative real part). Such an eigenvalue for $\Delta_V$ would correspond to an instability in the gauge-fixing condition, and could presumably be cured by making a more judicious choice of $\tilde g_{\mu\nu}$. (For example, if $\tilde g_{\mu\nu}$ happens to equal the fixed-point metric $g_{\mu\nu}$, then $\Delta_V = -\frac12\nabla^2$, which has a positive spectrum.) In our experiments we have not so far encountered such an instability. On the other hand, negative eigenvalues of $\Delta_L$ are an invariant property of the metric, which occur for many types of Euclidean black holes that one might wish to find numerically, including the ones we will study in Section 4.
Naively, one might discard Ricci-flow as an unuseful algorithm for finding such solutions. However, we will now argue that, given enough information about the nature of the flows for different initial metrics, this problem can be overcome by supplementing the Ricci-flow with a root-finding algorithm on the space of initial data.

Let us first set up some useful notation. Let $\mathcal{M}$ be the space of diffeomorphism classes of metrics on $M$ satisfying the given boundary conditions. (By a slight abuse of terminology, we will refer to these diffeomorphism classes as metrics.) Let $g^{(R)}$ be the Ricci-flat metric we are searching for. The tangent space to $\mathcal{M}$ at $g^{(R)}$ is (isomorphic to) $\mathcal{H}/\mathcal{H}_l$. For simplicity we assume that there are no zero modes of $\Delta_L$ about $g^{(R)}$, and in particular that there is not a moduli space of solutions. Let $\Sigma$ be the set of points in $\mathcal{M}$ through which a flow passes that asymptotes to $g^{(R)}$ in the future ($\lambda\to\infty$), and $\mathcal{F}$ the set of points through which a flow passes that asymptotes to $g^{(R)}$ in the past ($\lambda\to-\infty$). The tangent space to $\Sigma$ at $g^{(R)}$ is spanned by the positive modes of $\Delta_L$, and the tangent space to $\mathcal{F}$ by its negative modes. Hence, if $N$ is the number of negative modes, then $\mathcal{F}$ is $N$-dimensional and $\Sigma$ is $N$-codimensional. Starting from a point close to but not on $\Sigma$, the flow will pass close to $g^{(R)}$ before being diverted by the negative modes and asymptotically approaching the surface $\mathcal{F}$. (The flow may alternatively hit the boundary of $\mathcal{M}$ at finite $\lambda$ due to the formation of a metric singularity, but it will be exponentially close to $\mathcal{F}$ when this happens.)

The easiest case is $N=1$. Then $\Sigma$ is a hypersurface in $\mathcal{M}$, and (at least locally) divides it into two regions, while $\mathcal{F}$ consists of two flows emanating in opposite directions from $g^{(R)}$. A concrete example is provided by the ``small" Euclidean Schwarzschild black hole in a spherical cavity. The Ricci-flows of this system, for $D=4$, were studied in the paper \cite{Headrick:2006ti}. Let us briefly review that work. The small black hole is a Ricci-flat metric on the manifold $B^2\times S^2$, with a $U(1)\times SO(3)$ isometry group. Its negative mode shares those isometries and is localized near the horizon (the fixed locus of the $U(1)$ action). In \cite{Headrick:2006ti} the two flows emanating from the small black hole were found numerically. For one of them, the horizon shrinks to zero size in finite flow time, creating a singularity, while in the other the horizon expands monotonically and the metric asymptotically approaches the ``large" black hole solution, which is of the order of the size of the cavity. The flow starting from any metric sufficiently close to, but not on, $\Sigma$ will exhibit one of these two behaviors. Thus it is straightforward to determine---even automatically---on which side of the $\Sigma$ surface a given metric lies. If the small black hole solution were not known, we could find it starting with any one-parameter family of initial metrics that intersects $\Sigma$ by using a bracketing method to find that intersection point.

For small localized black holes in Kaluza-Klein theory, as we will discuss later in detail, we expect a similar situation. There exists one negative mode, which either shrinks the horizon sphere (forming a singularity) or expands it (also forming a singularity since the analogous large black hole cannot be reached without a topology change). Thus one does not need to know the explicit form of the negative modes of $\Delta_L$. One can simply deduce which side of $\Sigma$ a flow lies on, and use a bracketing procedure.

As we see later the non-uniform strings and some localized solutions have two negative modes. Then the situation is more interesting as the surface $\Sigma$ is codimension-2. Whilst here we have not attempted to solve this shooting problem, presumably it is still possible to do so, although it would be more complicated as one cannot use simple physical arguments to deduce what flow one is tending towards on the 2-dimensional surface $\mathcal{F}$. One would first tune the initial data to remove the dominant negative mode (that with highest absolute eigenvalue), and then tune again to remove the second negative mode. 

There are (at least) three potential difficulties to be addressed in the discussion above. Firstly, the picture presented presupposes we can construct a family of initial data $\mathcal{I}$ that intersects $\Sigma$. Whilst an $N$ dimensional surface may generically intersect a codimension $N$ surface, it may also not. Thus one cannot naively take an $N$ dimensional family of initial geometries. One must use some physical intuition to determine how these must vary in order for the family to contain a point in $\Sigma$. As we will see later in the Kaluza-Klein example this is certainly possible to do for one negative mode, but we have not tried for more negative modes, and presumably it becomes harder to find such data. Secondly, having found data $\mathcal{I}$ that intersects $\Sigma$, one must consider points on $\mathcal{I}$ that are sufficiently close that they are initially attracted to the fixed point of interest. Points far from $\Sigma$ may flow directly to other fixed points or to the boundary of $\mathcal{M}$ (by formation of singularities). In any algorithm that requires an initial guess there is always a basin of attraction around a fixed point, and the difficulty here is that one must lie in the basin of attraction for a multi-parameter set of initial data, rather than just finding one point in that basin. The third difficulty is that for a complicated black hole we may not know a priori how many negative modes exist. This is a more severe problem. One approach would be to start with a one-dimensional family of initial data, and if that doesn't work move to two and then three. However, given the above-mentioned problem that without knowing how to intersect $\Sigma$ one might simply miss it, it seems unlikely that one could use this method in practice without some physical intuition on the number or nature of the negative modes.

\subsection{Newton's method}

We now turn to \eqref{Newton}, in which we iteratively solve the linearized equation. Actually, it is useful to refine \eqref{Newton} a bit to allow for a variable step size:
\begin{equation}\label{Newton2}
h_{\mu\nu} = -\epsilon\Delta_H^{-1}R^H_{\mu\nu}\,,
\end{equation}
where $\epsilon$ is a parameter (generally between 0 and 1) that may be chosen independently at each step. This is useful because otherwise---particularly if the inital metric is not close to a solution---some steps may be extremely large, and in fact may lead outside the space of (positive-definite) metrics. How to choose $\epsilon$ optimally is a standard problem in the implementation of Newton's method; in fact for the Kaluza-Klein examples we exhibit later we were always able to take $\epsilon = 1$.

The utility of the method lies in the fact that there exist direct methods, such as biconjugate gradient, to solve the linear equation $\Delta_Hh_{\mu\nu} = -\epsilon R^H_{\mu\nu}$. In a real-space discretization, the operator $\Delta_H$ is sparse, and these methods are extremely fast. They work well as long as $\Delta_H$ is well conditioned (does not have eigenvalues that are very small in absolute value). In contrast to the Ricci-flow method, they are not in the least affected by the presence of negative eigenvalues. Hence any fixed point will possess a genuine basin of attraction in the space of initial metrics---no root-finding algorithm is required.

Another difference with the Ricci-flow method is that, in \eqref{Newton2}, $h_{\mu\nu}$ depends on the choice of background metric $\tilde g_{\mu\nu}$, in a way that cannot be removed by a diffeomorphism. In this sense the method is less ``inherently geometrical" than Ricci-flow. The latter describes a continuous curve in $\mathcal{M}$ (the space of diffeomorphism classes on $M$) depending only on the initial metric, with an evolution that is local on ${M}$. Newton's method, on the other hand, jumps around in the space of metrics in a way that is neither diffeomorphism-invariant nor local. Thus, unlike Ricci-flow, it neither requires nor provides information on the global structure of $\mathcal{M}$. It simply homes in on the nearest fixed point. In practice we found it to be a more robust algorithm than Ricci-flow in the case where negative modes of $\triangle_L$ occur for the solution of interest.

Ricci-flow and Newton's methods also differ in their rate of convergence. Whilst the approach to a fixed point under Ricci-flow is governed by the spectrum of the operator $\triangle_L$, with short wavelengths converging in flow time quicker than long, for Newton's method the convergence in iteration time is uniform in all wavelengths. This is a potential advantage of the method when considering non-compact manifolds, where one might truncate the manifold for numerical purposes, but then removing this regulating truncation $\triangle_L$ will have modes with increasingly small eigenvalue which would converge very slowly in Ricci-flow time.\footnote{Note that we are referring to Ricci-flow time and the iteration time for Newton's method, rather than the actual computer time. For example, in practice the time taken to solve the linear system required for each iteration of Newton's method may depend on the condition number of the linear system which will become worse if very small eigenvalue modes occur.}

%
\section{Application to black holes in Kaluza-Klein theory}
\label{sec:KK}
%

We will now use the example of black holes in $5$-dimensional Kaluza-Klein theory in order to illustrate the two algorithms we have presented here to solve the Einstein-DeTurck equation. To this end we begin with a brief review of the various black hole types we consider, before moving on to show how to apply the methods and present results. 

\subsection{Review of black holes in Kaluza-Klein theory}

We will be concerned only with single component horizons. The geometries of interest will asymptote to the flat Euclidean metric with compactified Euclidean time $\tau$, with proper length $\beta$, and Kaluza-Klein circle direction with length $L$. Thus the geometry asymptotes to the product $S^1_\beta \times \mathbb{R}^{3} \times S^1_L$ where $S^1_\beta$ is the Euclidean time circle and $S^1_L$ is the Kaluza-Klein compact dimension, with metric,
\begin{eqnarray}
ds^2 = d\tau^2 + dy^2 + y^2 d\Omega^2 + dz^2
\end{eqnarray}
using angular coordinates $\tau \in [0, \beta)$ and $z \in [0, L)$, and we have used spherical coordinates on $ \mathbb{R}^{3}$ with $d\Omega^2$ the line element on a unit $S^{2}$. Since we are interested in static black holes then $\partial / \partial \tau$ will be a Killing vector field and generate a $U(1)$ isometry, and will vanish smoothly at the horizon.

The known black holes preserve the $SO(3)$ axisymmetry manifest in the asymptotic geometry above, and thus to a lower dimensional observer they appear spherically symmetric. Hence they can be written in the form,
\begin{eqnarray}
ds^2 = T d \tau^2 + A dx^2 + B dy^2 + 2 F dx dy + S d\Omega^2
\label{eq:metric}
\end{eqnarray}
where $T,A,B,F,S$ are functions of two coordinates $x, y$. The solutions fall into 3 classes;\\

\noindent {\bf Localized black holes}\\
This one parameter family of solutions have a horizon topology $S^{3}$. For small horizon area the geometry near the horizon appears similar to the $4$-dimensional Schwarzschild solution. As the area is increased the horizon geometry deforms as it `feels' its images in the extra dimension. Little is known about this branch of solutions in the highly deformed region. Previous numerical constructions for $D = 5,6$ have been given by \cite{Sorkin:2003ka,Kudoh:2003ki, Kudoh:2004hs}, and analytic approximations have been used to study the branch in the small area limit as an expansion in the horizon radius over $L$ \cite{Harmark:2003yz,Gorbonos:2004uc,Karasik:2004ds,Gorbonos:2005px,Chu:2006ce,Kol:2007rx}. We find later that the small localized black holes possess one negative mode of $\triangle_L$ as for $4$-dimensional Schwarzschild. As one moves along the branch of solutions, increasing their size, their temperature decreases. We term these solutions the \emph{small} localized solutions. A minimum temperature is then reached. At this point the tangent to the solutions gives a zero mode of $\triangle_L$. Proceeding further the solutions become hotter again, and the zero mode continues to a second negative mode. We term these solutions the \emph{large} localized solutions.
\\

\noindent {\bf Uniform black strings}\\
These are a one parameter family of solutions with horizon topology $S^{2}\times S^1$ which we may write explicitly as the product of $4$-dimensional Schwarzschild with a circle, length $L$;
\begin{eqnarray}
ds^2 = \left( 1 -  r_0/r  \right) d\tau^2 + \left( 1 - r_0/r \right)^{-1} dr^2 + r^2 d\Omega^2 + dz^2
\end{eqnarray}
where $r_0$ provides the parameter along the family. Due to the direct product structure of the metric these solutions inherit the negative mode of the $4$-dimensional Schwarzschild solution. Using the above coordinates the Schwarzschild negative mode is a transverse traceless mode of $\triangle_L$ preserving the $SO(3)$ isometries of the form, $A^{(r_c)}_{\mu\nu}(r)$, with eigenvalue $- \omega^2 r_0^2 \simeq 0.38$ \cite{GPY}. Then the negative mode inherited by the uniform string takes the form $h_{\mu\nu} = A^{(r_0)}_{\mu\nu}(r)$, $h_{\mu z} = h_{zz} = 0$, and has the same eigenvalue, so that it preserves all the isometries of the uniform string, including the translational invariance generated by $\partial / \partial z$.

Gregory and Laflamme demonstrated that for a critical $r_0$ which we call $r_c$, this solution has a zero mode of the Lichnerowicz operator \cite{GL}. Reall showed that this is simply related to the usual $4$-dimensional Schwarzschild negative mode \cite{Reall:2001ag} with the Gregory-Laflamme zero mode of the uniform string with radius $r_c$ being given by $A^{(r_c)}_{\mu\nu}(r) e^{i \omega z}$. For $r_0 < r_c$ we expect this zero mode to flow to a negative mode with similar harmonic $z$ dependence, and this is indeed what we see numerically -- in Lorentzian signature the zero mode extends to the dynamic  Gregory-Laflamme instability. Conversely for $r_0 > r_c$ the mode flows to a positive mode. Hence the uniform strings fall into those with $r_0 > r_c$ and one translationally invariant negative mode, and those with $r_0 < r_c$ for $r_0$ near $r_c$ which have in addition a negative mode with harmonic $z$ dependence. In fact, as one decreases $r_0$ further one obtains additional higher harmonic negative modes, associated with higher harmonic zero modes. \\

\noindent {\bf Non-uniform black strings}\\
The zero mode of a uniform string with $r_0 = r_c$ provides a linear Ricci-flat deformation of the background solution. This linear deformation can be lifted to a full non-linear deformation, as illustrated by Gubser's order by order analysis \cite{Gubser:2001ac}. The new branch of solutions, which is inhomogenous in the circle direction, are hence termed non-uniform strings. Numerical constructions of this branch can be found in \cite{Wiseman:2002zc,Sorkin:2006wp,Kleihaus:2006ee}. 
We find later that the non-uniform strings emerging from the uniform branch have two negative modes. One is the continuation of the homogeneous uniform string negative mode. The second is the continuation of the zero mode at the branching point into a negative mode, agreeing with the Morse theory picture of Kol \cite{Kol:2002xz,Kol:2004ww}. \\

The uniform strings are the only solution known analytically, and therefore the task is to employ our numerical method to construct the localized and non-uniform string solutions about which relatively little is known. 
A natural conjecture is that the localized black holes and non-uniform string branches meet at a `merger point' \cite{Harmark:2002tr,Kol:2002xz,Kol:2005vy,Asnin:2006ip}. In particular Kol has made a specific proposal for the singular solution that enables the change in topology \cite{Kol:2002xz}. Studies of the branches from both side indicate consistency with such a `merger' \cite{Wiseman:2002ti,Kudoh:2004hs}, and local studies of the increasingly curved region of the geometry that is supposed to become singular also indicates consistency \cite{Kol:2003ja, Sorkin:2006wp}. However there are, as yet, no conclusive studies, largely because the localized branch of solutions has proven more difficult to construct than the non-uniform solutions which have been recently constructed to high precision \cite{Sorkin:2006wp}.

\subsection{Characterizing the Kaluza-Klein black holes}

We consider our static vacuum black hole spacetimes as smooth  Riemannian manifolds with a $U(1)$ Killing vector that vanishes at the horizon. Then we are naturally led to work with the canonical ensemble for the black hole, with the black hole Hawking temperature $T = 1/\beta$ being the control parameter, or classically, the asymptotic size $\beta$ of the circle generated by the Killing vector. In addition to this control parameter we have the asymptotic Kaluza-Klein circle size $L$. Due to the classical scale invariance of the vacuum Einstein equations we may, without loss of generality, choose to fix $L$ and work in units where $L = 1$. Solutions with any other value of $L$ may then be trivially obtained by the appropriate scaling,

Since the static black hole geometries are smooth with only an asymptotic boundary, then fixing $L$ we have that $\beta$ is the only control parameter available. However previous numerical results \cite{Kudoh:2004hs} indicate (although the effect appears subtle so they are inconclusive) that the localized black holes have a minimum temperature as one moves along the one parameter branch of solutions. This is indeed confirmed by our numerical results here. For generic static black hole solutions one might expect that branches of solutions have maxima and minima of the temperature. 
This raises two important questions. Firstly how does one find the solution of interest given that two or more solutions may have the same topology and temperature, ie. there is no black hole uniqueness? Secondly, suppose in a given topology there are no solutions below a certain temperature -- as we will find later for the localized black holes. Then how do we see that no solution exists?

The answer to these questions depends on the method being used. Let us first discuss Newton's method. For a given temperature multiple solutions may exist. However, the basis of attraction of the two solutions will be different. Hence one must find an appropriate initial guess geometry that selects the solution of interest by being initially sufficiently close to it. We will see this in practice later for the localized black holes. For Ricci-flow, black hole fixed points may not be attractors due to negative modes. However qualitatively the same answer is true, so that after appropriately tuning initial data to arrive at a fixed point,  the family of initial data we tune will determine the fixed point reached.

If no solution exists at a given temperature then Newton's method will find no solution. However, as we have discussed above, Newton's method does not give a flow in the space of geometries (diffeomorphism equivalence classes) so if one does not find a solution one cannot simply conclude that one does not exist. Since the Ricci-flow is a flow in the space of geometries, one might hope that its global behaviour does give an indication of whether a solution exists or not. Unfortunately this is not so clear cut as the tuning of the family of initial data requires that that family intersect the finite codimension hypersurface $\Sigma$ which is closed under Ricci-flow and attracted to the fixed point. Failure to find a solution can simply indicate that as the control parameter changes, the family of initial data chosen -- now a function of the tuning parameters and this control parameter -- fails to intersect $\Sigma$.

Thus we control which black hole we wish to locate by choosing the appropriate topology (say to differentiate strings from localized black holes) and then specify the inverse temperature $\beta$, and having done that choose an appropriate initial guess. Having found solutions there are various quantities of interest that we will display later.

Firstly at large radius, $S \sim y^2 \rightarrow \infty$, the time and Kaluza-Klein circle sizes become homogeneous in $x$, and these homogeneous components give rise to 2 asymptotic charges, the mass (which is conserved under time evolution in the Lorentzian setting) and relative tension (which is not conserved). Following \cite{Traschen:2001pb, Townsend:2001rg, Kol:2003if, Harmark:2003dg} this leading homogeneous correction to the asymptotic metric takes the form,
\begin{eqnarray}
ds^2 \simeq \left( 1 - \frac{2 a}{y} \right) d\tau^2 + \left( 1 + \frac{2 c}{y} \right) dy^2 + y^2 d\Omega^2 + \left( 1 +\frac{2 b}{y} \right) dx^2 
\end{eqnarray}
choosing the radial coordinate as the sphere radius. Then, choosing units where $G_N = 1$ the mass $M$ and tension $\tau$ are defined as,
\begin{eqnarray}
M =  a -  \frac{1}{2} \, b \; , \qquad
\tau \, L = \frac{1}{2} \, a -  b 
\end{eqnarray}
and give rise to a first law and Smarr relation,
\begin{eqnarray}
dM  =  T dS + \tau d L \; , \qquad
2 M   =  3 \, T S + \tau L
\label{eq:smarr}
\end{eqnarray}
where $S$, the entropy, is given in our units by $S = A_{horiz}/4$ for horizon area $A_{horiz}$.
As usual the free energy is given by $F = M - T S$.

Secondly the geometry of the solutions, particularly near the horizon is of interest. For the localized solutions we define $R_{eq}$ to be the round equatorial $2$-sphere radius of the horizon, and $L_{polar}$ to be the geodesic distance from the equator to the pole. The axis of symmetry is exposed in the localized solutions, and we define $2 L_{axis}$ to be the proper length of the geodesic between the poles of the horizon that follows this axis about the Kaluza-Klein circle. We define the eccentricity of the horizon, $\epsilon$, to be given in terms of the ratio of the area of the round equatorial horizon $2$-sphere, $A_{eq}$ and the geodesic $2$-sphere in the horizon that contains both poles $A_{pol}$, as $\epsilon = A_{pol}/A_{eq} - 1$. We denote $R_{\tau}$ to be the proper length divided by $(2 \pi)$ -- the `radius' -- of the time circle at the point on the symmetry axis equidistant from the horizon poles. Finally we may regard the time circle as being fibered over the axis of symmetry to give s $2$-sphere. We call this the time $2$-sphere, and call its area $A_{time}$. For the non-uniform solutions we denote $R_{max}, R_{min}$ as the maximum and minimum round $2$-sphere radii within the horizon -- for the uniform strings we take these both equal to the horizon radius. We define $2 L_{horiz}$ to be the proper length of a closed geodesic within the horizon that wraps the Kaluza-Klein circle once. 

We will also consider the geometry of the horizon directly by embedding it into Euclidean space later. However the geometry near the horizon is also of interest, and one probe of this that we will consider is the eigenvalue  of the negative modes of $\triangle_L$ that the metric possesses.

%
\subsection{Details of numerical construction}
%

We now discuss some pertinent details that underlie the numerical construction, and are important in order to correctly interpret the results.

We use second order accurate real space finite differencing to approximate the Einstein-DeTurck equations. In order to implement the Kaluza-Klein asymptotics we introduce a boundary at large but finite $2$-sphere radius and fix the induced metric on the boundary to be that of the round sphere, radius $R >> L$, with thermal circle length $\beta$ and Kaluza-Klein circle with length $L$. Thus we are making two approximations, firstly the finite differencing and secondly the finite asymptotic boundary. In appendix \ref{app:conv} we discuss convergence testing and dependence of our results on the physical boundary position, justifying the approximations used here. In addition we demonstrate that whilst we are solving the Einstein-DeTurck equations, the vector field $\xi$ is vanishing in the solution which is, therefore, Ricci flat as required.

\begin{figure}[htbp]
\centerline{\includegraphics[width=2.5in,height=5in]{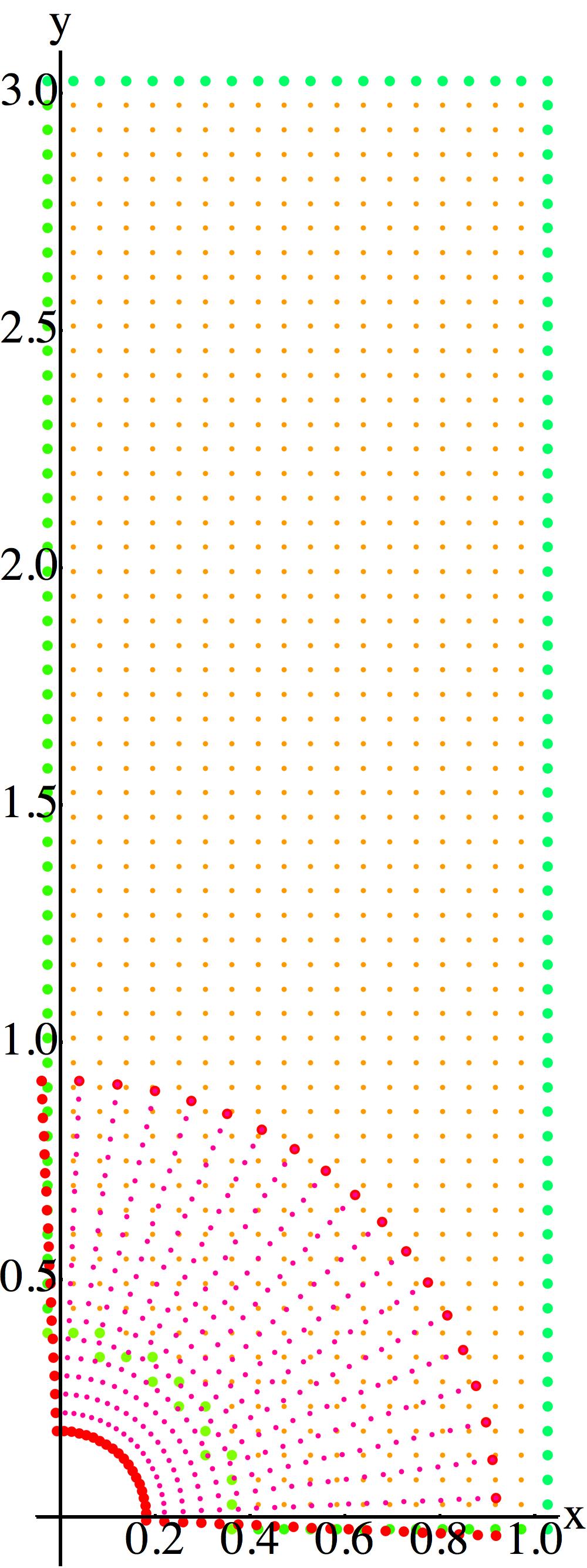}}
\caption{
The geometry of the two overlapping charts for the localized black holes. The points drawn are for the lowest resolution considered here, $20 \times 60$ with $R = 3 L$ where we have chosen $L = 1$. Boundary points are shown enlarged as are points that must be interpolated from the other chart.}
\label{fig:charts}
\end{figure}

For the non-uniform strings we may choose one coordinate chart with coordinates as above \eqref{eq:metric}, where we take $x$ to be the circle direction and $y$ the radial direction. Then $y = 0$ and $R$ give the horizon and truncated asymptotic boundary, and $x = 0$ and $L$ have mirror boundary conditions to implement the periodic Kaluza-Klein circle. It is an important point that the generalized harmonic coordinates we use have sufficient freedom to choose the boundaries of the chart to be at these locations. For the localized black holes we use two overlapping coordinate charts, one adapted to the asymptotic region, given by \eqref{eq:metric}, and the other to the horizon region, with coordinates, 
\begin{eqnarray}
ds^2 = T' d \tau^2 + A' dr^2 + B' da^2 + 2 F' dr da + S' d\Omega^2
\label{eq:metric2}
\end{eqnarray}
where the charts are related by $x = (r + r_c) \cos{a}$, $y = (r + r_c) \sin{a}$ and we have taken $r_c = 0.2$ for the results shown here. 
Figure \ref{fig:charts} shows the geometry of this chart. Second order interpolation is used to translate the metric from one chart into the other. The majority of results shown later are for boundary radius $R = 3 L$, which is sufficient to give good results for the size of black holes we are interested in. Data is shown mainly for the resolution $80 \times 240$, the first number being the points in the circle direction $x$, the latter the points in the radial direction $y$. For the localized solution the horizon chart has $80 \times 80$ points respectively. For convergence testing we also consider $20 \times 60$ and $40 \times 120$ lattices. The resolution $80 \times 240$ is still quite modest, and we expect to improve this in future work. However, it is interesting to note that already our results, particularly for the localized black holes, are comparable in reach to those performed using other methods at higher resolutions such as in \cite{Kudoh:2004hs}.

At the asymptotic boundary, chosen at constant $y$, we have fixed the induced metric by specifying the metric functions $T,A,S$ but must still determine boundary conditions for $B$ and $F$. As discussed earlier we impose the geometric boundary conditions $\xi_x = \xi_y = 0$ which yield conditions for $B$ and $F$. Although the charts give the illusion of having boundaries, for example at $y = 0$ and $y = R$, the metric represented is a smooth Riemannian manifold. Hence the boundary conditions of the Einstein-DeTurck equations in a given chart are simply determined from requiring this smoothness. On the boundaries $x = 0$ and $x = L$ one can show that all metric functions have Neumann boundary conditions, except $F$ which must vanish. At the axis of symmetry of the localized black holes again the functions must be Neumann, except $F$ which again vanishes, but in addition regularity of the geometry enforces that $S$ vanishes as $S = B \, y^2$. On the horizon the metric functions must again be Neumann, except for $T$ which vanishes as $T = \kappa^2 y^2 A$ for strings and $\kappa^2 r^2 A$ for localized solutions, where $\kappa$ is a constant over the horizon and is related to the temperature as $\beta = 2 \pi \sqrt{T_{\infty}} / \kappa$ where $T_{\infty}$ gives the asymptotic value of $T$.

There are two classes of behaviour asymptotically in Kaluza-Klein theory. The modes with non-trivial harmonic dependence in the circle direction, which behave as massive fields from the point of view of the dimensionally reduced theory, decay exponentially at large radius. The modes that are constant on the circle, which reduce to light lower dimensional degrees of freedom, decay with a power law asymptotically. One subtlety is that taking $R = 3 L$ for the asymptotic boundary is far enough for all the `massive' modes to decay away. However, for the `massless' modes which determine the asymptotic value of $L$ and $\beta$ the values they reach at $R = 3 L$ are not very good approximations to their asymptotic values. Thus we find solutions using our domain, applying asymptotic boundary conditions at $R = 3 L$ by specifying $T$ and $A$ there -- let us call these $T_{asym}, A_{asym}$. We then take the solution near the boundary, which is essentially independent of $x$, and use this as initial data for the ODEs governing the massless modes which we then evolve out to a much larger $R = 1000 L$ to ensure that the massless modes are in their asymptotic regime. Then we extract $T_{\infty}, A_{\infty}$ from the limiting values of $T$ and $A$ there, and in addition the coefficients of the $1/r$ terms in order to compute mass and entropy. The difference between $T_{\infty}, A_{\infty}$ and our control parameters $T_{asym}, A_{asym}$ is small, of order a few percent, but for accurate determination of physical quantities it is important to perform this `asymptotic improvement'. Since this difference is so small the actual asymptotic values of $T, A$ differ little from their values specified by the control parameters -- if this were not the case it would be difficult to specify the solution of interest by fixing the control parameters.
Checking the Smarr law is satisfied then  gives a gauge as to whether this asymptotic extraction of the charges is functioning correctly. A similar procedure was used in \cite{Wiseman:2002zc}.

Since we are exploiting the various isometries of the metric in order to reduce the effective dimension of the problem,
the equations contain naively singular terms such as $(S - B y^2) / y^2$ at the axis of symmetry $y = 0$, which derive from the curvature of the sphere which shrinks there. Analysis of the equations shows that in addition to the smooth behaviour we require there are also singular behaviours. The previous approach to solving the static Einstein equations in this context suffered from developing the singular behaviour unless great care was taken. However, in both the Ricci-flow and Newton method approach we use here we find a very robust behaviour at the axis of symmetry, with no instability issues arising there. For both methods, since the approach manifestly is based on the geometry being smooth, the potential singular behaviour seems not to surface. For Ricci-flow one can understand this as the Ricci-flow attempting to locally diffuse away curvature. 

In Newton's method we must construct the linearization of $R^H$ to give the operator $\triangle_H$ and solve the linear system $\triangle_H^{-1} R^H$. It is crucial for the method that the ability to invert the operator $\triangle_H$ does not require it to be positive. In practice after the finite differencing we map the interior points of our charts into elements of a finite dimensional vector. Then we may represent the metric and its curvature $R^H$ as vectors. The linearization of $R^H$ is then computed by a numerical functional differentiation of $R^H$ with respect to the metric, to yield the matrix operator $\triangle_H$. 
Care is taken to implement the boundary conditions and interpolation from other patches during this numerical differentiation.
Since we have used real space finite differencing, the operator $\triangle_H$ is very sparse. Then solving the finite dimensional linear system $\triangle_H^{-1} R^H$ can be efficiently achieved using the biconjugate gradient method. In practice this stably solves the sparse system, and being based on Krylov methods which utilize the finite dimensional nature of the linear problem rather than any form of iteration, is totally insensitive to whether eigenfunctions are positive or negative. Only zero modes will present problems for this method, and near zero modes may potentially destabilize the method.

%
\subsection{Ricci-flow applied to localized black holes}
%

%
%

We now demonstrate use of the Ricci-flow method to find small localized $5$-d black holes. Whilst the Ricci-flow method can be made to work in practice, as we shall see it is less elegant than Newton's method if one simply wants to find solutions. The first task when constructing these solutions is to choose a suitable one parameter family of smooth initial data that pierces $\Sigma$. We have used the following family,
\begin{eqnarray}
ds^2 &= & \left( I(r+r_c;d_{min},d_{max}) \; {g_{1}}_{\mu\nu} + \left( 1-I(r+r_c;d_{min},d_{max}) \right) {g_{2}}_{\mu\nu} \right) dx'^\mu dx'^\nu \nonumber \\
&& {g_{1}}_{\mu\nu} dx'^\mu dx'^\nu = T_0 d\tau^2 + dr^2 +  ( r + r_c )^2 \left( da^2 + \sin^2{a} d \Omega^2_2 \right) \nonumber \\ 
&& \qquad \qquad \qquad = T_0 d\tau^2 + dx^2 +  dy^2 + y^2 d \Omega^2_2 \nonumber \\
&& {g_{2}}_{\mu\nu} dx'^\mu dx'^\nu = \frac{r^2}{r_0^2} d\tau^2 + dr^2 + ( r^2 + r_0^2 ) \left( da^2 + \sin^2{a} d \Omega^2_2 \right)  
\label{eq:init}
\end{eqnarray}
which gives a linear interpolation between the near horizon behaviour $g_2$ and the flat metric $g_1$, where we use the interpolation function,
\begin{eqnarray}
I(x;x_{min},x_{max}) = \frac{1}{2} \tanh \tan \left( \frac{\frac{1}{2}(x_{max}+x_{min}) - x}{x_{max} - x_{min}} \pi \right) .
\end{eqnarray}
We note that this data satisfies all relevant boundary conditions, with $\kappa = 1/r_0$. The family  possesses two parameters, $T_0, r_0$. One combination determines the inverse temperature as $\beta = (2 \pi / \kappa) \sqrt{T_0}$. This is then fixed by the boundary conditions, and we are left with one parameter to vary. Let us take this to be $r_0$.

Consider a value of $\beta$ where a small localized solution exists. Since these solutions possess one negative mode, the surface $\Sigma$ is codimension one, and we must identify if we are on the `+' side or `-' side. It is a question of geometric interest where the `+' and `-' flows from the fixed point go. We might expect the `-' flow to behave in a similar manner to that of the usual Schwarzschild solution and this is a good guess. Indeed the horizon shrinks to zero size in finite flow time, and in a singular manner (with curvature invariants such as the Ricci scalar becoming singular there). Presumably if one resolves the singularity the flow continues to hot  Kaluza-Klein  space. More subtle is the what happens to the `+' flow. For Schwarzschild in a finite cavity the flow goes to the large black hole. In the asymptotically flat case the flow goes in the same direction as the cavity but eats up all of the space,  as there is no fixed point to reach. However in the case of a localized black hole, in a cavity with boundary at $R >> L$ one would not expect a large localized solution, as it would be highly deformed in the Kaluza-Klein circle direction. Instead the relevant end point is likely to be a large uniform string. In order to reach this solution the flow would have to change the topology of the manifold from the localized to string topology. The likeliest conjecture then is that the `+' mode also flows to a singularity in finite time, although this singularity arises from the time circle on the symmetry axis $y = 0$ shrinking to zero size -- if resolved this would then flow to the large black string. In the simulations this is indeed what we see.

\begin{figure}[htbp]
\centerline{\includegraphics[width=3.5in,height=2.5in]{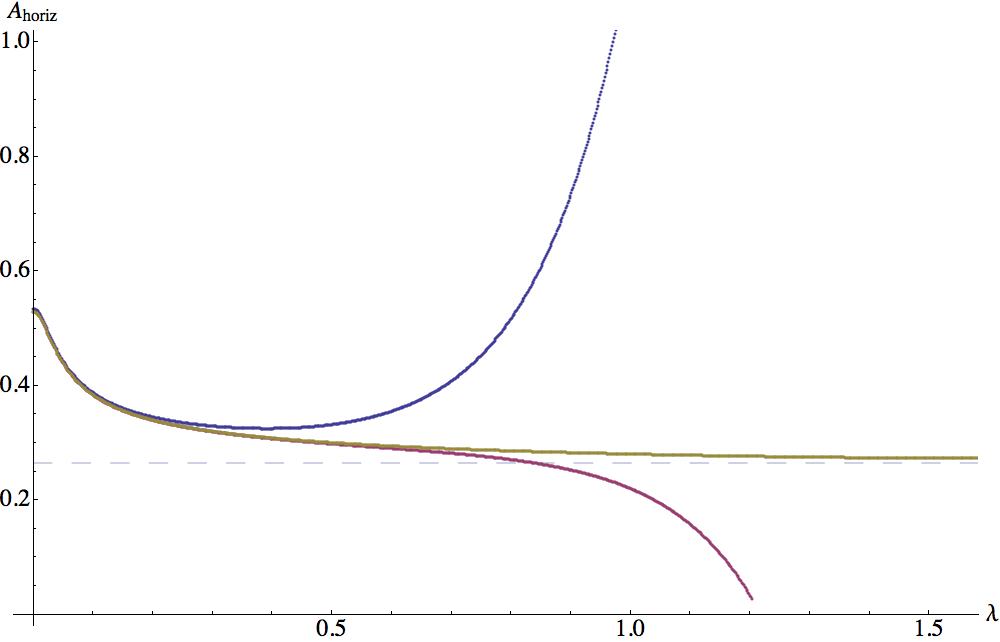}\includegraphics[width=3.5in,height=2.5in]{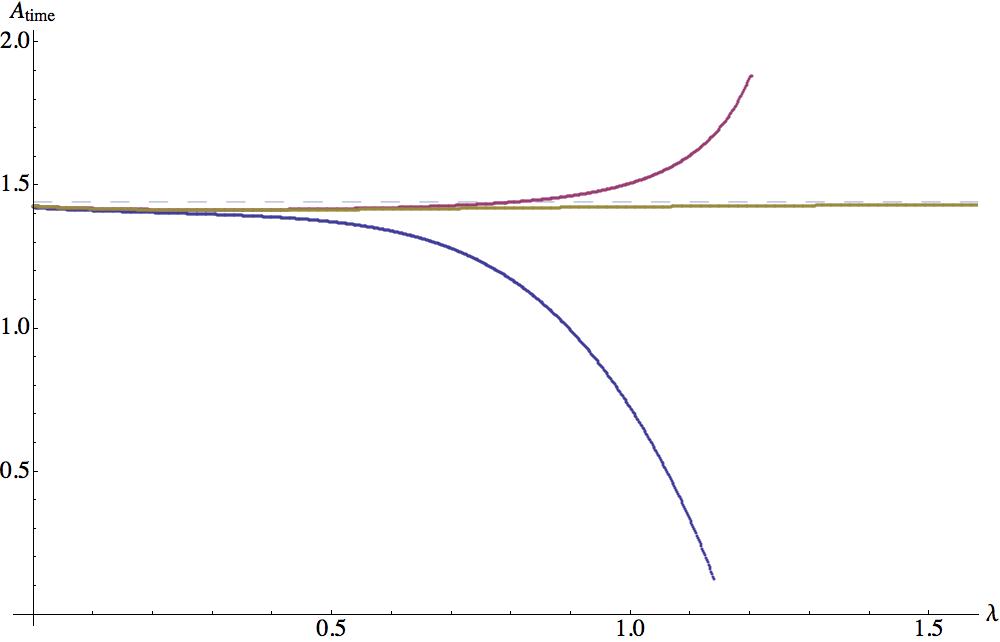}}
\caption{
Behaviour of the area of the horizon (\emph{left}) and area of the time circle 2-sphere (\emph{right}) as a function of Ricci-flow time for 3 initial data. In blue, $r_0$ is slightly greater than the critical value, in red, $r_0$ is slightly less, and the yellow gives the critical value. The red and blue curves diverge from the yellow due to the presence of the single negative mode. The dashes horizontal lines give the value of the area at the fixed point, calculated using Newton's method, and thus we see agreement with the critical flow.
}
\label{fig:Flowarea}
\end{figure}

\begin{figure}[htbp]
\centerline{\includegraphics[width=3.5in,height=2.in]{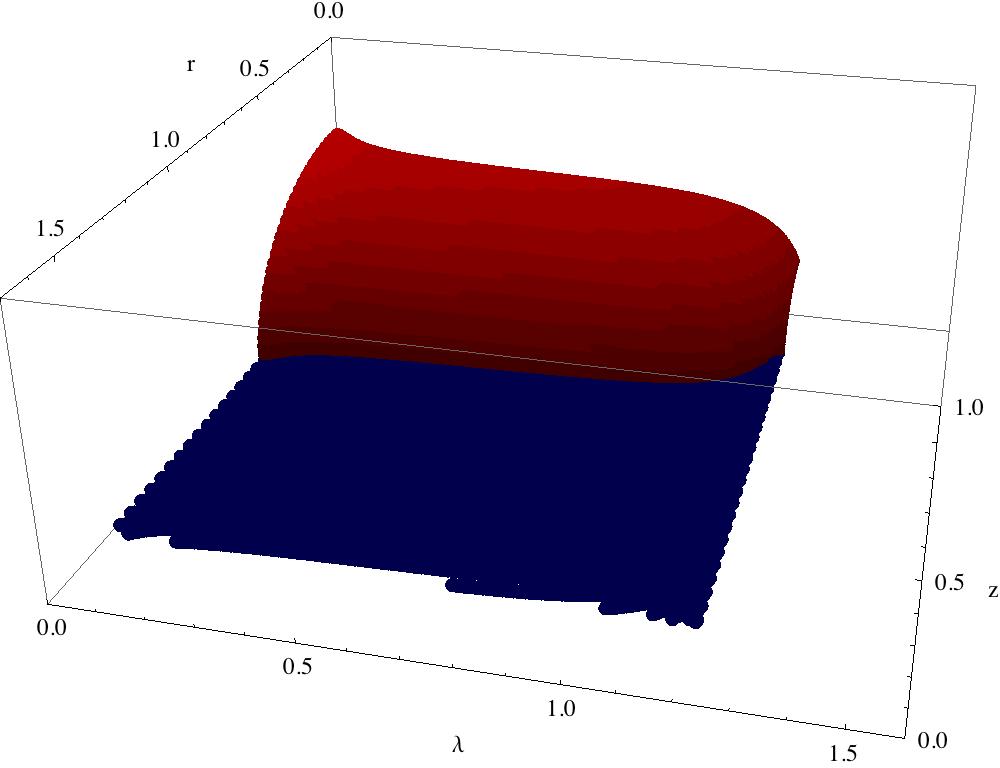}}
\centerline{\includegraphics[width=3.5in,height=2.in]{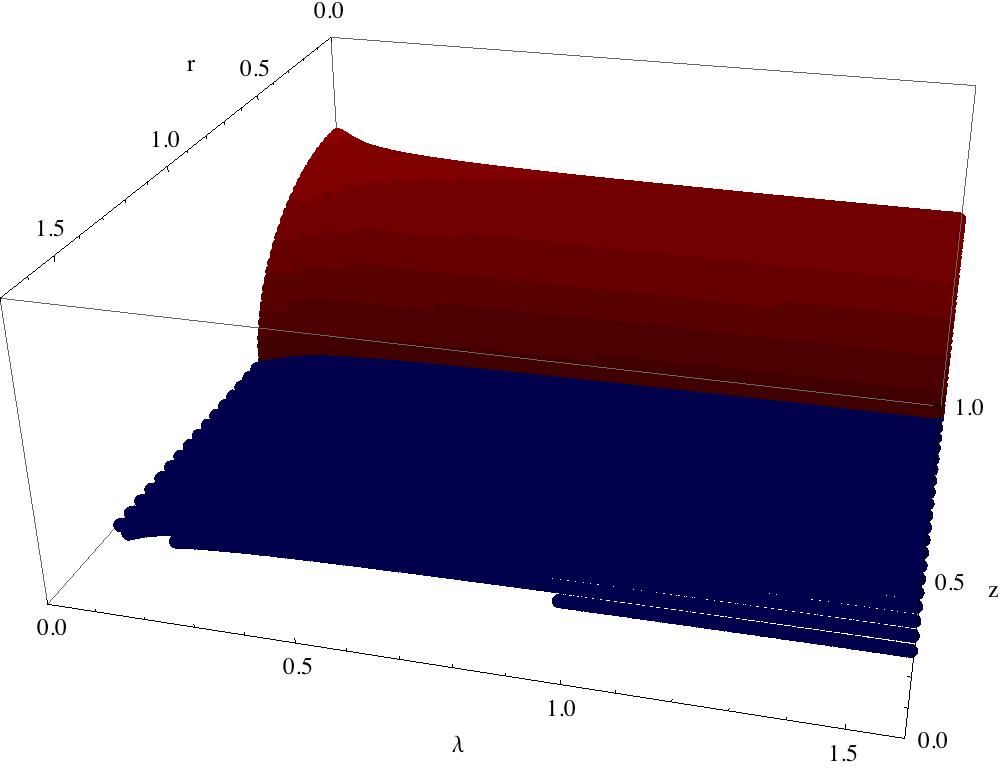}}
\centerline{\includegraphics[width=3.5in,height=2.in]{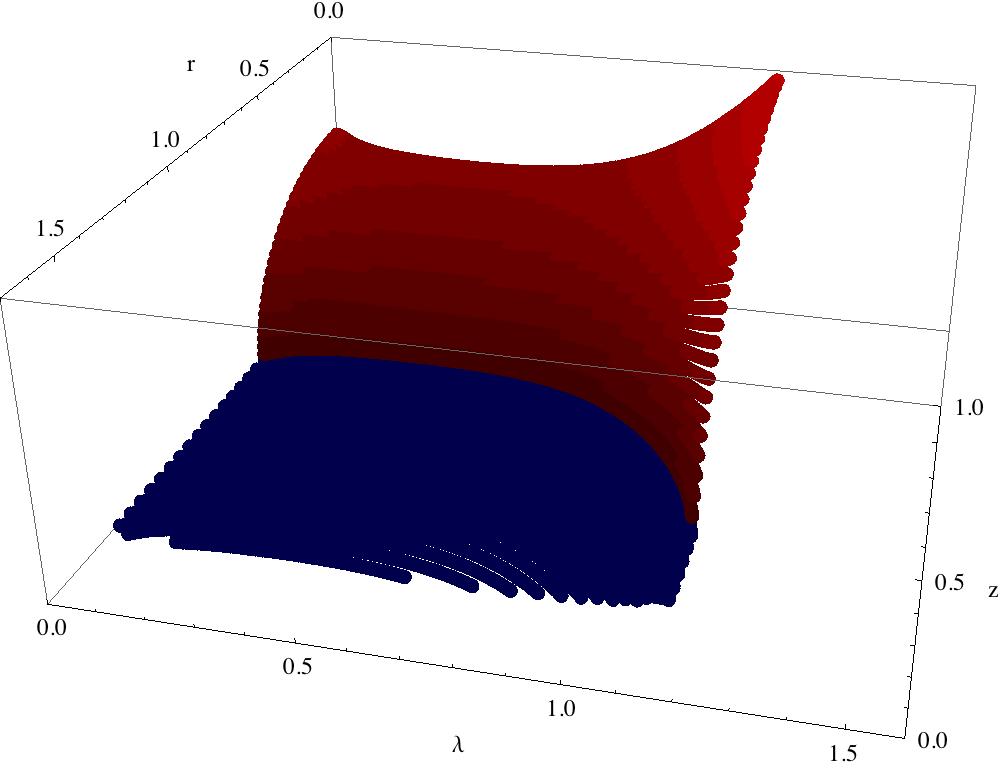}}
\caption{
Cross-sections of proper embeddings of the horizon are shown in red for $L = 1$ against Ricci-flow time for the same 3 flows as in the previous figures. The axis of symmetry is also shown in blue with its proper length. 
The top figure corresponds to the red curves, $r_0$ less than the critical value, the middle figure gives the critical flow, and the bottom figure corresponds the blue curves so $r_0$ is greater than the critical value.
}
\label{fig:Flowembed}
\end{figure}

We simulated the flows using an explicit implementation of the DeTurck flow, and chose the background metric $\tilde{g}$ simply to be the same as the initial metric. We have only explored the low resolution $20 \times 60$ as we are not so much interested in the solutions themselves here, but rather the mechanics of finding them. In figure \ref{fig:Flowarea} we show the horizon area and time circle $2$-sphere area for 3 flows with $\beta = 1.5$. We find that tuning $r_0$ a critical value, $r_{critical} \simeq 0.598123$, is required to approach the fixed point. This critical flow is shown, in addition to two other flows, one with a lower $r_0 = 0.598$ where the horizon sphere shrinks to zero size in finite time -- the `-' behaviour -- and the other for $r_0 = 0.600$ where the time sphere shrinks and the horizon expands -- the `+' behaviour . We see the flow $r_0 = r_{critical}$  appears to stabilize at a fixed point. Interestingly for the `-' and `+' behaviours when the sphere shrinks, the time circle expands, and vice versa. In figure \ref{fig:Flowembed} we show full proper embeddings of the horizon and symmetry axis against flow time for these three flows, and we see that the entire horizon remains round as it shrinks while the symmetry axis length seems to be unaffected.

\begin{figure}[htbp]
\centerline{\includegraphics[width=3.5in,height=2.5in]{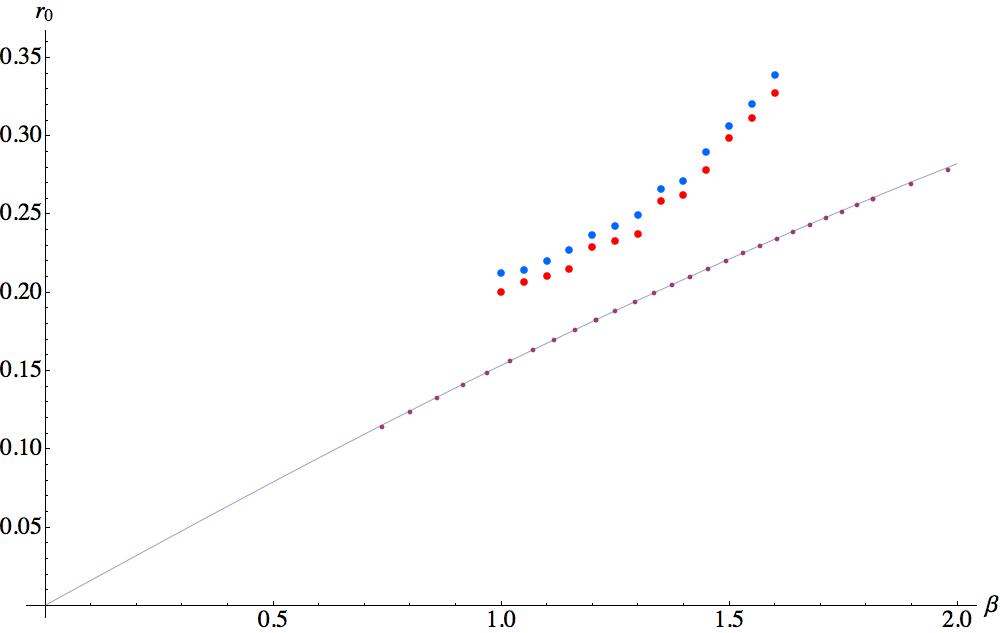}}
\caption{
Brackets for $r_0$ over a range of $\beta$. For the lower end of the bracket the sphere collapses, and for the upper end the time circle shrinks in finite flow time. Outside this range, higher resolution is required to accurately simulate the flows. The value $r_0$ gives the initial radius of the horizon for the flow.
We also plot the equatorial radius of the horizon for solutions shown as small dots, together with the small localized approximation for this radius which is rather accurate over this range of $\beta$.
}
\label{fig:Flowtuning}
\end{figure}

Since the nature of the two possible singularities is straightforward to determine, it is simple to automate a bracketing algorithm that finds the solutions. In figure \ref{fig:Flowtuning} we show brackets in $r_0$ that include the value of $r_{critical}$ against $\beta$ for a range of $\beta$. The value of $r_0$ in the initial data gives the initial size of the horizon. In the plot we also show the size of the horizon of the fixed point.
 We see that as $\beta$ increases, the size of the horizon increases relatively more quickly than the size of the horizon of the fixed point. Indeed we see that for quite small $\beta \simeq 1.6$ the initial horizon diameter (ie, $2 r_0$) is already becoming comparable to the size of the Kaluza-Klein circle. We find that to accurately simulate the flow we require higher resolutions, and it appears that the initial data used may develop a coordinate pathology during the flow. Presumably this could be resolved with more effort, but we have no attempted to do so here. Even doing this, it is not clear how high $\beta $ could be pushed while allowing a critical solution -- recall that our initial data does not have to intersect $\Sigma$ for all $\beta$.
We see in the next section that Newton's method also doesn't find localized solutions above $\beta \sim 3.39$. Thus the range of $\beta$ we have probed here is substantially less than that which is possible, and given the growing size of $r_0$ relative to the fixed point size, we suspect that our family does fail to intersect $\Sigma$ before this maximum $\beta$.

We have only attempted to find the small localized solutions with their single negative mode. Whilst in principle we believe that one can find the large localized solutions and non-uniform strings with two negative modes by finding an appropriate 2 parameter family of initial data, we have not tried it. The principle complication is identifying how to tune the initial data, given the late time behaviour of the flow, and it would be interesting to examine whether there is interesting geometry involved, as in the one negative mode case, or whether this becomes purely a technical problem.

Needless to say the fixed points found agree with those found by Newton's method -- for example the areas shown for the critical flows in figure \ref{fig:Flowarea} agree with the values found from Newton's method -- and hence we defer showing the properties of the solutions until the next section. We see the Ricci-flow algorithm we present does work in practice. It is simple to implement the explicit flow, and it would be relatively straightforward to implement an implicit algorithm to allow larger time steps to be taken than in the explicit case although whether this would improve the speed is unclear as this depends on the details of solving the implicit system. 

Whilst the flow itself is easy to construct, and the tuning process easy to implement, at least for codimension one $\Sigma$, it appears that the tuning must be done rather precisely to closely approach the fixed point. This is to be expected due to the asymptotically flat boundary condition. The approach to a stable fixed point is governed by the lowest eigenmode. We truncate our asymptotic region to have a boundary at finite radius. Then we expect the lowest mode is goverened by that radius. Taking the radius to be larger will reduce the eigenvalue of the lowest mode, and removing the truncation will yield a continuous spectrum down to zero eigenvalue. In practice the size of the bracket must be systematically reduced until the flow approaches and remains near the fixed point for some considerable flow time. In our case, for $R = 3 L$, tuning $r_0$ to $10^{-6}$ yields flows that remain near the fixed point only for $1.0< \lambda < 2.0$ before it develops a `-' or `+' singularity. Taking the solution around $\lambda = 1.5$ then gives a good approximation to the fixed point solution, but this is an additional source of systematic error beyond the finite differencing and infinity truncation,
and removing the infinity regulator makes this problem worse.

Thus whilst conceptually and practically simple to implement, the Ricci-flow method does suffer from the problem that it is difficult to attain good accuracy of the fixed point solution by tuning a family of initial data. For low resolutions such as the $20 \times 60$ used here the error in the resolution and finite infinity truncation may outweigh this accuracy error. However, for higher resolutions, and for removing the regulator clearly this is presents a serious deficiency of the algorithm that limits its usefulness.

Such a precision tuning problem could be avoided by modifying the algorithm as follows. A crude bracket is constructed using the initial family of data above. Each side of the bracket flows close to the fixed point for some time and then diverges away in the direction of the negative mode. However, we may take the metrics given by flowing each side of the bracket for a time so that the flow has approached the fixed point, and not yet diverged away. Let us denote these $g_{1,2}$. Then one may construct a new one parameter family of metrics $g(\lambda) = \lambda \, g_1 + (1 - \lambda) g_2$ which again bracket the fixed point for $\lambda \in (0, 1)$. One then proceeds to use this new family to refine the bracket. Since one is closer to the fixed point, for a given precision of tuning in the parameter $\lambda$ one should get closer to the fixed point than for an equivalent tuning of the initial family parameter $r_0$. Clearly this procedure can be iterated. Whilst this procedure will work, it is somewhat ad hoc. Deciding how long to run the flow for, and how accurately to find the bracket, are just some of the points that one would have to specify.  

In essence the Ricci-flow can get close to a solution by the simple tuning of an initial family of data, but once one is near the fixed point, there are more efficient methods to hone closely in on that fixed point. Newton's method described next represents precisely such a method of honing in on the fixed point, and thus one use of the Ricci-flow method could be to provide initial data for Newton's method, where this was difficult to construct. In the case of the solutions we discuss here we have analytic initial data in the basin of attraction of the solutions of interest for Newton's method, and once one solution on a branch is found, one can easily move along the branch. However in more complicated situations it might be unclear how to write down initial data in the basin of attraction of a branch of solutions, particularly if this branch were disconnected from other known branches. Then Ricci-flow might provide a valuable tool to construct metrics close to the fixed point.

We note that the Ricci-flow, being a flow in the space of diffeomorphism classes, does give global geometric information about the space of metrics as the flow proceeds. Newton's method, while simpler to use in practice, gives no geometric information other than the solution it may find. For example, suppose we were not interested in the details of a solution, but merely its existence -- black holes on Randall-Sundrum branes are an interesting problem of this variety \cite{Tanaka:2002rb,Emparan:2002px,Emparan:2002jp}. Then a relatively coarse bracketing of the `+' and `-' behaviours gives strong evidence of existence, even though the flows would not go very near the fixed point and therefore the details of the fixed point solution could not be read off. 
We may regard the results above as giving a `numerical proof' of existence of localized black hole solutions -- although in this case there is no reason to doubt their existence. With Newton's method one must test existence by studying convergence with discretization scale. However with the Ricci-flow, since we are clearly probing the global structure of the space of geometries, rather than just finding a single solution, we may regard our brackets above as providing convincing evidence for existence.

As we will see shorty, in the case of the localized solutions Newton's method clearly demonstrates that the small localized branch reach a maximum $\beta$ which can be continued to large localized branch, so whilst $\beta$ has a maximum the solution branch is smooth there. The Ricci-flow method is compatible with this in the sense that we have not found solutions for higher $\beta$. However, one might imagine a branch of solutions could actually end at some $\beta$. In this case it might be difficult to know from Newton's method why solutions were not being found. With the Ricci-flow method, one might be able to gain some additional global information from the nature of the flows.

%
\subsection{Newton method applied to localized and non-uniform string solutions}
%

%

The implementation of Newton's method is technically considerably more complicated than the Ricci-flow, as the Einstein-DeTurck equations and metric must be recast as vectors, the operator $\triangle_H$ must be computed -- we use a numerical functional differentiation -- and then the large but sparse linear system must be solved -- in our case using biconjugate gradient -- in order to provide the basis for Newton's method. However, once this technology is in place the method works extremely well in practice.

The first step in finding the solutions of interest is to find appropriate initial data in the basin of attraction of that solution. This is not trivial, but in practice is manageable with a little trial and error. For example, in the case of $5$-d localized solutions, we find that the initial data given by equation \eqref{eq:init} above with $r_0  = 0.15$ gives a good initial metric and background metric to find the $\beta = 1.2$ small localized solution. Once a solution on small is found it is very simple to find neighbouring ones simply by perturbing the solution metric and background metric to change $\beta$ to a new nearby value. For example, we use the perturbation, $\delta g = I(y,L,2 L) d\tau^2$ to implement this. For a small perturbation we remain in the basin of attraction of the solution at the new value of $\beta$, and that solution is quickly found.

A subtlety then arises when a maximum/minimum in $\beta$ arises along a branch of solutions. For the example of small localized $5$-d black holes we start with $\beta  = 1.2$ and then gradually increase $\beta$ but find no convergence above $\beta = 3.39$. However, plotting physical quantities against $\beta$ one sees that this is not the end of the branch of solutions, but merely a maximum in $\beta$ so that $\beta_{max} = 3.39$. 
In order to pass this maximum to find the large localized solutions one must again find initial data with $\beta < \beta_{max}$ but that is in the basin of attraction of the large localized branch rather than the small localized one. This may be achieved by taking solutions found for small branch near the suspected maximum $\beta_{max}$, and using them to extrapolate to an initial guess with $\beta$ less than the maximum which should lie on large branch. Suppose we have small branch solutions $g_i$, with $i = 1, \ldots N$, and corresponding $\beta_i$ such that $\beta_i < \beta_j$ if $i < j$. In order to extrapolate we must find a quantity which varies monotonically through $\beta_{max}$, rather than having a maximum/minimum there. For example, we may take the quantity $L_{axis}$ which appears to decrease monotonically along the small branch near $\beta_{max}$ and across the join to the large branch. Thus at the maximum, $\beta$ varies quadratically with $L_{axis}$. Then we use the values of $(L_{axis})_i$ for the solutions, and extrapolate the data to find an initial guess $g_{N+1}$ for a value $(L_{axis})_{N+1}$ which is small enough to correspond to a large localized solution. $\beta_{N+1}$ is then simply determined from $g_{N+1}$. In practice this procedure works well and gives good results, allowing us to cross to the large localized solutions, having found the small localized branch.

Finding initial data for the non-uniform strings is less straightforward than for the small localized solutions. It is very straightforward to find uniform strings, using an analogous scheme to the small localized solutions -- simple analytic initial data can be found in the basin of attraction. In particular in $5$-d we use,
\begin{eqnarray}
ds^2 & = & = \frac{r_0^2 \, y^2}{y^2 + r_0^2} d\tau^2 + dx^2 + dy^2 + \left( r_0^4 ( 1 - I(y,0,1) ) + y^4 \right)^{1/2} d \Omega^2_2 
\label{eq:initUS}
\end{eqnarray}
to find uniform strings, and choosing $r_0  = 0.28$ yields a solution near the marginal point which is at $\beta_{GL}  = 1.752$. The challenge however is to find initial data which is in the basin of attraction of the non-uniform strings. In $5$-d we know from Gubser's work \cite{Gubser:2001ac} that for weakly non-uniform strings we have $\beta > \beta_{GL}$. Thus we perturb the metric and background metric for this marginal solution to lower its temperature,
\begin{eqnarray}
\delta A &= & A \left( 1 - 0.02 (1 - I(y,0,L) ) \right)  
\end{eqnarray}
and perturb only the metric (not the background) with a non-uniform profile,
\begin{eqnarray}
\delta B &= & B\left( 1 - 0.05 \cos{\pi x} \left(1 - I(y,0,L) \right) \right)  \nonumber \\
\delta S &= & S \left( 1 - 0.05 \cos{\pi x} \left(1 - I(y,0,L) \right) \right) 
\label{eq:initNUBS}
\end{eqnarray}
We find this does indeed relax to a weakly non-uniform string. We may then traverse the non-uniform branch by adjusting the temperature as for the localized black holes. Note that it is convenient to keep the background metric uniform as then it is straightforward to distinguish between the uniform and non-uniform branches by simply looking at whether the metric is uniform.

However a more systematic method is to find the uniform string at $\beta_{GL}$, and construct the Gregory-Laflamme zero mode. This is simple in practice as Newton's method already requires the linearization $\triangle_H$ to be computed, and it is straightforward to compute the low lying eigenfunctions of a sparse matrix such as this. Having done so, we take a uniform solution with $\beta > \beta_{GL}$ but still near the marginal point, and then add some amount of the zero mode to it. We note that the linear perturbation of the $\beta = \beta_{GL}$, marginal solution preserves most of the boundary conditions even when added to a solution with a different $\beta > \beta_{GL}$.\footnote{Only the boundary condition for $\xi = 0$ at the asymptotic boundary is not preserved, and this is quickly restored during Newton's method.} Adding very little, one remains in the basin of attraction of the uniform solution. However, when sufficient amplitude of this perturbation is added, one is moved into the basin of attraction of the non-uniform solution. Having found one non-uniform string, one may explore the branch by perturbing the temperature as for the localized solutions.

A potentially attractive feature of Newton's method is that in principle the convergence is uniform in the wavelength of the deformation from the fixed point. Removing the asymptotic truncation for Ricci-flow presents the problem that the long wavelength modes evolve slowly, whereas for Newton's method this is not the case. However in practice the time does depend on how long the biconjugate gradient method takes to find solutions, and this can depend on the spectrum of $\triangle_H$. We have not investigated this here, and other methods to solve the linear system may behave differently, but it is a point to note.

\begin{figure}[htbp]
\centerline{\includegraphics[width=3.5in,height=2.5in]{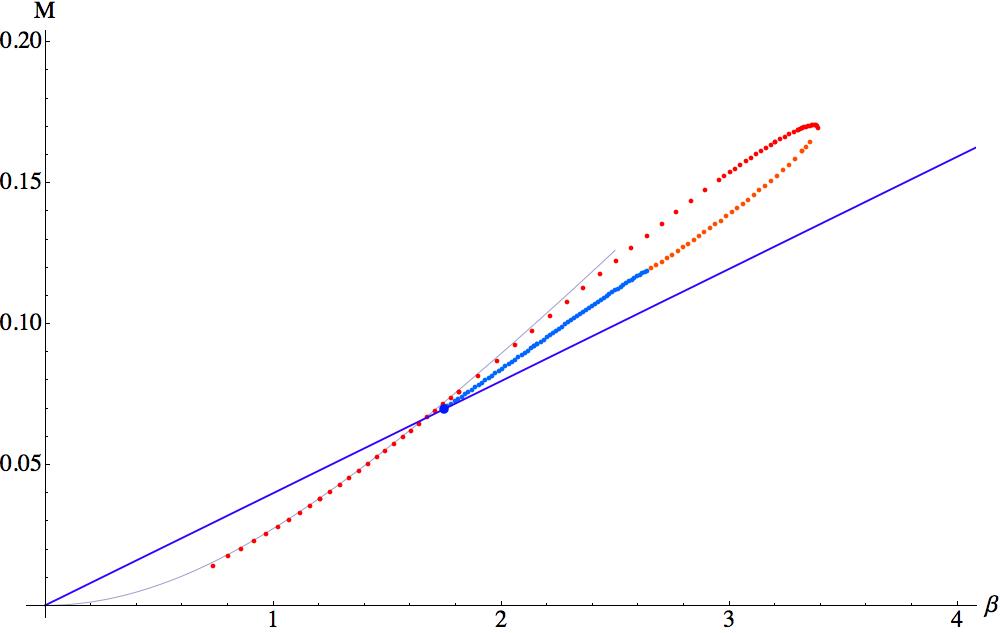}\includegraphics[width=3.5in,height=2.5in]{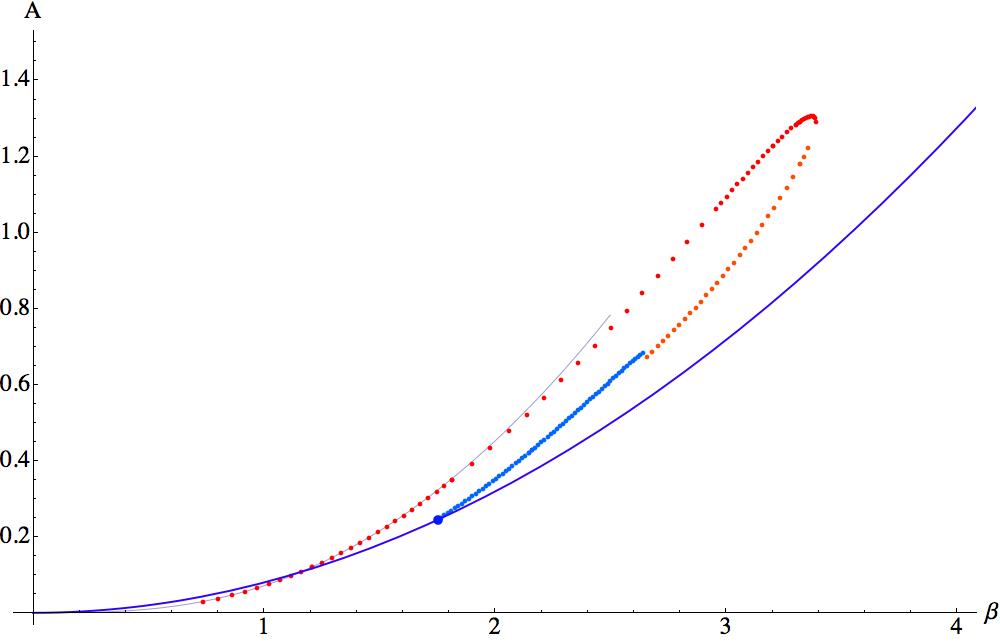}}
\centerline{\includegraphics[width=3.5in,height=2.5in]{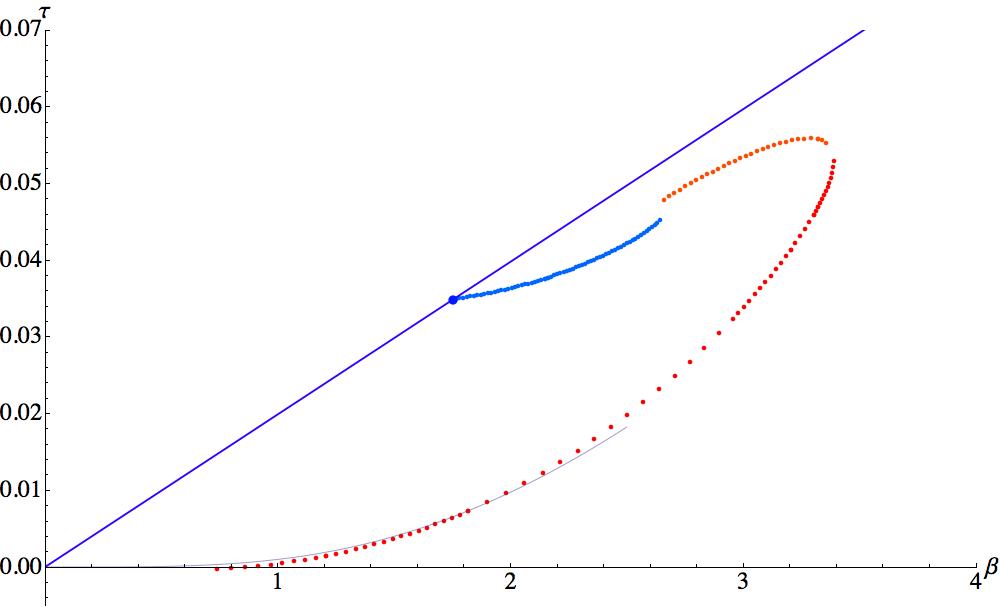},\includegraphics[width=3.5in,height=2.5in]{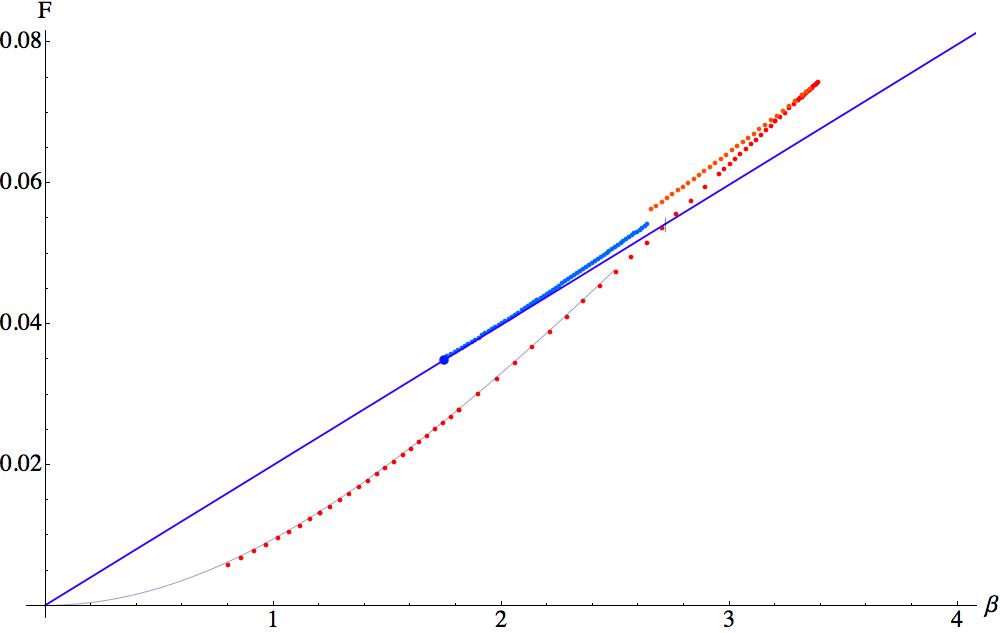}}
\caption{
Thermodynamic behaviour of the various static Kaluza-Klein black holes for $L = 1$.
\emph{Top left:} Plot of mass, $M$, against $\beta$. The small localized branch are shown in red, the large localized branch in orange, and the non-uniform solutions are in light blue. All are for $80\times240$ resolution with $R = 3 L$. The uniform branch is given by the dark blue line, and the small localized analytic approximations are shown by a thin line. The solid blue disc indicates the Gregory-Laflamme marginal point. Similar conventions are used to plot the horizon area $A_{horiz}$ (\emph{top right}), tension $\tau$ (\emph{bottom left}) and free energy $F$ (\emph{bottom right}).
}
\label{fig:thermo}
\end{figure}

\begin{figure}[htbp]
\centerline{\includegraphics[width=3.5in,height=2.5in]{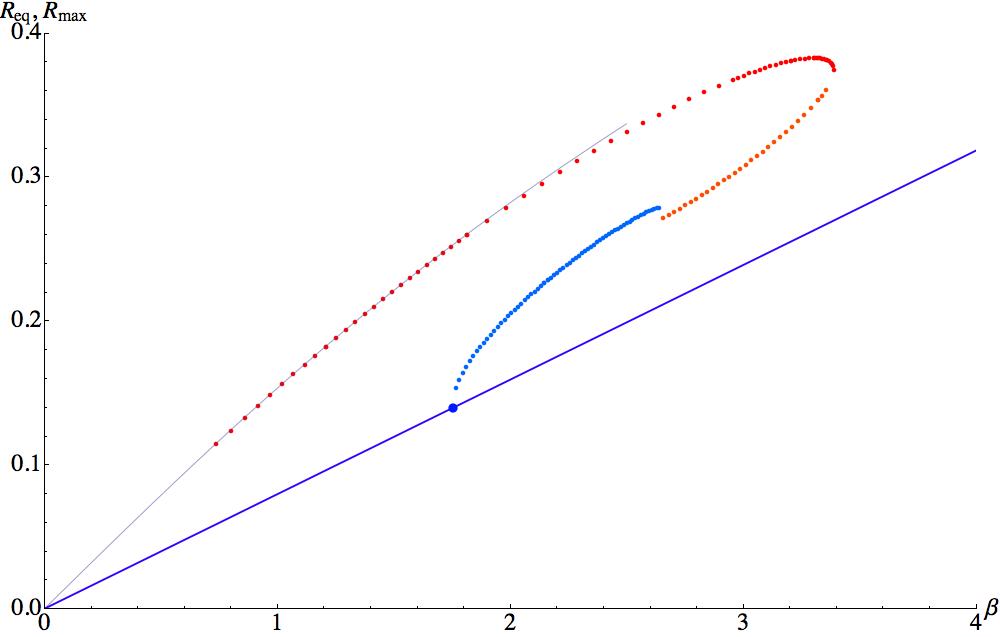}\includegraphics[width=3.5in,height=2.5in]{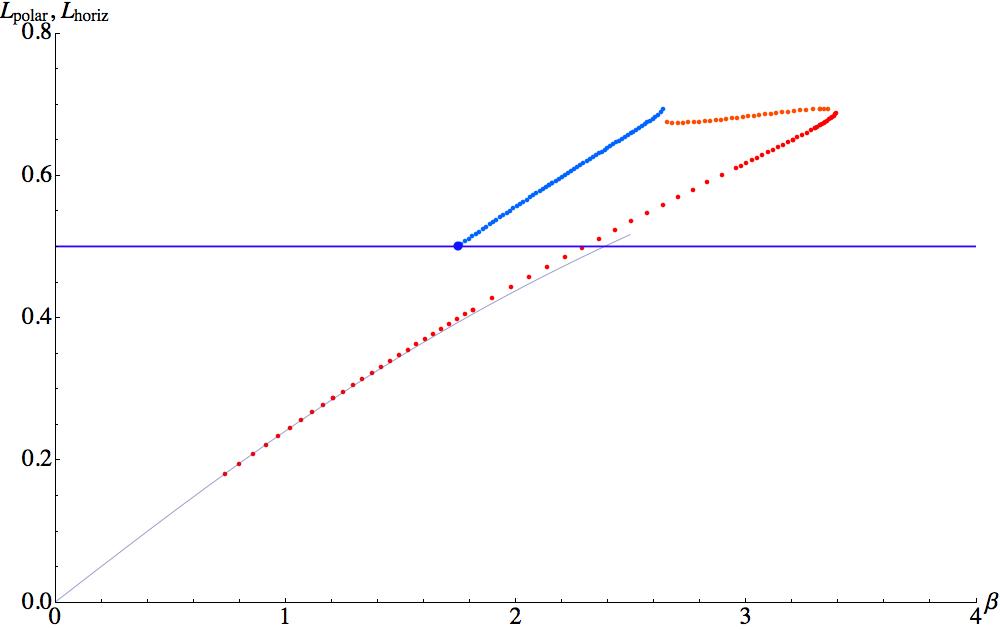}}
\centerline{\includegraphics[width=3.5in,height=2.5in]{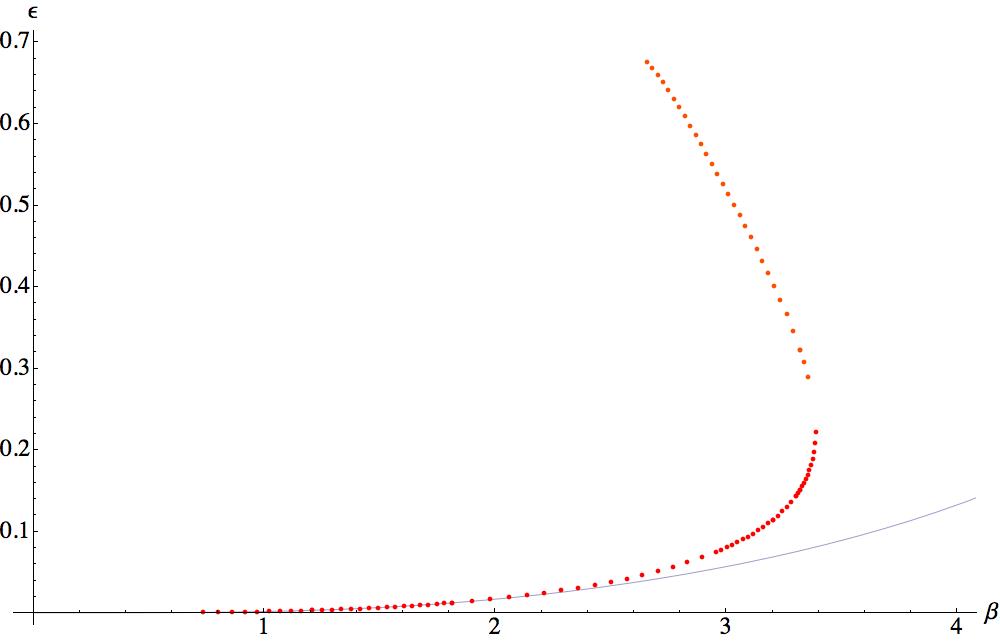}}
\caption{
Aspects of the geometry of the various static Kaluza-Klein black holes. The same conventions are used as in previous figure. \emph{Top left:} Plot of equatorial horizon radius, $R_{eq} $ for the small and large localized solutions and maximum horizon radius $R_{max} $ for the uniform and non-uniform strings against $\beta$. 
 \emph{Top right:} Plot of polar horizon length, $L_{polar}$,  for the localized solutions and horizon length, $L_{horiz}$, for the strings against $\beta$.  \emph{Bottom:} Plot of area eccentricity, $\epsilon$, for the small and large localized solutions against $\beta$.
}
\label{fig:geom}
\end{figure}

\begin{figure}[htbp]
\centerline{\includegraphics[width=3.5in,height=2.5in]{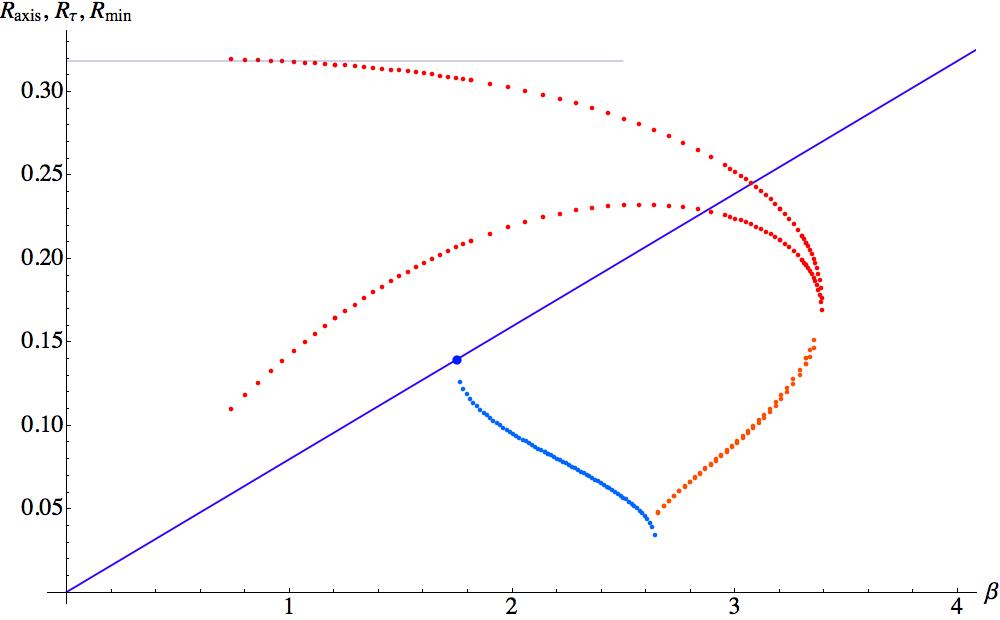}}
\caption{ 
For the small and large localized solutions we show the axis length, $L_{axis}$, and mirror plane time circle size, $R_{\tau} $, against $\beta $. For the strings we plot minimum horizon radius $R_{min} $. These curves characterize how near to the suspected `merger point' the solutions constructed reach. The same conventions are used as for previous figures. 
}
\label{fig:shrink}
\end{figure}

In figures \ref{fig:thermo}, \ref{fig:geom} and \ref{fig:shrink}, we show various how physical quantities vary against $\beta$ for the $5$-d localized and non-uniform strings. Data is shown for the highest resolution used, $80 \times 240$. Where available we give the perturbative results for localized solutions (computed up second order \cite{Karasik:2004ds}) which compare well up to $\beta  \sim 2$ where the perturbative expansion at different orders starts to differ significantly, signaling the non-perturbative regime. The limited data for $5$-d localized solutions previously existing agrees \cite{Kudoh:2004hs,Sorkin:2003ka}, and the previous calculations of $5$-d non-uniform strings also agree \cite{Kleihaus:2006ee,Sorkin:2006wp}.
In appendix  \ref{app:conv} we estimate the numerical error in the results to be of order $\sim 2 \%$ or less by considering data computed at different resolutions and for different boundary positions. In that appendix we also check that the solutions are consistent with being Ricci-flat rather than non-trivial Ricci solitons.

It was previously observed for $6$-d that such physical quantities were consistent with a merger of the localized and non-uniform branches \cite{Kudoh:2004hs}. Our $5$-dimensional data is indeed consistent with such a merger at $\beta_{merger}  \simeq 2.64$, where the behaviour of various quantities plotted is consistent with being continuous, although not smooth. The figures show agreement in all quantities, taking into account the percent level systematic errors.
We find experimentally that increasing the resolution allows solutions to be constructed closer to this potential merger point, with $80 \times 240$ being sufficient to approach quite closely. We note that we are able to find localized solutions closer to the merger point than previous numerical approaches which employed higher resolutions \cite{Kudoh:2004hs}.

In figure  \ref{fig:thermo} we show the various quantities related to thermodynamic behaviour. The free energy curve shows the expected first order phase transition between small localized solutions at high temperature, and uniform strings at low temperature, with the transition being at $\beta  \simeq 2.72$. The free energy shows a clear cusp behaviour at $\beta_{max}$. The mass, area and tension appear to have a smooth behaviour at $\beta_{max}$ and we return to this interesting point later. In figure \ref{fig:geom} we plot various geometric properties of the solutions and in figure \ref{fig:shrink} we show geometric quantities that characterize distance to a possible merger. For the non-uniform solutions distance from the merger can be characterized by the minimum radius of the horizon, $R_{min}$. The smallest value this reaches for our data corresponds to $\lambda \sim 3.6$, where $\lambda$ is the quantity used by Gubser to characterize the non-uniformness, $\lambda = ( R_{max} / R_{min} - 1)/2$ \cite{Gubser:2001ac}. On the localized side we may assess the distance from merger either by measuring the proper length of the symmetry axis $L_{axis}$, or by measuring the maximum radius of the Euclidean time circle on that axis which, for these solutions, is given by $R_{\tau}$. As we see from the figure both these shrink, where we have plotted $R_{axis} = L_{axis}/2 \pi$ so that this may be treated as a radius and all three quantities -- $R_{min}, R_{axis}$ and $R_{\tau}$ -- may be directly compared in value. We see that the localized branch is not found quite as near to the merger point as the strings, but still they are close. We note that $R_{axis} \sim R_{\tau}$ as they decrease, indicating that the 2-sphere formed by the fibration of the time circle over the symmetry axis is becoming round as it shrinks, which we would expect in the type of merger discussed by Kol \cite{Kol:2002xz}.

\begin{figure}[htbp]
\centerline{\includegraphics[width=3.5in,height=2.5in]{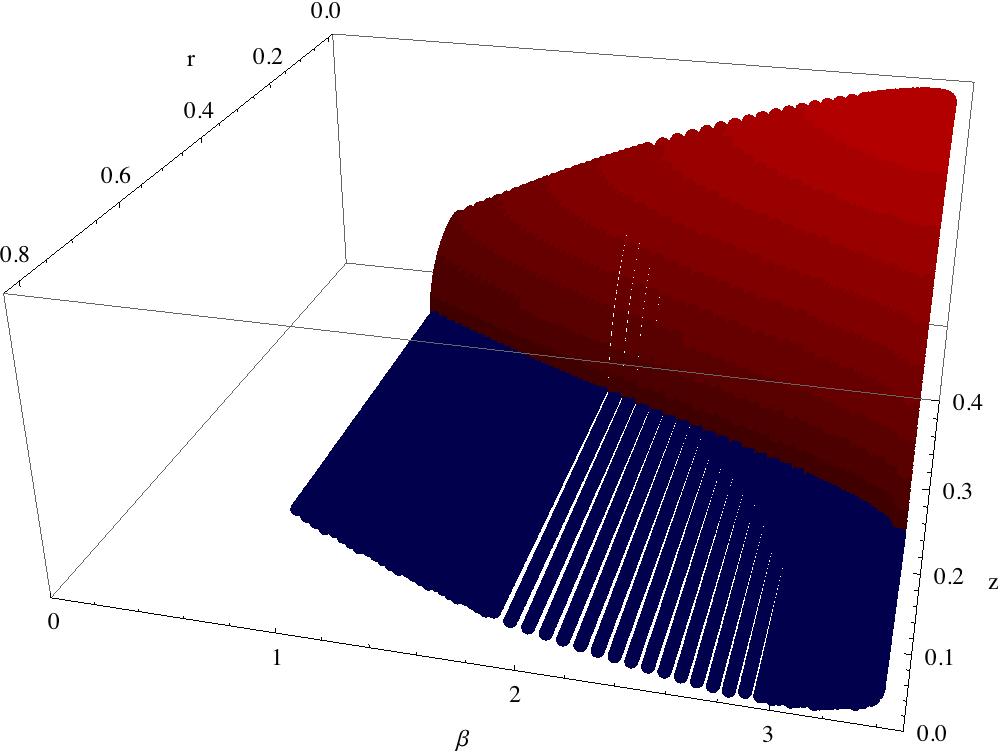}}
\centerline{\includegraphics[width=3.5in,height=2.5in]{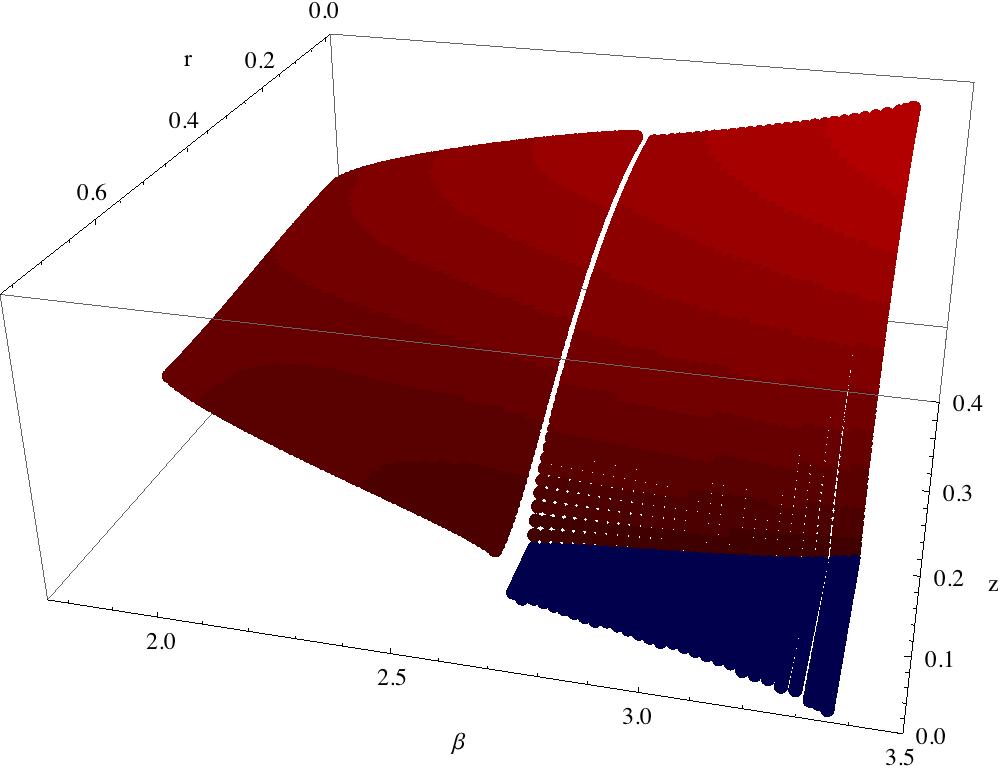}}
\caption{
Cross-sections of proper embeddings for the horizons of the various black holes against $\beta$ for $L = 1$. The upper frame shows the small localized solutions, and the lower frame the large localized and non-uniform string solutions. The horizon is shown in red, and for the localized solutions the axis of symmetry is also shown with its proper length in blue. 
}
\label{fig:embed}
\end{figure}

In figure \ref{fig:embed} we show a plot of the horizon embeddings as a function of $\beta$, and again see consistency with a merger, where we recall that there is percent level systematic error in the geometry of the embedding, and the value of $\beta$.  It would be interesting to attempt to check if the conical geometry proposed by Kol \cite{Kol:2002xz} near the merger point is emerging on the localized side, as has been checked on the non-uniform side \cite{Kol:2003ja,Sorkin:2006wp}. We leave this for future work.

\begin{figure}[htbp]
\centerline{\includegraphics[width=4.5in,height=3.5in]{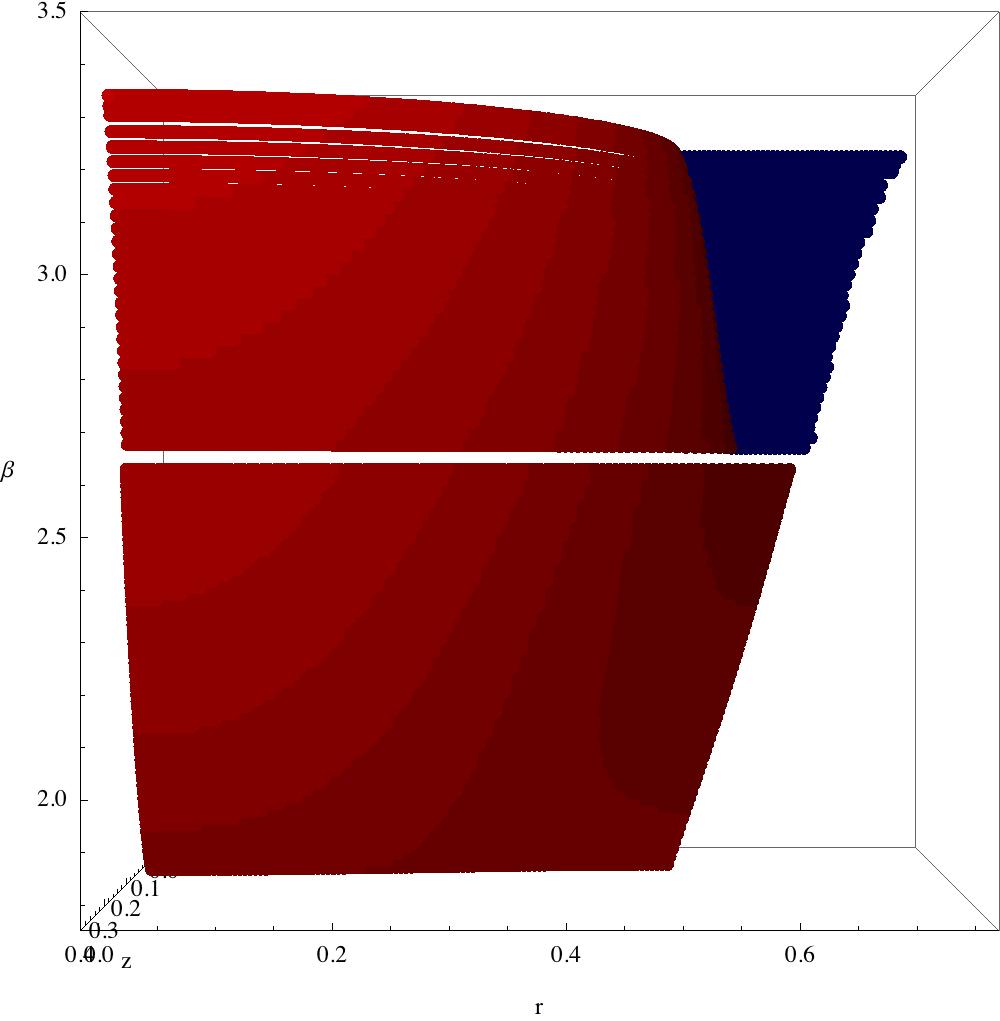}}
\caption{
This figure shows another view of the previous embedding figure. 
}
\end{figure}

\begin{figure}[htbp]
\centerline{\includegraphics[width=4.5in,height=3.in]{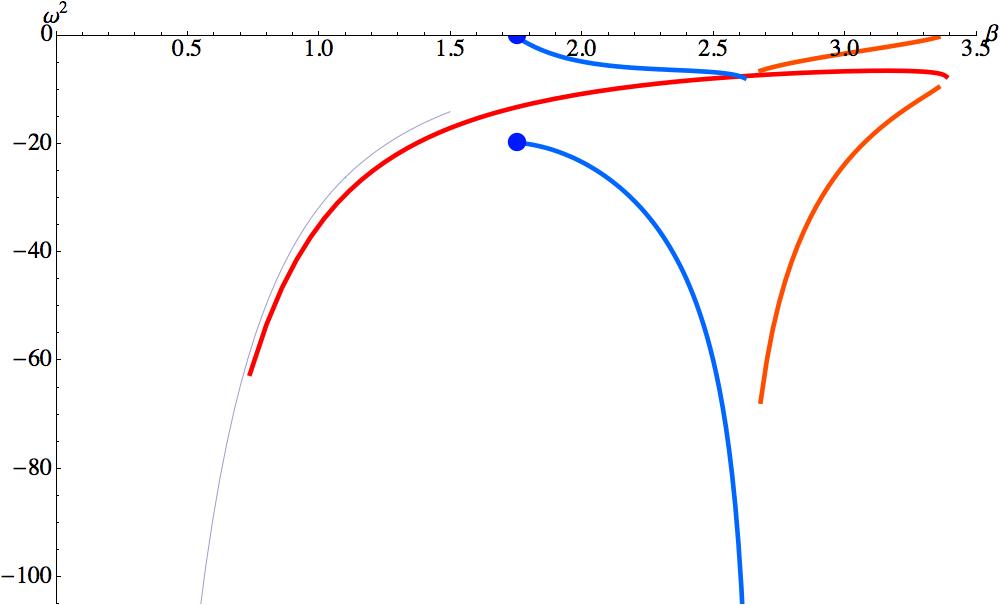}}
\caption{
The negative eigenvalues, $\omega^2$, of the operator $\triangle_L$ which are $U(1)\times SO(3)$ invariant are shown plotted against $\beta$ for small and large localized, and non-uniform string solutions. 
The solid discs show the results for a uniform string at the marginal Gregory-Laflamme point, where we have an zero mode and one negative mode. We observe that the zero mode continues to a negative mode. The line gives the leading order approximation for small localized solutions derived from the negative mode of $5$-d Schwarzschild (computed in \cite{Kol:2004pn}). 
}
\label{fig:negevals}
\end{figure}

\begin{figure}[htbp]
\centerline{\includegraphics[width=4.5in,height=3.in]{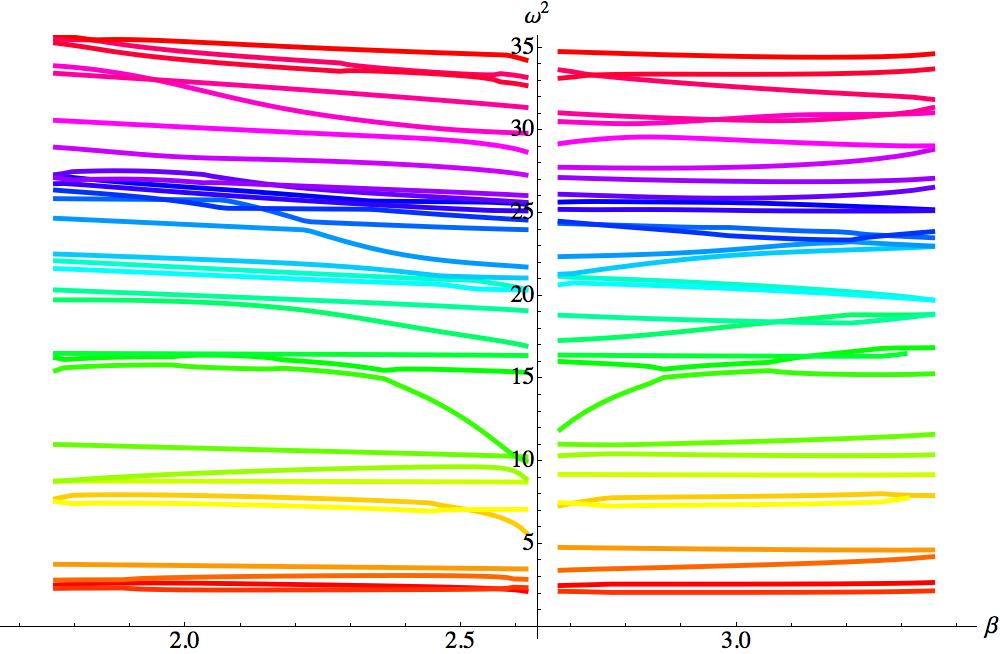}}
\caption{
The 30 lowest positive $U(1)\times SO(3)$ invariant eigenvalues of $\triangle_H$ for the large localized (left of the axis) and non-uniform solutions (right of the axis) in the region of the potential merger for asymptotic boundary truncation $R = 3 L$. This truncation discretizes the spectrum, which can then be used as a probe of the potential merger.
For both localized and non-uniform solutions the eigenvalues are coloured in the order given by their value. We observe that there appears to be a good correspondence between the spectra of the localized and non-uniform solutions near the merger, consistent with the spectra being continuous (although presumably not smooth) across the merger.
}
\label{fig:posevals}
\end{figure}

Since we are constructing the linearized operator $\triangle_H$, it is straightforward to find the low lying eigenvalues and functions of this sparse matrix. However, in practice we have utilized the isometries of the solution, and so only compute the operator $\triangle_H$ restricted to action on modes that are constant on the time circle and $2$-sphere. Hence we may only find eigenmodes that are singlets under the $U(1)\times SO(3)$ isometry group. In principle a good way to characterize the merger is to compute the spectrum of this operator and examine its behaviour on both sides of the merger point, as the low lying spectra provide a probe of the geometry globally.
Firstly we compute the negative eigenvalues showing these in figure \ref{fig:negevals}. As noted earlier in section \ref{sec:spectrum} these negative eigenvalues should correspond to negative eigenmodes of the operator $\triangle_L$. We see that as expected the small  localized solutions have one negative mode in the class of $U(1)\times SO(3)$ invariant eigenfunctions, and we find that the large localized solutions have two. At the join of these branches, the tangent to the space of solutions will precisely give a normalizable (ie. in our case, boundary condition preserving) zero mode, and it is this which continues to the second negative mode for the large localized branch. We find the non-uniform strings have two $U(1)\times SO(3)$ invariant negative modes. In principle additional negative modes might develop nearer to the merger point than we have accessed. However it is interesting that for the range of solutions explored here we see two $U(1)\times SO(3)$ invariant negative modes as the merger point is approached from both sides. In particular a new result is that over the range of $\beta$ tested we see no zero modes  develop on either branch that would correspond to that branch joining with some exotic new branch of static solutions with $U(1) \times SO(3)$ isometry. The only zero modes are the one at the Gregory-Laflamme marginal point, and that at $\beta = \beta_{max}$ which is understood simply as a maximum in $\beta$, and not as a mode that may be exponentiated to generate an independent branch of solutions. We see that one negative mode appears to become increasingly negative near the merger point and appears to be diverging to minus infinity -- presumably there is an associated critical behaviour that might be studied. The other negative eigenvalue appears to remain finite, and is consistent with a merger. 

In figure \ref{fig:posevals} we show the lowest 30 positive $U(1)\times SO(3)$ invariant  eigenmodes for the large localized and non-uniform solutions near the merger for $R = 3 L$. These modes correspond to physical eigenmodes (as opposed to short wavelength modes which are sensitive to the discretization) of the various operators discussed in section \ref{sec:spectrum}, namely $\triangle_H$ and $\triangle_V$. Without the asymptotic boundary regulator this spectrum would be continuous down to zero. However with the regulator $R = 3 L$ it becomes discrete and can be used as a global probe of the geometry, and in particular can be used to test for the potential merger. A minor point is that due to the finite boundary truncation and discretization a few of the modes have a small imaginary component in their eigenvalue, and thus it is the real part of the eigenvalue that we are plotting. We estimate the error  in the eigenvalues in this figure due to discretization to be quite small, of order $\pm 1$ in absolute value,  from comparison between different resolutions. We see a good correspondence of the spectra at the potential merger point, which again lends considerable weight to the idea that these branches merge. We note that assuming a merger occurs, it appears unlikely that any of these positive eigenmodes  have the opportunity to become negative closer to the merger point than we have probed on either side, although of course this possibility cannot be ruled out. 

\begin{figure}[htbp]
\centerline{\includegraphics[width=3.5in,height=2.5in]{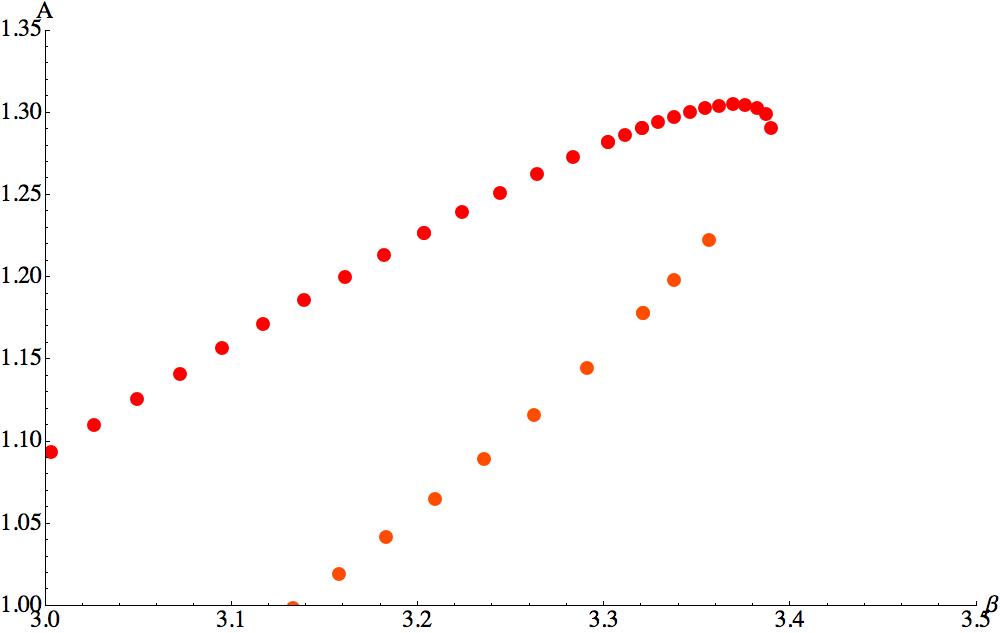}\includegraphics[width=3.5in,height=2.5in]{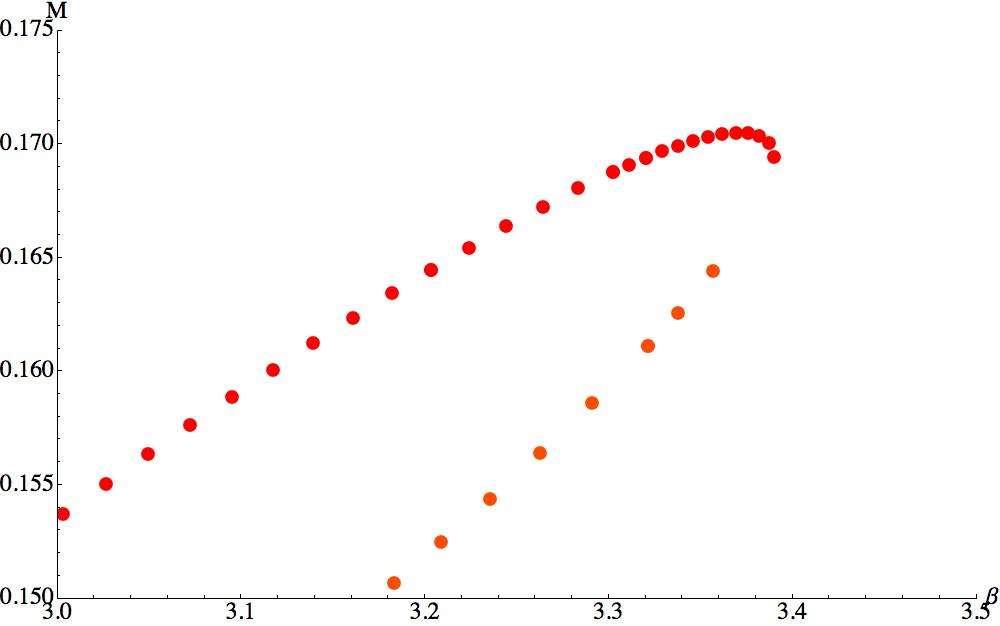}}
\centerline{\includegraphics[width=3.5in,height=2.5in]{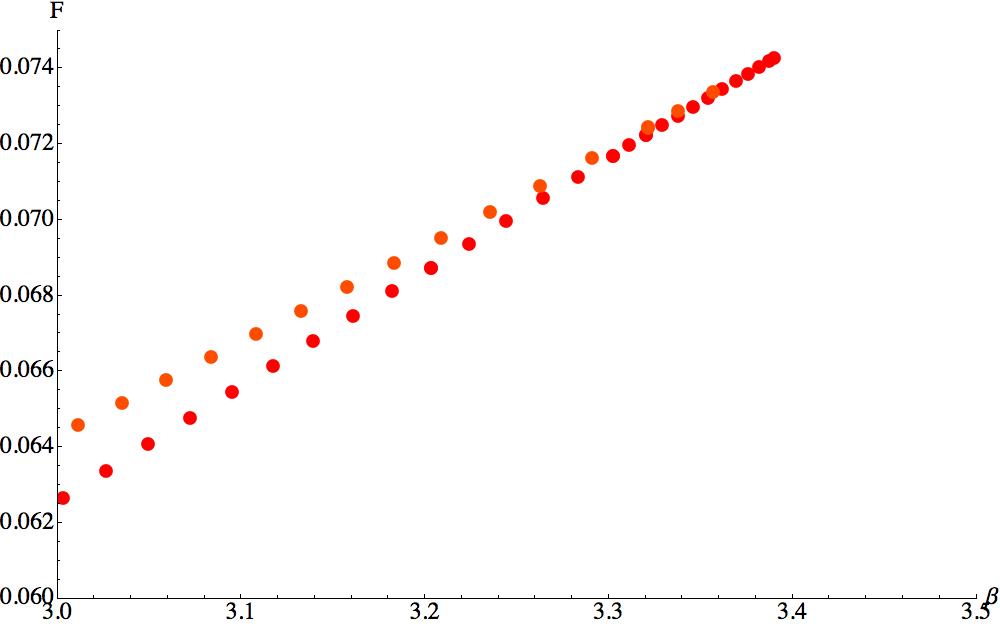}}
\caption{
Plots of horizon area (top left), mass (top right) and free energy (bottom), which are reproductions of previous figures but focussing  on the join between the small and large localized solutions. In particular the mass and area curves appear smooth at the maximum $\beta$ point, and therefore indicate a positive specific heat for a narrow range of the small localized solutions. In contrast the free energy curve is cuspy at the maximum $\beta$.
}
\label{fig:specificheat}
\end{figure}

In order for the small and large localized branches to join smoothly -- and there is no reason to expect this join not to be smooth in  the space of metrics -- there must be a maximum in the mass at some $\beta \le \beta_{max}$. The question is whether the maximum lies exactly at $\beta_{max}$ or below it. This is a very interesting question as if the maximum lies exactly at $\beta_{max}$ then the specific heat capacity of all the solutions shown will be negative. Note that this would mean the mass against $\beta$ relation would be cuspy at $\beta_{max}$ -- as for example for the free energy -- but this does not imply a lack of smoothness in the space of metrics.
However, if it lies below it, then for some range of $\beta$ the localized solutions may have positive specific heat, as the gradient of energy with temperature -- ie. mass with inverse $\beta$ -- will be positive. 
From the previous figures we see that it is a close call. In figure \ref{fig:specificheat} we reproduce the area, mass and free energy plots focussed near the join of the small and large branches. 
The mass and area appear to be smooth at $\beta_{max}$, whereas the free energy, shown for contrast, clearly shows a cuspy behaviour.

From a purely geometric point of view it would have been surprising if the maximum in mass coincided exactly with a maximum in $\beta$. One might wonder then why this is true for free energy? This follows from the fact that the free energy is related to the action. 
At the maximum $\beta$ the tangent to the space of solutions precisely gives a normalizable zero mode. Note 
that the tangent to a solution branch always gives a zero mode, but not a normalizable (or boundary condition preserving) one, as it will generically result in a perturbation of $\beta$.
Since the Ricci-flatness condition derives from the variation of the Euclidean Einstein-Hilbert action (plus boundary terms), $I$, this implies that $I$ should be stationary at $\beta_{max}$ as moving along a solution branch will change $I$ only by terms involving the change in boundary conditions. It is precisely a normalizable zero mode that leaves $I$ invariant.  Since the free energy $F$ is simply related to $I$ by $I = \beta F$ for these solutions, we should expect it to be stationary at $\beta_{max}$ -- resulting in a cusp behaviour for $F$ against $\beta$ -- and this is indeed what we see. However there is no analogous argument that the mass should be stationary.

We see that actually the maximum mass lies on the small localized branch a little below the maximum  $\beta$ point. Let us denote this $\beta = \beta_{mass}$. Then we find that $\beta_{mass}/\beta_{max} \simeq 0.995 \pm 0.001$, where the error is estimated by comparing data for $20 \times 60$, $40 \times 120$ and $80 \times 240$ resolutions, each of which sees the same effect, and from comparison of $R = 3 L$, $R = 6 L$ and $R = 10 L$ data, which again all see this effect. This maximum mass point coincides with the maximum area point, as dictated by the First Law. 
Whilst the effect we see is robust to varying resolution and boundary truncation, the $\beta$ range from $\beta_{mass}$ to $\beta_{max}$ is rather small, and so one might worry whether our numerics are correctly resolving this detail sufficiently well. That said, it does appear that the curves for mass and area against $\beta$ -- in contrast to the free energy -- look rather smooth, and this alone implies $\beta_{mass} < \beta_{max}$. Clearly this is an issue for future work to confirm. However, if confirmed, this narrow window of small localized solutions, from $\beta_{mass}$ to $\beta_{max}$ would represent the first examples of static black holes that have positive specific heat, but contain normalizable negative modes. Whilst it has been shown that there is a relation between local thermodynamic instability (for static vacuum solutions, a negative specific heat) and the existence of negative modes \cite{Whiting:1988qr,Whiting:1988ge,Prestidge:1999uq,Reall:2001ag}, the solutions exhibited here represent a counter-example to any argument that suggests the converse might be true. This would be an interesting complement to the recent analytic arguments that AdS-Kerr provides a stationary counter-example \cite{Monteiro:2009tc}.

%
\section{Summary}
\label{sec:conclusion}
%

We have provided a new general approach to solving static and Euclidean Einstein equations in vacuum. Whilst we haven't included matter, the methods should generalize trivially, as the stress tensor terms will not change the ellipticity property of the Einstein-DeTurck equations, and the matter equations can be phrased as elliptic equations too \footnote{Scalar field equations will already be elliptic, and gauge fields can be made elliptic using a similar trick to that of DeTurck.}. Then the Ricci-flow method naturally generalizes to a diffusion flow of the Einstein-DeTurck-matter system, and Newton's method extends trivially.

Our method works for general cohomogenity metrics, although it allows isometries to be used to reduce the effective dimension of the problem. For a stable fixed point our Ricci-flow method simply reduces to DeTurck flow, used successfully previously to study Calabi-Yau and del Pezzo geometries. It would be interesting to return to these problems and try applying our Newton method to compare with Ricci-flow. However, the new feature of our approach is its ability to deal with Euclidean solutions with negative modes -- either by tuning data and using Ricci-flow, or by using Newton's method. We have seen that in the case of Kaluza-Klein black holes the latter approach provides the more practical route, and yields excellent results. However the Ricci-flow method does have the virtue of being a flow in the space of geometries, and can potentially be used to give global information about the space of solutions, and in particular a `numerical proof' of existence.

Our example application to Kaluza-Klein black holes has provided some interesting new results in itself;
\begin{itemize}
\item We have seen that the localized solutions have a minimum temperature and hence divide into the two branches we term `small' and `large'. 
\item We have seen consistency with a merger between the localized and non-uniform strings, having constructed localized solutions closer to the potential merger point than previously found. 
\item We have computed the Euclidean negative modes of these solutions which are constant on the time circle and 2-sphere, finding the small localized solutions have one, and the large localized and non-uniform solutions constructed have two. One of these two has finite eigenvalue at the potential merger, the other apparently diverges to minus infinity. 
\item Consideration of the low lying positive spectrum of $\triangle_H$ which is constant on the time circle and 2-sphere also shows consistency with a merger, and indicates that there are no additional zero modes likely near the merger point which preserve the $U(1) \times SO(3)$ isometry. The existence or not of additional zero modes preserving the isometries is important when considering the various merger scenarios \cite{Kol:2002xz} and global structure of the phase diagram -- for example when including multi-black hole configurations \cite{Dias:2007hg}. 
\item A more speculative but rather interesting result is our evidence for an intriguingly narrow window of small localized solutions with positive specific heat but which possess two normalizable Euclidean negative modes. Whilst local thermodynamic instability appears to imply the existence of negative modes \cite{Whiting:1988qr,Whiting:1988ge,Prestidge:1999uq,Reall:2001ag,Gregory:2001bd}, such solutions would then represent counterexamples to any reverse claim.
We caution that this window is surprisingly narrow and thus this is a somewhat subtle numerical issue that will require future work to confirm with confidence.
\end{itemize}

We began by stating that the problem of directly  finding general stationary, static and Euclidean solutions in a unified framework was a problem of interest. We have managed to find a way to unify the static and Euclidean approach for geometries with and without a horizon. However our method clearly does not easily generalize to the stationary case, where an analytic continuation to a (real) Euclidean geometry cannot be performed. Thus an interesting question for future work  is how to unify the stationary case with the other two.

\acknowledgments

We would like to thank Gustav Delius and the York mathematical physics group for a very interesting discussion on the Newton-Rapheson method. TW would like to thank the organizers of the Jerusalem 2007 workshop on ``Higher Dimensional GR'', the Niels Bohr institute 2008 workshop ``Mathematical Aspects of General Relativity'', and the Valencia 2008 workshop ``Quantum Black Holes, Braneworlds and Holography'' for providing a stimulating environment to develop these ideas. 
MH is supported by DOE grant No. DE-FG02-92ER40706. SK is supported by an STFC studentship, and TW by an STFC advanced fellowship and Halliday award.

\appendix

%
\section{Convergence and boundary tests}
\label{app:conv}
%

\begin{figure}[htbp]
\centerline{\includegraphics[width=3.5in,height=2.5in]{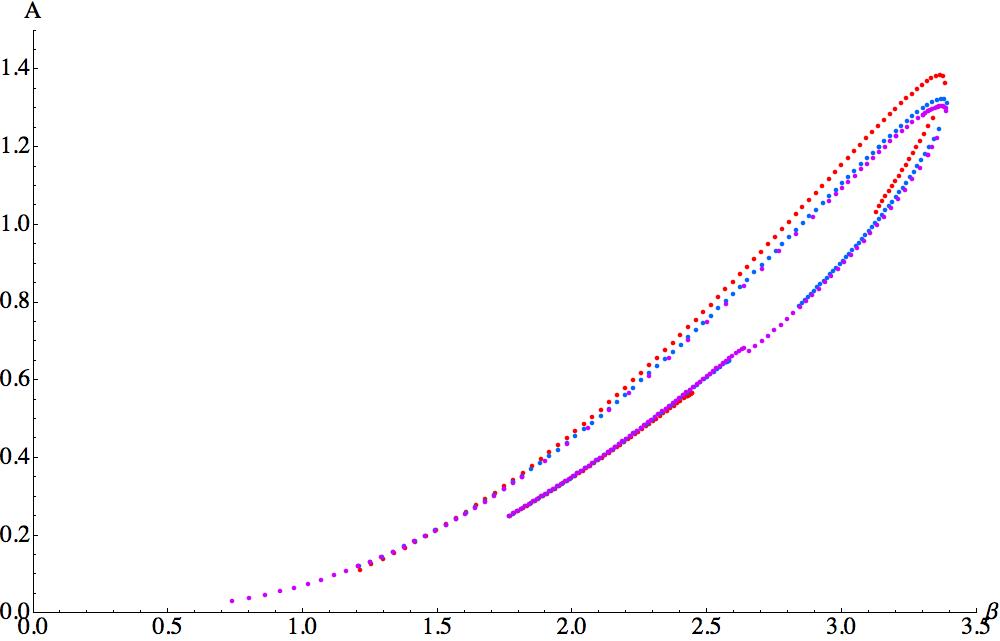}\includegraphics[width=3.5in,height=2.5in]{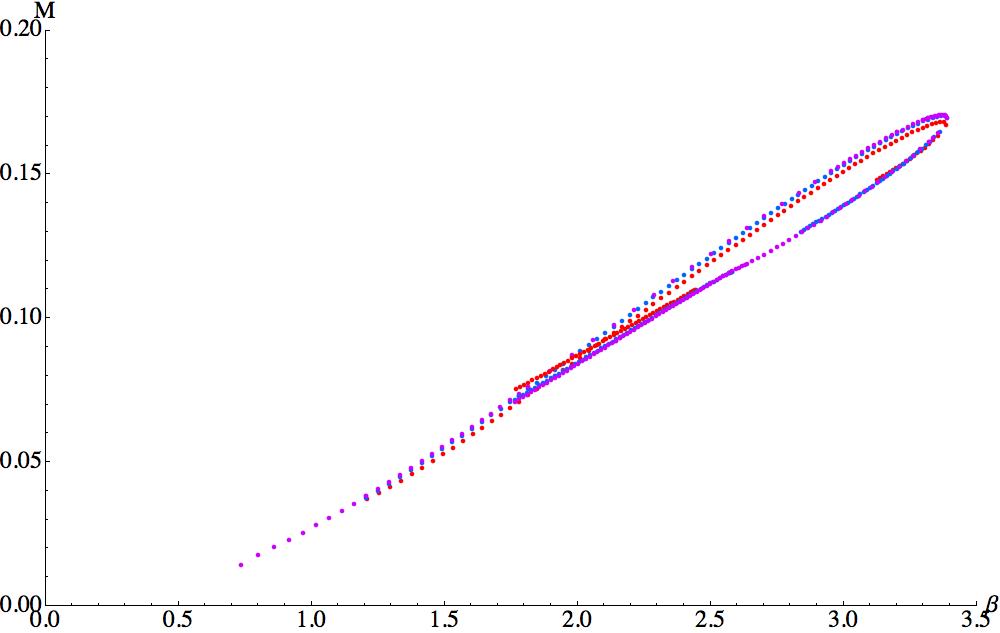}}
\caption{
Horizon area (left) and mass (right) for 3 resolutions $20 \times 60$ (red), $40 \times 120$ (blue) and $80 \times 240$ (purple), all with $R = 3 L$. Correct second order scaling to the continuum is observed, and the $80 \times 240$ data shown elsewhere in the paper is estimated to have a percent level finite resolution error. 
}
\label{fig:conv}
\end{figure}

Here we estimate finite resolution error and boundary truncation dependence. In figure \ref{fig:conv} we show the horizon area and mass curves giving data for 3 resolutions; $20 \times 60$, $40 \times 120$ and $80 \times 240$, all for $R = 3 L$. We see correct scaling to the continuum in these data -- and all other data shown in the paper -- and we estimate the error in the $80 \times 240$ data relative to the extrapolated continuum at approximately the percent level for the various quantities shown in this paper. We find that we may obtain non-uniform solutions and large localized solutions closer to the possible merger point as we go to higher resolution, and therefore it is likely that at a given resolution this error increases near to the merger point.

\begin{figure}[htbp]
\centerline{\includegraphics[width=3.5in,height=2.5in]{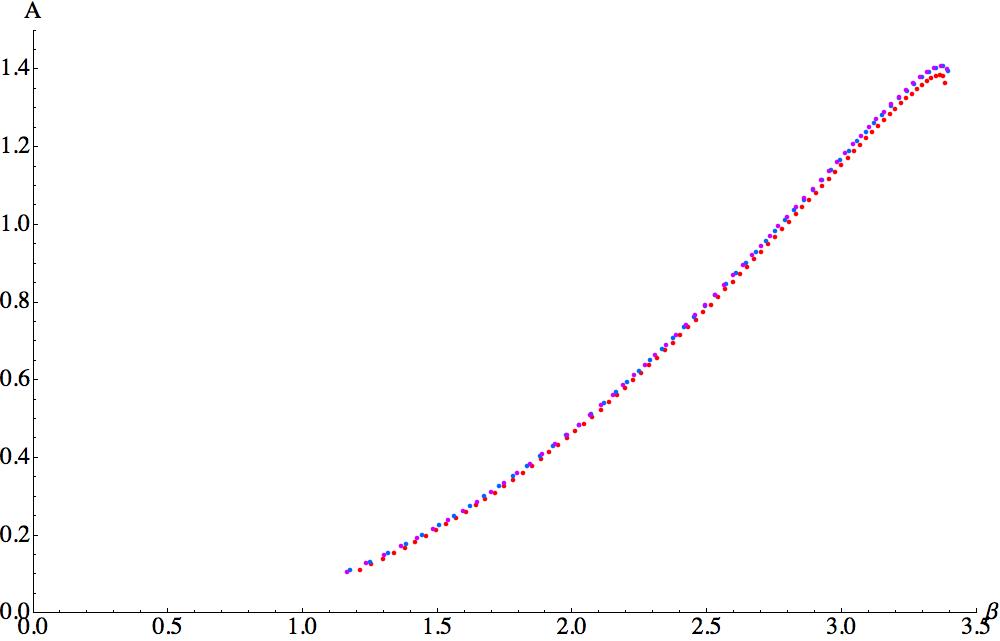}}
\caption{
Horizon area for 3 boundary truncations -- $R = 3 L$ (red), $R = 6 L$ (blue) and $R = 10 L$ (purple), all with resolution $20 \times 60$. We observe that already for $R = 3 L$ the error introduced by the boundary truncation appears to be small, of order $< 2 \%$. This is true for other observables studied.
}
\label{fig:asym}
\end{figure}

\begin{figure}[htbp]
\centerline{\includegraphics[width=3.5in,height=2.5in]{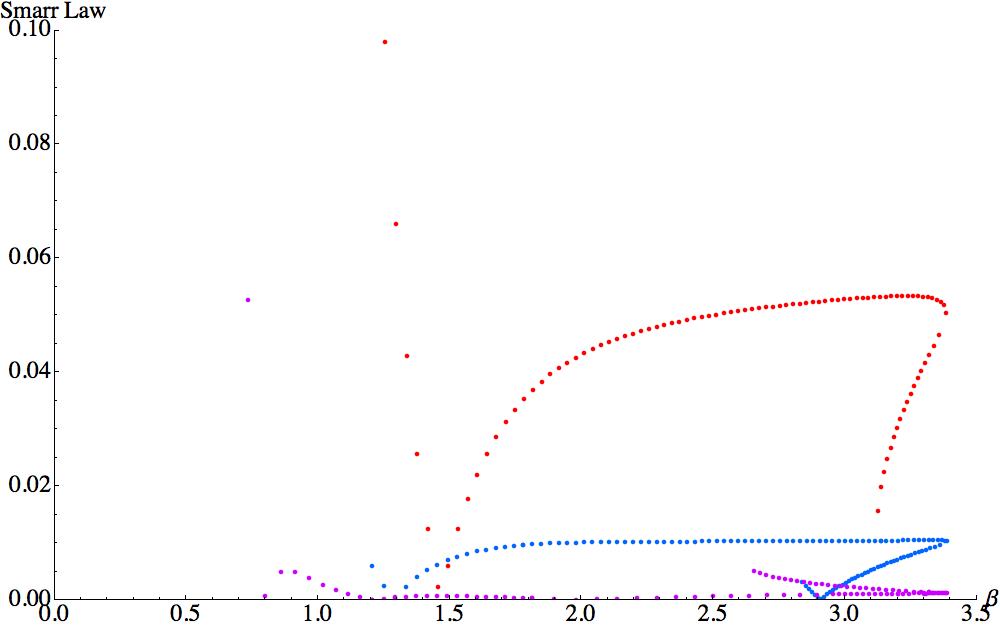},\includegraphics[width=3.5in,height=2.5in]{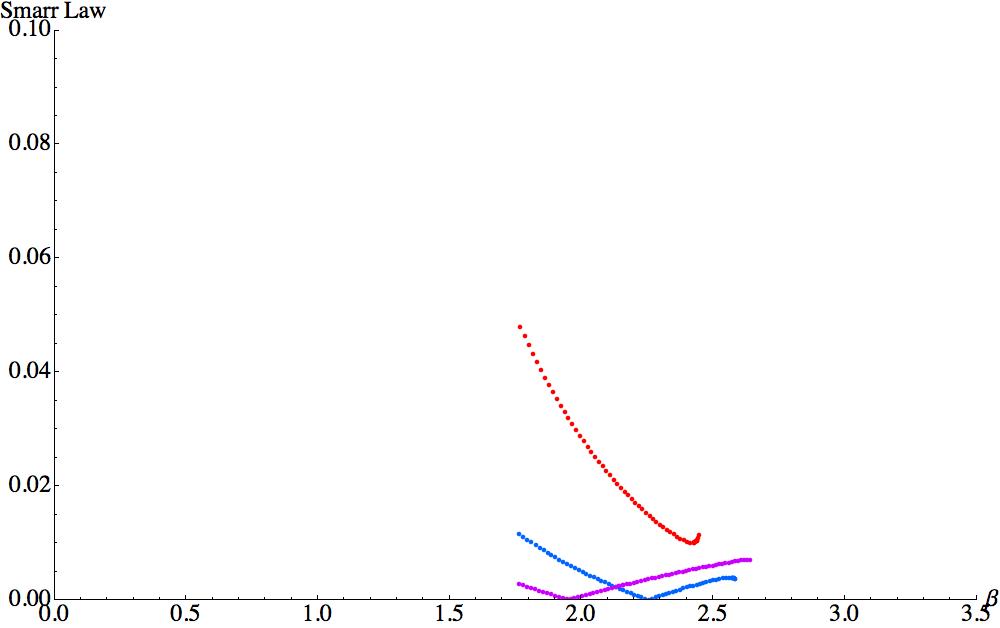}}
\caption{
Fractional violation of the Smarr relation for the localized solutions (left) and non-uniform strings (right) for 3 resolutions $20 \times 60$ (red), $40 \times 120$ (blue) and $80 \times 240$ (purple), all with $R = 3 L$.  
}
\label{fig:smarr}
\end{figure}

In figure \ref{fig:asym} we show the effect of changing the finite boundary truncation, giving the horizon area for $R = 3 L$ (the value used elsewhere in the paper), $R = 6 L$ and $R = 10 L$. We see that the systematic error in the $R = 3 L$ results are less than $2 \%$. We find the same behaviour for other measured quantities. The Smarr relation in equation \eqref{eq:smarr} allows another global check of error. In figure \ref{fig:smarr} we plot the fractional error in this relation for the localized black holes,
\begin{eqnarray}
\mathrm{Frac} \; \mathrm{Error}  & = & \frac{3 T S + \tau L}{2 M}
\end{eqnarray}
and see that for the $80 \times 240$ resolution this fractional error is less than $1 \%$ for all solutions found. We note that there are two contributions to error; the discretization error, and the finite boundary error. We expect that the discretization error will dominate, particularly for small localized solutions, and the data is consistent with this, with increasing resolution decreasing the error. For the $80 \times 240$ data, we expect that the major source of error is the finite boundary truncation, and for the non-uniform strings the error is at a similar level to that estimated from the finite truncation above. For the localized solutions the error is less than this, but we suspect this is coincidental.

\begin{figure}[htbp]
\centerline{\includegraphics[width=3.5in,height=2.5in]{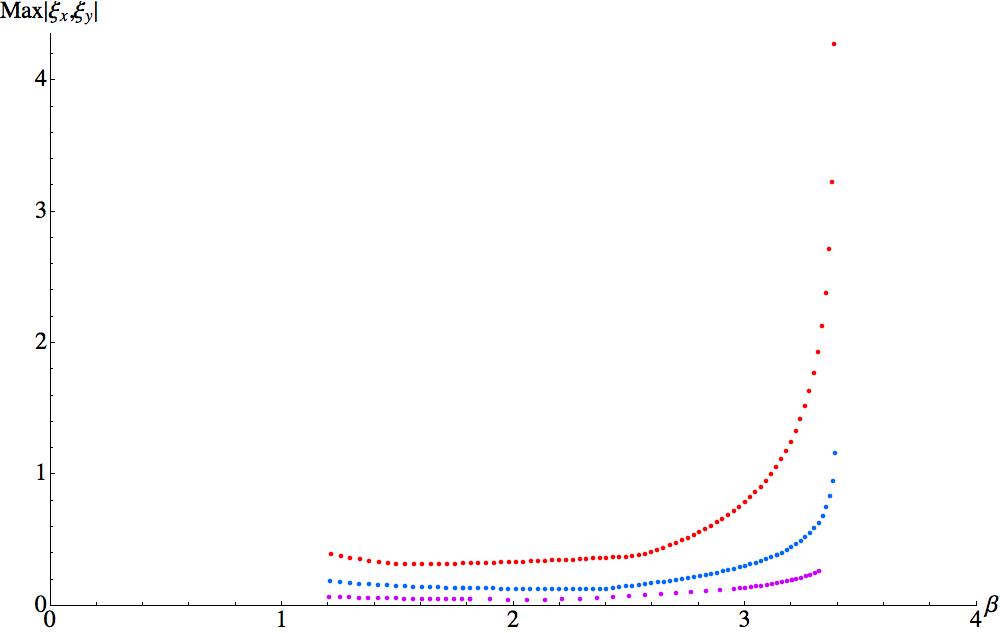}
\includegraphics[width=3.5in,height=2.5in]{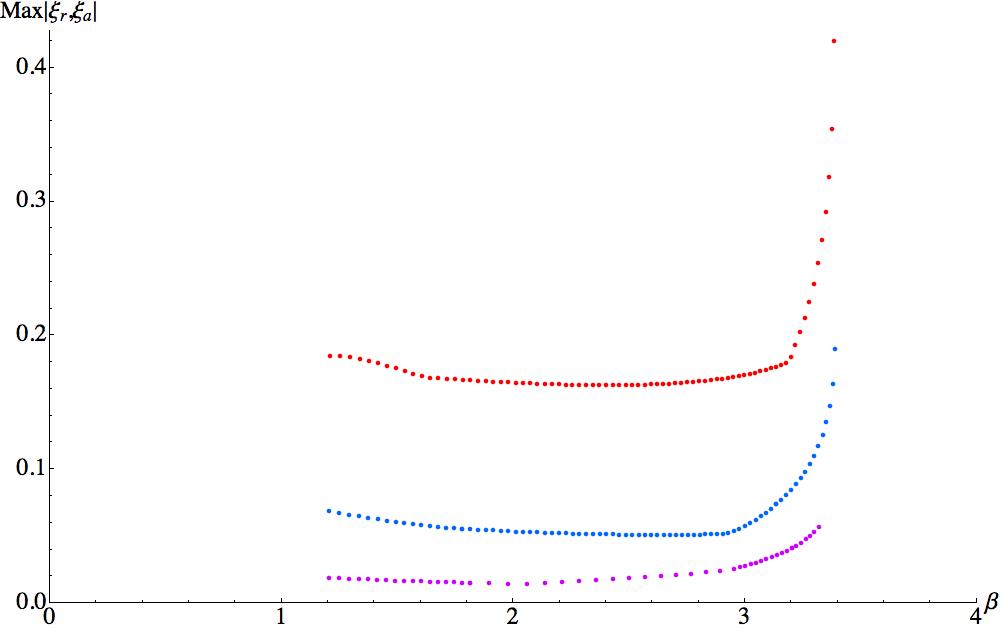}}
\caption{
Maximum absolute value of the vector $\xi$ for the small localized solutions for 
 3 resolutions $20 \times 60$ (red), $40 \times 120$ (blue) and $80 \times 240$ (purple), all with $R = 3 L$.  The left frame shows $\max{| \xi_x, \xi_y|}$ taken over the Cartesian chart, the right frame shows $\max{| \xi_r, \xi_a|}$ taken over the near horizon chart. We are interested in the behaviour of these maximum absolute values as the resolution increases -- the values themselves are not coordinate invariant quantities and therefore carry no significance. We see that at a given $\beta$ the maximum value of $|\xi_\mu|$ decreases in accordance with second order scaling to zero in the continuum. 
}
\label{fig:xi}
\end{figure}

Finally, in figure \ref{fig:xi} we show that the vector field $\xi$ is indeed consistent with vanishing in the continuum, confirming that we are finding a Ricci-flat solution, rather than a Ricci soliton. The maximum absolute value of the components of the vector field $\xi$ in the two charts is plotted for the small localized solutions for the 3 resolutions $20 \times 60$, $40 \times 120$ and $80 \times 240$. For all $\beta$ the maximum absolute value is seen to decrease consistent with second order scaling to zero in the continuum -- the values themselves, being coordinate dependent, are not physical. We see the same behaviour for the large localized and non-uniform strings.

%
\bibliographystyle{JHEP}
\bibliography{ref}

\providecommand{\href}[2]{#2}\begingroup\raggedright\begin{thebibliography}{10%
0}

\bibitem{ADD}
N.~Arkani-Hamed, S.~Dimopoulos, and G.~R. Dvali, {\it {The hierarchy problem
  and new dimensions at a millimeter}},  {\em Phys. Lett.} {\bf B429} (1998)
  263--272, [\href{http://xxx.lanl.gov/abs/hep-ph/9803315}{{\tt
  hep-ph/9803315}}].

\bibitem{ADD2}
I.~Antoniadis, N.~Arkani-Hamed, S.~Dimopoulos, and G.~R. Dvali, {\it {New
  dimensions at a millimeter to a Fermi and superstrings at a TeV}},  {\em
  Phys. Lett.} {\bf B436} (1998) 257--263,
  [\href{http://xxx.lanl.gov/abs/hep-ph/9804398}{{\tt hep-ph/9804398}}].

\bibitem{Randall:1999ee}
L.~Randall and R.~Sundrum, {\it {A large mass hierarchy from a small extra
  dimension}},  {\em Phys. Rev. Lett.} {\bf 83} (1999) 3370--3373,
  [\href{http://xxx.lanl.gov/abs/hep-ph/9905221}{{\tt hep-ph/9905221}}].

\bibitem{Randall:1999vf}
L.~Randall and R.~Sundrum, {\it {An alternative to compactification}},  {\em
  Phys. Rev. Lett.} {\bf 83} (1999) 4690--4693,
  [\href{http://xxx.lanl.gov/abs/hep-th/9906064}{{\tt hep-th/9906064}}].

\bibitem{AdSCFTreview}
O.~Aharony, S.~S. Gubser, J.~M. Maldacena, H.~Ooguri, and Y.~Oz, {\it {Large N
  field theories, string theory and gravity}},  {\em Phys. Rept.} {\bf 323}
  (2000) 183--386, [\href{http://xxx.lanl.gov/abs/hep-th/9905111}{{\tt
  hep-th/9905111}}].

\bibitem{Tangherlini:1963bw}
F.~R. Tangherlini, {\it {Schwarzschild field in n dimensions and the
  dimensionality of space problem}},  {\em Nuovo Cim.} {\bf 27} (1963)
  636--651.

\bibitem{Gibbons:2002av}
G.~W. Gibbons, D.~Ida, and T.~Shiromizu, {\it {Uniqueness and non-uniqueness of
  static black holes in higher dimensions}},  {\em Phys. Rev. Lett.} {\bf 89}
  (2002) 041101, [\href{http://xxx.lanl.gov/abs/hep-th/0206049}{{\tt
  hep-th/0206049}}].

\bibitem{Gibbons:2002bh}
G.~W. Gibbons, D.~Ida, and T.~Shiromizu, {\it {Uniqueness and non-uniqueness of
  static vacuum black holes in higher dimensions}},  {\em Prog. Theor. Phys.
  Suppl.} {\bf 148} (2003) 284--290,
  [\href{http://xxx.lanl.gov/abs/gr-qc/0203004}{{\tt gr-qc/0203004}}].

\bibitem{MP}
R.~C. Myers and M.~J. Perry, {\it {Black Holes in Higher Dimensional
  Space-Times}},  {\em Ann. Phys.} {\bf 172} (1986) 304.

\bibitem{Ring}
R.~Emparan and H.~S. Reall, {\it {A rotating black ring in five dimensions}},
  {\em Phys. Rev. Lett.} {\bf 88} (2002) 101101,
  [\href{http://xxx.lanl.gov/abs/hep-th/0110260}{{\tt hep-th/0110260}}].

\bibitem{Elvang:2007rd}
H.~Elvang and P.~Figueras, {\it {Black Saturn}},  {\em JHEP} {\bf 05} (2007)
  050, [\href{http://xxx.lanl.gov/abs/hep-th/0701035}{{\tt hep-th/0701035}}].

\bibitem{Elvang:2007hg}
H.~Elvang, R.~Emparan, and P.~Figueras, {\it {Phases of Five-Dimensional Black
  Holes}},  {\em JHEP} {\bf 05} (2007) 056,
  [\href{http://xxx.lanl.gov/abs/hep-th/0702111}{{\tt hep-th/0702111}}].

\bibitem{Emparan:2003sy}
R.~Emparan and R.~C. Myers, {\it {Instability of ultra-spinning black holes}},
  {\em JHEP} {\bf 09} (2003) 025,
  [\href{http://xxx.lanl.gov/abs/hep-th/0308056}{{\tt hep-th/0308056}}].

\bibitem{Emparan:2007wm}
R.~Emparan, T.~Harmark, V.~Niarchos, N.~A. Obers, and M.~J. Rodriguez, {\it
  {The Phase Structure of Higher-Dimensional Black Rings and Black Holes}},
  {\em JHEP} {\bf 10} (2007) 110,
  [\href{http://xxx.lanl.gov/abs/0708.2181}{{\tt 0708.2181}}].

\bibitem{Kol:2004ww}
B.~Kol, {\it {The phase transition between caged black holes and black strings:
  A review}},  {\em Phys. Rept.} {\bf 422} (2006) 119--165,
  [\href{http://xxx.lanl.gov/abs/hep-th/0411240}{{\tt hep-th/0411240}}].

\bibitem{Harmark:2005pp}
T.~Harmark and N.~A. Obers, {\it {Phases of Kaluza-Klein black holes: A brief
  review}},  \href{http://xxx.lanl.gov/abs/hep-th/0503020}{{\tt
  hep-th/0503020}}.

\bibitem{Harmark:2007md}
T.~Harmark, V.~Niarchos, and N.~A. Obers, {\it {Instabilities of black strings
  and branes}},  {\em Class. Quant. Grav.} {\bf 24} (2007) R1--R90,
  [\href{http://xxx.lanl.gov/abs/hep-th/0701022}{{\tt hep-th/0701022}}].

\bibitem{Hubeny:2002xn}
V.~E. Hubeny and M.~Rangamani, {\it {Unstable horizons}},  {\em JHEP} {\bf 05}
  (2002) 027, [\href{http://xxx.lanl.gov/abs/hep-th/0202189}{{\tt
  hep-th/0202189}}].

\bibitem{Delsate:2008kw}
T.~Delsate, {\it {Perturbative non uniform black strings in ${AdS}_6$}},  {\em
  Phys. Lett.} {\bf B663} (2008) 118--124,
  [\href{http://xxx.lanl.gov/abs/0802.1392}{{\tt 0802.1392}}].

\bibitem{Witten:1998zw}
E.~Witten, {\it {Anti-de Sitter space, thermal phase transition, and
  confinement in gauge theories}},  {\em Adv. Theor. Math. Phys.} {\bf 2}
  (1998) 505--532, [\href{http://xxx.lanl.gov/abs/hep-th/9803131}{{\tt
  hep-th/9803131}}].

\bibitem{Susskind:1997dr}
L.~Susskind, {\it {Matrix theory black holes and the Gross Witten transition}},
   \href{http://xxx.lanl.gov/abs/hep-th/9805115}{{\tt hep-th/9805115}}.

\bibitem{Li:1998jy}
M.~Li, E.~J. Martinec, and V.~Sahakian, {\it {Black holes and the SYM phase
  diagram}},  {\em Phys. Rev.} {\bf D59} (1999) 044035,
  [\href{http://xxx.lanl.gov/abs/hep-th/9809061}{{\tt hep-th/9809061}}].

\bibitem{Martinec:1998ja}
E.~J. Martinec and V.~Sahakian, {\it {Black holes and the SYM phase diagram.
  II}},  {\em Phys. Rev.} {\bf D59} (1999) 124005,
  [\href{http://xxx.lanl.gov/abs/hep-th/9810224}{{\tt hep-th/9810224}}].

\bibitem{Aharony:2004ig}
O.~Aharony, J.~Marsano, S.~Minwalla, and T.~Wiseman, {\it {Black hole-black
  string phase transitions in thermal 1+1- dimensional supersymmetric
  Yang-Mills theory on a circle}},  {\em Class. Quant. Grav.} {\bf 21} (2004)
  5169--5192, [\href{http://xxx.lanl.gov/abs/hep-th/0406210}{{\tt
  hep-th/0406210}}].

\bibitem{Harmark:2004ws}
T.~Harmark and N.~A. Obers, {\it {New phases of near-extremal branes on a
  circle}},  {\em JHEP} {\bf 09} (2004) 022,
  [\href{http://xxx.lanl.gov/abs/hep-th/0407094}{{\tt hep-th/0407094}}].

\bibitem{Aharony:2005bm}
O.~Aharony, S.~Minwalla, and T.~Wiseman, {\it {Plasma-balls in large N gauge
  theories and localized black holes}},  {\em Class. Quant. Grav.} {\bf 23}
  (2006) 2171--2210, [\href{http://xxx.lanl.gov/abs/hep-th/0507219}{{\tt
  hep-th/0507219}}].

\bibitem{Cardoso:2006ks}
V.~Cardoso and O.~J.~C. Dias, {\it {Gregory-Laflamme and Rayleigh-Plateau
  instabilities}},  {\em Phys. Rev. Lett.} {\bf 96} (2006) 181601,
  [\href{http://xxx.lanl.gov/abs/hep-th/0602017}{{\tt hep-th/0602017}}].

\bibitem{Caldarelli:2008mv}
M.~M. Caldarelli, O.~J.~C. Dias, R.~Emparan, and D.~Klemm, {\it {Black Holes as
  Lumps of Fluid}},  {\em JHEP} {\bf 04} (2009) 024,
  [\href{http://xxx.lanl.gov/abs/0811.2381}{{\tt 0811.2381}}].

\bibitem{star}
T.~Wiseman, {\it {Relativistic stars in Randall-Sundrum gravity}},  {\em Phys.
  Rev.} {\bf D65} (2002) 124007,
  [\href{http://xxx.lanl.gov/abs/hep-th/0111057}{{\tt hep-th/0111057}}].

\bibitem{Wiseman:2002zc}
T.~Wiseman, {\it {Static axisymmetric vacuum solutions and non-uniform black
  strings}},  {\em Class. Quant. Grav.} {\bf 20} (2003) 1137--1176,
  [\href{http://xxx.lanl.gov/abs/hep-th/0209051}{{\tt hep-th/0209051}}].

\bibitem{Kudoh:2003ki}
H.~Kudoh and T.~Wiseman, {\it {Properties of Kaluza-Klein black holes}},  {\em
  Prog. Theor. Phys.} {\bf 111} (2004) 475--507,
  [\href{http://xxx.lanl.gov/abs/hep-th/0310104}{{\tt hep-th/0310104}}].

\bibitem{Sorkin:2003ka}
E.~Sorkin, B.~Kol, and T.~Piran, {\it {Caged black holes: Black holes in
  compactified spacetimes. II: 5d numerical implementation}},  {\em Phys. Rev.}
  {\bf D69} (2004) 064032, [\href{http://xxx.lanl.gov/abs/hep-th/0310096}{{\tt
  hep-th/0310096}}].

\bibitem{Kleihaus:1996vi}
B.~Kleihaus and J.~Kunz, {\it {Static axially symmetric solutions of
  Einstein-Yang-Mills- dilaton theory}},  {\em Phys. Rev. Lett.} {\bf 78}
  (1997) 2527--2530, [\href{http://xxx.lanl.gov/abs/hep-th/9612101}{{\tt
  hep-th/9612101}}].

\bibitem{Kleihaus:1997ic}
B.~Kleihaus and J.~Kunz, {\it {Static black hole solutions with axial
  symmetry}},  {\em Phys. Rev. Lett.} {\bf 79} (1997) 1595--1598,
  [\href{http://xxx.lanl.gov/abs/gr-qc/9704060}{{\tt gr-qc/9704060}}].

\bibitem{Kleihaus:2000kg}
B.~Kleihaus and J.~Kunz, {\it {Rotating hairy black holes}},  {\em Phys. Rev.
  Lett.} {\bf 86} (2001) 3704--3707,
  [\href{http://xxx.lanl.gov/abs/gr-qc/0012081}{{\tt gr-qc/0012081}}].

\bibitem{Kudoh:2003xz}
H.~Kudoh, T.~Tanaka, and T.~Nakamura, {\it {Small localized black holes in
  braneworld: Formulation and numerical method}},  {\em Phys. Rev.} {\bf D68}
  (2003) 024035, [\href{http://xxx.lanl.gov/abs/gr-qc/0301089}{{\tt
  gr-qc/0301089}}].

\bibitem{Kudoh:2003vg}
H.~Kudoh, {\it {Thermodynamical properties of small localized black hole}},
  {\em Prog. Theor. Phys.} {\bf 110} (2004) 1059--1069,
  [\href{http://xxx.lanl.gov/abs/hep-th/0306067}{{\tt hep-th/0306067}}].

\bibitem{Kudoh:2004kf}
H.~Kudoh, {\it {6-dimensional localized black holes: Numerical solutions}},
  {\em Phys. Rev.} {\bf D69} (2004) 104019,
  [\href{http://xxx.lanl.gov/abs/hep-th/0401229}{{\tt hep-th/0401229}}].

\bibitem{Sorkin:2006wp}
E.~Sorkin, {\it {Non-uniform black strings in various dimensions}},  {\em Phys.
  Rev.} {\bf D74} (2006) 104027,
  [\href{http://xxx.lanl.gov/abs/gr-qc/0608115}{{\tt gr-qc/0608115}}].

\bibitem{Kleihaus:2006ee}
B.~Kleihaus, J.~Kunz, and E.~Radu, {\it {New nonuniform black string
  solutions}},  {\em JHEP} {\bf 06} (2006) 016,
  [\href{http://xxx.lanl.gov/abs/hep-th/0603119}{{\tt hep-th/0603119}}].

\bibitem{Kleihaus:2007cf}
B.~Kleihaus and J.~Kunz, {\it {Interior of Nonuniform Black Strings}},  {\em
  Phys. Lett.} {\bf B664} (2008) 210--213,
  [\href{http://xxx.lanl.gov/abs/0710.1726}{{\tt 0710.1726}}].

\bibitem{Yoshino:2008rx}
H.~Yoshino, {\it {On the existence of a static black hole on a brane}},  {\em
  JHEP} {\bf 01} (2009) 068, [\href{http://xxx.lanl.gov/abs/0812.0465}{{\tt
  0812.0465}}].

\bibitem{Delsate:2009bd}
T.~Delsate, {\it {Non Uniform Black Strings and Critical Dimensions in
  $AdS_d$}},  \href{http://xxx.lanl.gov/abs/0904.2149}{{\tt 0904.2149}}.

\bibitem{Headrick:2005ch}
M.~Headrick and T.~Wiseman, {\it {Numerical Ricci-flat metrics on K3}},  {\em
  Class. Quant. Grav.} {\bf 22} (2005) 4931--4960,
  [\href{http://xxx.lanl.gov/abs/hep-th/0506129}{{\tt hep-th/0506129}}].

\bibitem{Donaldson}
S.~Donaldson, {\it Some numerical results in complex differential geometry},
  \href{http://xxx.lanl.gov/abs/math.DG/0512625}{{\tt math.DG/0512625}}.

\bibitem{Douglas:2006rr}
M.~R. Douglas, R.~L. Karp, S.~Lukic, and R.~Reinbacher, {\it {Numerical
  Calabi-Yau metrics}},  {\em J. Math. Phys.} {\bf 49} (2008) 032302,
  [\href{http://xxx.lanl.gov/abs/hep-th/0612075}{{\tt hep-th/0612075}}].

\bibitem{Douglas:2006hz}
M.~R. Douglas, R.~L. Karp, S.~Lukic, and R.~Reinbacher, {\it {Numerical
  solution to the hermitian Yang-Mills equation on the Fermat quintic}},  {\em
  JHEP} {\bf 12} (2007) 083,
  [\href{http://xxx.lanl.gov/abs/hep-th/0606261}{{\tt hep-th/0606261}}].

\bibitem{Braun:2008jp}
V.~Braun, T.~Brelidze, M.~R. Douglas, and B.~A. Ovrut, {\it {Eigenvalues and
  Eigenfunctions of the Scalar Laplace Operator on Calabi-Yau Manifolds}},
  {\em JHEP} {\bf 07} (2008) 120,
  [\href{http://xxx.lanl.gov/abs/0805.3689}{{\tt 0805.3689}}].

\bibitem{Doran:2007zn}
C.~Doran, M.~Headrick, C.~P. Herzog, J.~Kantor, and T.~Wiseman, {\it {Numerical
  Kaehler-Einstein metric on the third del Pezzo}},  {\em Commun. Math. Phys.}
  {\bf 282} (2008) 357--393,
  [\href{http://xxx.lanl.gov/abs/hep-th/0703057}{{\tt hep-th/0703057}}].

\bibitem{Keller}
J.~Keller, {\it {Ricci iterations on Kahler classes}},  {\em Journal of the
  Institute of Mathematics of Jussieu} (2009)
  [\href{http://xxx.lanl.gov/abs/arXiv:0709.1490v2}{{\tt arXiv:0709.1490v2}}].

\bibitem{MR697987}
D.~M. DeTurck, {\it Deforming metrics in the direction of their {R}icci
  tensors},  {\em J. Differential Geom.} {\bf 18} (1983), no.~1 157--162.

\bibitem{Headrick:2006ti}
M.~Headrick and T.~Wiseman, {\it {Ricci flow and black holes}},  {\em Class.
  Quant. Grav.} {\bf 23} (2006) 6683--6708,
  [\href{http://xxx.lanl.gov/abs/hep-th/0606086}{{\tt hep-th/0606086}}].

\bibitem{Whiting:1988qr}
B.~F. Whiting and J.~York, James~W., {\it {Action Principle and Partition
  Function for the Gravitational Field in Black Hole Topologies}},  {\em Phys.
  Rev. Lett.} {\bf 61} (1988) 1336.

\bibitem{Whiting:1988ge}
B.~F. Whiting, {\it Black holes and thermodynamics},  {\em Class. Quant. Grav.}
  {\bf 7} (1990) 15--18.

\bibitem{Prestidge:1999uq}
T.~Prestidge, {\it {Dynamic and thermodynamic stability and negative modes in
  Schwarzschild-anti-de Sitter}},  {\em Phys. Rev.} {\bf D61} (2000) 084002,
  [\href{http://xxx.lanl.gov/abs/hep-th/9907163}{{\tt hep-th/9907163}}].

\bibitem{Reall:2001ag}
H.~S. Reall, {\it {Classical and thermodynamic stability of black branes}},
  {\em Phys. Rev.} {\bf D64} (2001) 044005,
  [\href{http://xxx.lanl.gov/abs/hep-th/0104071}{{\tt hep-th/0104071}}].

\bibitem{Gregory:2001bd}
J.~P. Gregory and S.~F. Ross, {\it {Stability and the negative mode for
  Schwarzschild in a finite cavity}},  {\em Phys. Rev.} {\bf D64} (2001)
  124006, [\href{http://xxx.lanl.gov/abs/hep-th/0106220}{{\tt
  hep-th/0106220}}].

\bibitem{Holzegel:2007zz}
G.~Holzegel, C.~Warnick, and T.~Schmelzer, {\it {Ricci flows connecting
  Taub-Bolt and Taub-NUT metrics}},  {\em Class. Quant. Grav.} {\bf 24} (2007)
  6201--6217.

\bibitem{Kudoh:2004hs}
H.~Kudoh and T.~Wiseman, {\it {Connecting black holes and black strings}},
  {\em Phys. Rev. Lett.} {\bf 94} (2005) 161102,
  [\href{http://xxx.lanl.gov/abs/hep-th/0409111}{{\tt hep-th/0409111}}].

\bibitem{Hawking:1982dh}
S.~W. Hawking and D.~N. Page, {\it {Thermodynamics of Black Holes in anti-De
  Sitter Space}},  {\em Commun. Math. Phys.} {\bf 87} (1983) 577.

\bibitem{York:1986it}
J.~York, James~W., {\it {Black hole thermodynamics and the Euclidean Einstein
  action}},  {\em Phys. Rev.} {\bf D33} (1986) 2092--2099.

\bibitem{Monteiro:2009tc}
R.~Monteiro, M.~J. Perry, and J.~E. Santos, {\it {Thermodynamic instability of
  rotating black holes}},  {\em Phys. Rev.} {\bf D80} (2009) 024041,
  [\href{http://xxx.lanl.gov/abs/0903.3256}{{\tt 0903.3256}}].

\bibitem{Wald}
R.~M. Wald, {\em General relativity}.
\newblock University of Chicago Press, 1984.

\bibitem{Garfinkle:2001ni}
D.~Garfinkle, {\it {Harmonic coordinate method for simulating generic
  singularities}},  {\em Phys. Rev.} {\bf D65} (2002) 044029,
  [\href{http://xxx.lanl.gov/abs/gr-qc/0110013}{{\tt gr-qc/0110013}}].

\bibitem{Pretorius:2005gq}
F.~Pretorius, {\it {Evolution of Binary Black Hole Spacetimes}},  {\em Phys.
  Rev. Lett.} {\bf 95} (2005) 121101,
  [\href{http://xxx.lanl.gov/abs/gr-qc/0507014}{{\tt gr-qc/0507014}}].

\bibitem{EellsLemaire}
J.~Eells and L.~Lemaire, {\it Another report on harmonic maps},  {\em Bull.
  London Math. Soc.} {\bf 20} (1988) 385--524.

\bibitem{EellsSampson}
J.~Eells and J.~Sampson, {\it Harmonic mappings of {R}iemannian manifolds},
  {\em Amer. J. Math.} {\bf 86} (1964) 109--160.

\bibitem{EellsBook}
J.~Eells, {\em Harmonic maps: selected papers of James Eells and
  collaborators}.
\newblock World Scientific, 1992.

\bibitem{Perelman1}
G.~Perelman, {\it The entropy formula for the {R}icci flow and its geometric
  applications},  \href{http://xxx.lanl.gov/abs/math.DG/0211159}{{\tt
  math.DG/0211159}}.

\bibitem{MR1145263}
N.~Koiso, {\it On rotationally symmetric {H}amilton's equation for
  {K}\"ahler-{E}instein metrics},  in {\em Recent topics in differential and
  analytic geometry}, vol.~18 of {\em Adv. Stud. Pure Math.}, pp.~327--337.
\newblock Academic Press, Boston, MA, 1990.

\bibitem{MR2084775}
X.-J. Wang and X.~Zhu, {\it K\"ahler-{R}icci solitons on toric manifolds with
  positive first {C}hern class},  {\em Adv. Math.} {\bf 188} (2004), no.~1
  87--103.

\bibitem{MR2243675}
H.-D. Cao, {\it Geometry of {R}icci solitons},  {\em Chinese Ann. Math. Ser. B}
  {\bf 27} (2006), no.~2 121--142.

\bibitem{HeadrickdP2}
M.~Headrick and T.~Wiseman, {\it Numerical {K\"a}hler-{R}icci soliton on the
  second del {P}ezzo},  \href{http://xxx.lanl.gov/abs/0706.2329 [math.DG]}{{\tt
  0706.2329 [math.DG]}}.

\bibitem{GPY}
D.~J. Gross, M.~J. Perry, and L.~G. Yaffe, {\it {Instability of Flat Space at
  Finite Temperature}},  {\em Phys. Rev.} {\bf D25} (1982) 330--355.

\bibitem{Hori:2001ax}
K.~Hori and A.~Kapustin, {\it Duality of the fermionic 2d black hole and {N} =
  2 {L}iouville theory as mirror symmetry},  {\em JHEP} {\bf 08} (2001) 045,
  [\href{http://xxx.lanl.gov/abs/hep-th/0104202}{{\tt hep-th/0104202}}].

\bibitem{Garfinkle:2003an}
D.~Garfinkle and J.~Isenberg, {\it {Critical behavior in Ricci flow}},
  \href{http://xxx.lanl.gov/abs/math/0306129}{{\tt math/0306129}}.

\bibitem{MR2115754}
D.~Garfinkle and J.~Isenberg, {\it Numerical studies of the behavior of {R}icci
  flow},  in {\em Geometric evolution equations}, vol.~367 of {\em Contemp.
  Math.}, pp.~103--114.
\newblock Amer. Math. Soc., Providence, RI, 2005.

\bibitem{Holzegel:2007ud}
G.~Holzegel, T.~Schmelzer, and C.~Warnick, {\it {Ricci Flow of Biaxial Bianchi
  IX Metrics}},  \href{http://xxx.lanl.gov/abs/0706.1694}{{\tt 0706.1694}}.

\bibitem{TianZhuConvergence}
G.~Tian and X.~Zhu, {\it Convergence of {K}\"ahler-{R}icci flow},  {\em J.
  Amer. Math. Soc.} {\bf 20} (2007) 675--699.

\bibitem{Harmark:2003yz}
T.~Harmark, {\it {Small black holes on cylinders}},  {\em Phys. Rev.} {\bf D69}
  (2004) 104015, [\href{http://xxx.lanl.gov/abs/hep-th/0310259}{{\tt
  hep-th/0310259}}].

\bibitem{Gorbonos:2004uc}
D.~Gorbonos and B.~Kol, {\it {A dialogue of multipoles: Matched asymptotic
  expansion for caged black holes}},  {\em JHEP} {\bf 06} (2004) 053,
  [\href{http://xxx.lanl.gov/abs/hep-th/0406002}{{\tt hep-th/0406002}}].

\bibitem{Karasik:2004ds}
D.~Karasik, C.~Sahabandu, P.~Suranyi, and L.~C.~R. Wijewardhana, {\it {Analytic
  approximation to 5 dimensional black holes with one compact dimension}},
  {\em Phys. Rev.} {\bf D71} (2005) 024024,
  [\href{http://xxx.lanl.gov/abs/hep-th/0410078}{{\tt hep-th/0410078}}].

\bibitem{Gorbonos:2005px}
D.~Gorbonos and B.~Kol, {\it {Matched asymptotic expansion for caged black
  holes: Regularization of the post-Newtonian order}},  {\em Class. Quant.
  Grav.} {\bf 22} (2005) 3935--3960,
  [\href{http://xxx.lanl.gov/abs/hep-th/0505009}{{\tt hep-th/0505009}}].

\bibitem{Chu:2006ce}
Y.-Z. Chu, W.~D. Goldberger, and I.~Z. Rothstein, {\it {Asymptotics of
  d-dimensional Kaluza-Klein black holes: Beyond the newtonian approximation}},
   {\em JHEP} {\bf 03} (2006) 013,
  [\href{http://xxx.lanl.gov/abs/hep-th/0602016}{{\tt hep-th/0602016}}].

\bibitem{Kol:2007rx}
B.~Kol and M.~Smolkin, {\it {Classical Effective Field Theory and Caged Black
  Holes}},  {\em Phys. Rev.} {\bf D77} (2008) 064033,
  [\href{http://xxx.lanl.gov/abs/0712.2822}{{\tt 0712.2822}}].

\bibitem{GL}
R.~Gregory and R.~Laflamme, {\it {Black strings and p-branes are unstable}},
  {\em Phys. Rev. Lett.} {\bf 70} (1993) 2837--2840,
  [\href{http://xxx.lanl.gov/abs/hep-th/9301052}{{\tt hep-th/9301052}}].

\bibitem{Gubser:2001ac}
S.~S. Gubser, {\it {On non-uniform black branes}},  {\em Class. Quant. Grav.}
  {\bf 19} (2002) 4825--4844,
  [\href{http://xxx.lanl.gov/abs/hep-th/0110193}{{\tt hep-th/0110193}}].

\bibitem{Kol:2002xz}
B.~Kol, {\it {Topology change in general relativity and the black-hole
  black-string transition}},  {\em JHEP} {\bf 10} (2005) 049,
  [\href{http://xxx.lanl.gov/abs/hep-th/0206220}{{\tt hep-th/0206220}}].

\bibitem{Harmark:2002tr}
T.~Harmark and N.~A. Obers, {\it {Black holes on cylinders}},  {\em JHEP} {\bf
  05} (2002) 032, [\href{http://xxx.lanl.gov/abs/hep-th/0204047}{{\tt
  hep-th/0204047}}].

\bibitem{Kol:2005vy}
B.~Kol, {\it {Choptuik scaling and the merger transition}},  {\em JHEP} {\bf
  10} (2006) 017, [\href{http://xxx.lanl.gov/abs/hep-th/0502033}{{\tt
  hep-th/0502033}}].

\bibitem{Asnin:2006ip}
V.~Asnin, B.~Kol, and M.~Smolkin, {\it {Analytic evidence for continuous self
  similarity of the critical merger solution}},  {\em Class. Quant. Grav.} {\bf
  23} (2006) 6805--6827, [\href{http://xxx.lanl.gov/abs/hep-th/0607129}{{\tt
  hep-th/0607129}}].

\bibitem{Wiseman:2002ti}
T.~Wiseman, {\it {From black strings to black holes}},  {\em Class. Quant.
  Grav.} {\bf 20} (2003) 1177--1186,
  [\href{http://xxx.lanl.gov/abs/hep-th/0211028}{{\tt hep-th/0211028}}].

\bibitem{Kol:2003ja}
B.~Kol and T.~Wiseman, {\it {Evidence that highly non-uniform black strings
  have a conical waist}},  {\em Class. Quant. Grav.} {\bf 20} (2003)
  3493--3504, [\href{http://xxx.lanl.gov/abs/hep-th/0304070}{{\tt
  hep-th/0304070}}].

\bibitem{Traschen:2001pb}
J.~H. Traschen and D.~Fox, {\it {Tension perturbations of black brane
  spacetimes}},  {\em Class. Quant. Grav.} {\bf 21} (2004) 289--306,
  [\href{http://xxx.lanl.gov/abs/gr-qc/0103106}{{\tt gr-qc/0103106}}].

\bibitem{Townsend:2001rg}
P.~K. Townsend and M.~Zamaklar, {\it {The first law of black brane mechanics}},
   {\em Class. Quant. Grav.} {\bf 18} (2001) 5269--5286,
  [\href{http://xxx.lanl.gov/abs/hep-th/0107228}{{\tt hep-th/0107228}}].

\bibitem{Kol:2003if}
B.~Kol, E.~Sorkin, and T.~Piran, {\it {Caged black holes: Black holes in
  compactified spacetimes. I: Theory}},  {\em Phys. Rev.} {\bf D69} (2004)
  064031, [\href{http://xxx.lanl.gov/abs/hep-th/0309190}{{\tt
  hep-th/0309190}}].

\bibitem{Harmark:2003dg}
T.~Harmark and N.~A. Obers, {\it {New phase diagram for black holes and strings
  on cylinders}},  {\em Class. Quant. Grav.} {\bf 21} (2004) 1709,
  [\href{http://xxx.lanl.gov/abs/hep-th/0309116}{{\tt hep-th/0309116}}].

\bibitem{Tanaka:2002rb}
T.~Tanaka, {\it {Classical black hole evaporation in Randall-Sundrum infinite
  braneworld}},  {\em Prog. Theor. Phys. Suppl.} {\bf 148} (2003) 307--316,
  [\href{http://xxx.lanl.gov/abs/gr-qc/0203082}{{\tt gr-qc/0203082}}].

\bibitem{Emparan:2002px}
R.~Emparan, A.~Fabbri, and N.~Kaloper, {\it {Quantum black holes as holograms
  in AdS braneworlds}},  {\em JHEP} {\bf 08} (2002) 043,
  [\href{http://xxx.lanl.gov/abs/hep-th/0206155}{{\tt hep-th/0206155}}].

\bibitem{Emparan:2002jp}
R.~Emparan, J.~Garcia-Bellido, and N.~Kaloper, {\it {Black hole astrophysics in
  AdS braneworlds}},  {\em JHEP} {\bf 01} (2003) 079,
  [\href{http://xxx.lanl.gov/abs/hep-th/0212132}{{\tt hep-th/0212132}}].

\bibitem{Kol:2004pn}
B.~Kol and E.~Sorkin, {\it {On black-brane instability in an arbitrary
  dimension}},  {\em Class. Quant. Grav.} {\bf 21} (2004) 4793--4804,
  [\href{http://xxx.lanl.gov/abs/gr-qc/0407058}{{\tt gr-qc/0407058}}].

\bibitem{Dias:2007hg}
O.~J.~C. Dias, T.~Harmark, R.~C. Myers, and N.~A. Obers, {\it {Multi-black hole
  configurations on the cylinder}},  {\em Phys. Rev.} {\bf D76} (2007) 104025,
  [\href{http://xxx.lanl.gov/abs/0706.3645}{{\tt 0706.3645}}].

\end{thebibliography}\endgroup
%

\end{document}